\documentclass[a4paper,12pt]{article}
\pdfoutput=1
\usepackage{graphicx,subfigure,amsmath,amssymb,multirow}
\usepackage{cite}

\newlength{\dinwidth}
\newlength{\dinmargin}
\def\be{\begin{equation}}
\def\ee{\end{equation}}
\def\ba{\begin{eqnarray}}
\def\ea{\end{eqnarray}}
 \def\la{ \langle}
  \def\ra{ \rangle}
     \def\e{ \epsilon}
      \def\r{ \gamma}
       \def\lbd{\lambda}
        \def \d {{\rm d}}
           \def\w{\omega} 
            \def\u{\mu}
              \def\a{\alpha}
  \def\b{\beta}
\def\v{\nu}
     \def\ve{ \varepsilon}  

\usepackage{color}
\def\tb{\textcolor{blue}}

\usepackage[utf8]{inputenc}
\usepackage[numbers,sort&compress]{natbib}

\allowdisplaybreaks

\begin{document}
\title{\vspace{-2.5cm} \bf \Large Self-consistency and covariance of light-front quark models: testing via $P$, $V$, and $A$ meson decay constants, and $P\to P$ weak transition form factors}

\author{Qin Chang$^{a,b}$\footnote{changqin@htu.edu.cn}, Xiao-Nan Li$^{a}$\footnote{lixnff@foxmail.com}, Xin-Qiang Li$^{b}$\footnote{xqli@mail.ccnu.edu.cn}, Fang Su$^{b}$\footnote{sufang@itp.ac.cn}, Ya-Dong Yang$^{b}$\footnote{yangyd@mail.ccnu.edu.cn}\\[0.2cm]
{ $^a$\small Institute of Particle and Nuclear Physics, Henan Normal University}\\[-0.2cm]
{ \small Henan 453007, People’s Republic of China}\\
{ $^b$\small Institute of Particle Physics and Key Laboratory of Quark and Lepton Physics~(MOE) }\\[-0.2cm]
{ \small Central China Normal University, Wuhan, Hubei 430079, China}}
\date{}
\maketitle

\begin{abstract}
\noindent In this paper, we test the self-consistencies of the standard and the covariant light-front quark model and study the zero-mode issue via the decay constants of pseudoscalar ($P$), vector ($V$) and axial-vector ($A$) mesons, as well as the $P\to P$ weak transition form factors. With the traditional type-I correspondence between the manifestly covariant and the light-front approach, the resulting $f_{V}$ as well as $f_{^1\!A}$ and $f_{^3\!A}$ obtained with the $\lbd=0$ and $\lbd=\pm$ polarization states are different from each other, which presents a challenge to the self-consistency of the covariant light-front quark model. However, such a  self-consistency problem can be ``resolved'' within the type-II scheme, which requires an additional replacement $M\to M_0$ relative to the type-I case. Moreover, the replacement $M\to M_0$ is also essential for the self-consistency of the standard light-front quark model. In the type-II scheme, the valence contributions to the physical quantities~(${\cal Q}$) considered in this paper are always the same as that obtained in the standard light-front quark model, $[{\cal Q}]_{\rm val.}=[{\cal Q}]_{\rm SLF}$, and the zero-mode contributions to $f_{V,^1\!A,^3\!A}$ and $f_-(q^2)$ exist only formally but vanish numerically, which further implies that $[{\cal Q}]_{\rm val.}\dot{=} [{\cal Q}]_{\rm full}$. In addition, the manifest covariance of the covariant light-front quark model is violated in the traditional type-I scheme, but can be recovered by taking the type-II correspondence.
\end{abstract}

\newpage

\section{INTRODUCTION}

The standard light-front~(SLF) quark model~\cite{Terentev:1976jk,Berestetsky:1977zk,Jaus:1989au,Jaus:1989av} based on the light-front~(LF) formalism~\cite{Brodsky:1997de} provides a conceptually simple but phenomenologically feasible framework for calculating  the non-perturbative quantities of hadrons, such as the decay constants, transition form factors, distribution amplitudes and so on~\cite{Jaus:1991cy,Jaus:1996np,Cheng:1996if,ODonnell:1996sya,Cheung:1996qt,Choi:1996mq,Choi:1997iq,Choi:1998jd,DeWitt:2003rs,Choi:2007yu,Choi:2007se,Barik:1997qq,Hwang:2000ez,Hwang:2010hw,Hwang:2001zd,Hwang:2009cu,Geng:2001de,Chang:2016ouf,Chang:2017sdl,Chang:2018aut,Chang:2018mva,Wang:2017mqp}. In the SLF approach, the constituent quark and antiquark in a bound-state are required to be on their respective mass-shells, the physical quantities are computed directly in three-dimensional LF momentum space, and the plus component~($\mu=+$, the so-called ``good'' component) of the current matrix elements is usually taken in order to avoid the zero-mode contribution. Obviously, the Lorentz covariance of the matrix elements obtained in the SLF quark model is lost. Moreover, the usual recipe to avoid the zero-mode contributions by taking the plus component is in fact always invalid for many cases, for instance, the composite spin-1 systems~\cite{Jaus:1999zv}. While the zero-mode issue is highly nontrivial and deserves careful analyses~\cite{Choi:1998nf}, the SLF quark model is powerless for determining the zero-mode contributions by itself. Because of these shortcomings, the SLF quark model was soon superseded by the manifestly covariant light-front~(CLF) quark model. 

The CLF quark model was firstly exploited by Jaus~\cite{Jaus:1999zv}, Choi and Ji~\cite{Choi:1998nf}, as well as Cheng \textit{et al.}~\cite{Cheng:1997au}, with the help of the manifestly covariant Bethe-Salpeter~(BS) approach~\cite{Salpeter:1951sz,Salpeter:1952ib}, and has been further studied in Refs.~\cite{Jaus:2002sv,Cheng:2003sm,Bakker:2000pk,Bakker:2002mt,Bakker:2003up,Choi:2005fj,Choi:2013mda}. Compared to the SLF approach, the CLF quark model is characterized by the following two distinguished features: it provides a systematic way to explore the zero-mode effects; the results obtained are guaranteed to be covariant after the spurious contribution proportional to the lightlike four-vector $\omega=(0,2,{0}_\bot)$ is canceled by the inclusion of zero-mode contributions~\cite{Jaus:1999zv}. Because of these two advantages, the CLF quark model has been used extensively to study the weak and radiative decays, as well as the other features of hadrons; see, for instance, Refs.~\cite{Choi:2004ww,Choi:2009ai,Choi:2009ym,Choi:2010ha,Choi:2010zb,Choi:2011xm,Choi:2010be,Choi:2014ifm,Choi:2017uos,Choi:2017zxn,Ryu:2018egt,Hwang:2001hj,Hwang:2001wu,Hwang:2010iq,Cheung:2014cka,Wang:2008ci,Wang:2008xt,Shen:2008zzb,Wang:2009mi,Cheng:2017pcq,Kang:2018jzg,Verma:2011yw,Shi:2016gqt,Wang:2018duy}.  

However, we should notice that there still exist some debates about the self-consistency of the CLF quark model. A known example is the vector meson decay constant, $f_V$, for which the calculation has to be made with a given polarization state of the vector meson. Unfortunately, it was found that the resulting $[f_V]_{\rm CLF}^{\lbd=0}$ and $[f_V]_{\rm CLF}^{\lbd=\pm}$, extracted respectively with the longitudinal ($\lbd=0$) and the transverse ($\lbd=\pm$) polarization state, are not consistent with each other~\cite{Cheng:2003sm}, 
\begin{eqnarray}\label{eq:fvpuzzle1}
[f_V]_{\rm CLF}^{\lbd=0} \neq [f_V]_{\rm CLF}^{\lbd=\pm}\,,
\end{eqnarray}
because $[f_V]_{\rm CLF}^{\lbd=0}$ receives an additional contribution characterized by the coefficient $B_1^{(2)}$, which provides about $10\%$ correction to $f_V$ numerically~\cite{Jaus:2002sv}. This inconsistency can be easily found from the formulas and numerical results given, for instance, in Refs.~\cite{Jaus:1999zv,Jaus:2002sv}.

A possible resolution to the ``$[f_V]_{\rm CLF}$ inconsistency puzzle'' exhibited by Eq.~\eqref{eq:fvpuzzle1} has been discussed in Ref.~\cite{Choi:2013mda} by modifying the  correspondence between the manifestly covariant BS approach and the LF quark model. Traditionally, the LF covariant vertex function $\chi_V$ and the factor $D_{V,{\rm con}}=M+m_1+m_2$ appearing in the vertex operator are replaced by the wave function~(WF) $\psi_V$ and the factor $D_{V,{\rm LF}}=M_0+m_1+m_2$, respectively, via~\cite{Choi:2013mda}
\begin{align} \label{eq:type-I}
\sqrt{2N_c}\frac{\chi_V(x,k_{\bot})}{1-x} \;\to\;  \frac{ \psi_V(x,k_{\bot})}{\sqrt{x(1-x)} \hat{M}_0}\,,\qquad D_{V,{\rm con}}  \;\to\; D_{V,{\rm LF}}\,,\qquad (\text{type-I})
\end{align}
where $M_0$ is the kinetic invariant mass of the vector meson, and $\hat{M}_0\equiv \sqrt{M_0^2-(m_1-m_2)^2}$. The type-I correspondence has been widely used to connect the two different approaches in most of previous works\footnote{In some literatures~(see, for instance, Refs.~\cite{Jaus:1999zv,Jaus:2002sv,Cheng:2003sm}), the convention for the vertex function $h_V=\chi_V\hat{N}_V$, with $\hat{N}_V=x(M^2-M_0^2)$, is used instead.}, claiming that some results obtained in the SLF quark model, such as $[f_V]_{\rm SLF}$, are not trustworthy due to the lack of zero-mode contributions~\cite{Jaus:2002sv,Cheng:2003sm}. However, this correspondence would result in the inconsistency problem demonstrated by Eq.~\eqref{eq:fvpuzzle1}. 

The correspondence between $\chi$ and $\psi$ in Eq.~\eqref{eq:type-I} has been derived by matching the CLF expressions to the SLF ones for some quantities that are free of the zero-mode effects, such as the pseudoscalar meson decay constant $f_P$ and the weak transition form factor $f_{+}^{P\to P}(q^2)$~\cite{Jaus:1999zv,Cheng:2003sm,Choi:2010be}. However, the validity of the replacement $D_{V,{\rm con}} \to D_{V,{\rm LF}}$ has not yet been clarified explicitly in the same framework. This motivates Choi and Ji~\cite{Choi:2013mda} to advocate the replacement $M\to M_0$ in each and every term containing $M$ in the integrand of $[f_V]_{\rm CLF}^{\lbd=0}$ and $[f_V]_{\rm CLF}^{\lbd=\pm}$, rather than only in the $D$ factor. As a result, the correspondence given by Eq.~\eqref{eq:type-I} should be generalized to~\cite{Choi:2013mda}
\begin{align} \label{eq:type-II}
\sqrt{2N_c}\frac{\chi_V(x,k_{\bot})}{1-x} \;\to\;  \frac{ \psi_V(x,k_{\bot})}{\sqrt{x(1-x)} \hat{M}_0}\,,\qquad M  \;\to\; M_0\,.\qquad (\text{type-II})
\end{align}
It is interesting to note that, under this scheme for the replacements, one gets numerically~\cite{Choi:2013mda}
\begin{align}\label{eq:cslfV}
[f_V]_{\rm CLF}^{\lbd=0} \,=\, [f_V]_{\rm CLF}^{\lbd=\pm} \,=\, [f_V]_{\rm SLF}\,,
\end{align}
which implies that the ``$[f_V]_{\rm CLF}$ inconsistency puzzle'' can be ``resolved'' and, moreover, a valid connection is established between the CLF and the SLF quark model.

From the above observations, one may conclude that the type-II correspondence specified by Eq.~\eqref{eq:type-II} might provide a self-consistent scheme in connecting the manifestly covariant and the LF approach. However, before making such a solid conclusion, it is necessary to further test such an interesting scheme via other quantities in addition to $f_V$, such as the decay constants of axial-vector mesons $^3\!A$ and $^1\!A$ with quantum numbers $^{2S+1}\!L_J\,=\,^3\!P_1$ and $^1\!P_1$, as well as the weak transition form factor $f_-^{P\to P}(q^2)$, which are all plagued by the zero-mode effects. We shall show later that $f_{^3\!A}$ and $f_{^1\!A}$ in the traditional CLF quark model also suffer the self-consistency problem as in the case of $f_V$. In addition, it should be noted that $[f_V]_{\rm SLF}$ given in Eq.~\eqref{eq:cslfV} is actually obtained with $\lbd=0$, {i.e.}, $[f_V]_{\rm SLF}^{\lbd=0}$. So, in order to claim the self-consistencies of LF quark models, one should also check carefully whether the SLF results for $f_V$ obtained with $\lbd=0$ and $\lbd=\pm$, $[f_V]_{\rm SLF}^{\lbd=0}$ and $[f_V]_{\rm SLF}^{\lbd=\pm}$, are consistent with each other. In this paper, besides the issues mentioned above, we shall also investigate the covariance of the CLF quark model, which in fact is possibly violated when the LF vertex function is used~\cite{Jaus:1999zv}.

Our paper is organized as follows. In Sec. 2, we recapitulate the SLF and CLF quark models. In Sec.  3, our theoretical results are presented for the decay constants of pseudoscalar, vector and axial-vector mesons, as well as the $P\to P$ weak transition form factors; the self-consistencies, the zero-mode contributions, as well as the covariance of the LF quark models are also discussed in detail. Finally, our conclusions are made in Sec.  4. 

\section{THEORETICAL FRAMEWORK}
\label{sec:2}

\subsection{SLF quark model}

In this subsection, we give a brief overview of the SLF approach for calculating the current matrix elements, details of which could be found, for instance, in Refs.~\cite{Jaus:1989au,Jaus:1989av,Cheng:1996if}. 

For a meson bound-state with total momentum $p$ and consisting of a quark $q_1$ with mass $m_1$ and an antiquark $\bar{q}_2$ with mass $m_2$, we can represent the momenta of $q_1$ and $\bar{q}_{2}$ in terms of the LF relative momentum variables $(x, {k_{\bot}})$ as 
\begin{align}\label{eq:momk1}
&k_1^+=k^+=x p^+\,,\quad\,k_{1\bot}=xp_{\bot}+k_{\bot} \,,  \\[0.2cm]
&k_2^+=p^+-k_1^+=\bar{x}p^+ \,,\quad\, k_{2\bot}=\bar{x}p_{\bot}-k_{\bot}\,,
\end{align}
where $\bar{x}=1-x$, $k_{\bot}=(k^x,\,k^y)$, and $p_{\bot}=(p^x,\,p^y)$. One can take $p_{\bot}=0$ when assuming that the meson moves along the $z$ axis. In the LF formalism, such a meson bound-state can be expanded, in the leading Fock-state approximation, as 
\begin{equation}
|M\ra =  \sum_{h_1,h_2} \int \frac{\d k^+ \d^2{ k_{\bot}}}{(2\pi)^32\sqrt{k_1^+ k_2^+}} \Psi_{h_1,h_2}(k^+, { k_{\bot}})|q_1:k_1^+,k_{1\bot},h_1\ra|\bar{q}_2:k_2^+,k_{2\bot},h_2\ra\,,
\label{eq:Fockexp}
\end{equation}
where $h_{1(2)}$ denotes the helicity of the (anti)quark, $\Psi_{h_1,h_2}(k^+, {k_{\bot}})$ is the momentum-space WF, and the one-particle states $|q_1\ra$ and $|\bar{q}_2\ra$ are defined by $|q_1\ra=\sqrt{2k_1^+}b^{\dagger}|0\ra$ and $|\bar{q}_2\ra=\sqrt{2k_2^+}d^{\dagger}|0\ra$, respectively. The particle creation and annihilation operators satisfy the equal-LF-time anticommutation relations
\begin{equation}
\{b^{\dagger}_{h} (k), b_{h'} (k^{\prime}) \}=\{d^{\dagger}_{h} (k), d_{h'} (k^{\prime}) \}= (2\pi)^3  \delta(k^+-k'^{+})\delta^2({ k}_{\bot}-{ k}'_{\bot}) \delta_{h h'}.
\label{anticommutation}
\end{equation}

The momentum-space WF in Eq.~\eqref{eq:Fockexp} satisfies the normalization condition
\begin{eqnarray}\label{eq:norc}
\sum_{h_1,h_2}\int  \frac{\d x\,\d^2{k}_{\bot}}{2(2\pi)^3} |\Psi_{h_1,h_2}(x,{k}_{\bot})|^2
=1\,,
\end{eqnarray}
and can be expressed as~\cite{Jaus:1989au}
\begin{eqnarray}
\label{eq:LFWFP2}
\Psi_{h_1,h_2}(x,{k}_{\bot})=S_{h_1,h_2}(x,{k}_{\bot})\,\psi(x,{k}_{\bot}) \,,
\end{eqnarray}
where the radial WF $\psi(x,{k}_{\bot})$ describes the momentum distributions of the constituent quarks in the bound-state, and the spin-orbital one $S_{h_1,h_2}(x,{k}_{\bot})$ constructs a state with definite spin $(S,S_z)$ out of the LF helicity eigenstates $(h_1,h_2)$~\footnote{It should be noted that no state with a fixed number of constituents can be an eigenstate of the angular momentum operator $J^2$, because $J^2$ is interaction-dependent on the LF and not diagonal in particle number. For now, there is not a practical way to construct an eigenstate of $J^2$ purely in the LF dynamics without any approximation. As illustrated in Ref.~\cite{Bakker:1979eg}, the WF for a $q_1\bar{q}_2$ bound-state can be chosen to be a simultaneous eigenfunction of the mass operator $\hat{M}^2$ and the angular momentum operators $\hat{J}^2$ and $\hat{J}_3$, only when the angular condition~(Eq.~(2.20) in Ref.~\cite{Bakker:1979eg}) is satisfied. However, in order to satisfy this condition, the interaction part of the mass operator $\hat{M}^2$ must be a function of scalar products of momenta, relative angular momentum and spin operators, but otherwise is not restricted at all. For more details, the readers are referred to Refs.~\cite{Terentev:1976jk,Jaus:1989au,Bakker:1979eg,Chung:1988my}.}. For the former, we shall adopt the Gaussian-type WFs~\cite{Chung:1988mu} 
\begin{eqnarray}\label{eq:WFs}
\psi_s(x,k_{\bot}) &=&4\frac{\pi^{\frac{3}{4}}}{\beta^{\frac{3}{2}}} \sqrt{ \frac{\partial k_z}{\partial x}}\exp\left[ -\frac{k_z^2+k_\bot^2}{2\beta^2}\right]\,,\qquad\text{for s-wave meson}\\[0.2cm]
\label{eq:WFp}
\psi_{p}(x,k_{\bot}) &=&\frac{\sqrt{2}}{\beta}\psi_s(x,k_{\bot}) \,,\qquad\text{for p-wave meson}
\end{eqnarray}
where $\beta$ is the variational parameter and can be determined, for example, from the meson spectroscopy, while $k_z$ is the relative momentum in the $z$-direction and takes the form
\begin{eqnarray}
k_z=(x-\frac{1}{2})M_0+\frac{m_2^2-m_1^2}{2 M_0}\,,
\end{eqnarray}
with the kinetic invariant mass squared given by
\begin{eqnarray}
M_0^2=\frac{m_1^2+{k}_{\bot}^2}{x}+\frac{m_2^2+{k}_{\bot}^2}{\bar{x}}\,.
\end{eqnarray} 
The spin-orbital WF $S_{h_1,h_2}(x,{k}_{\bot}) $ can be obtained from the ordinary equal-time static one with assigned  quantum number $J^{PC}$ through the interaction-independent Melosh transformation~\cite{Jaus:1989au,Chung:1988my}. It is more convenient to use the covariant form, which, after applying the equations of motion on spinors, can be written explicitly as~\cite{Jaus:1989av,Cheng:2003sm,Hwang:2009cu,Hwang:2012nw}
\begin{eqnarray}\label{eq:defS2}
S_{h_1,h_2}=\frac{\bar{u}_{h_1}(k_1)\,\Gamma' \,v_{h_2}(k_2)}{\sqrt{2}\hat{M}_0}\,,
\end{eqnarray}
with $\hat{M}_0$ already defined below Eq.~\eqref{eq:type-I}, and the vertex operators $\Gamma'$ given, respectively, as
\begin{eqnarray}
&&\Gamma'_P=\r_5\,,\qquad\text{for $P$ meson}\nonumber\\[0.2cm]
&& \Gamma'_V=-\not\!\hat{\epsilon}+\frac{\hat{\epsilon}\cdot (k_1-k_2)}{D_{V,\rm LF}}\,,\quad D_{V,\rm LF}=M_0+m_1+m_2\,,\qquad\text{for $V$ meson}\nonumber\\[0.2cm]
&&\Gamma'_{{}^1\!A}=-\frac{\hat{\epsilon}\cdot (k_1-k_2)}{D_{^1\!A,\rm LF}} \r_5\,, \quad D_{^1\!A,\rm LF}=2\,,\qquad\text{for ${}^1\!A$ meson}\nonumber\\[0.2cm]
&& \Gamma'_{ {}^3\!A}=-\frac{\hat{M}_0^2}{2\sqrt{2} M_0}\left[  \not\!\hat{\epsilon}+\frac{\hat{\epsilon}\cdot (k_1-k_2)}{D_{^3\!A,\rm LF}} \right]\r_5 \,, \quad D_{^3\!A,\rm LF}=\frac{\hat{M}_0^2}{m_1-m_2}\,, \qquad\text{for ${}^3\!A$ meson}\label{eq:vSLF}
\end{eqnarray} 
where the longitudinal and transverse polarization vectors are given, respectively, by
\begin{align}
\hat{\epsilon}^{\mu}_{\lbd=0}&=\frac{1}{M_0}\left(p^+,\frac{p_{\bot}^2-M_0^2}{p^+},p_{\bot}\right)\,,\\[0.2cm]
\hat{\epsilon}^{\mu}_{\lbd=\pm}&=\left(0,\frac{2}{p^+}\epsilon_{\bot}\cdot p_{\bot}, \epsilon_{\bot}\right)\,, 
\quad {\rm with}\quad \epsilon_{\bot}\equiv \mp \frac{(1,\pm i)}{\sqrt{2}}\,.
\end{align}  

Equipped with all the formulas given above, we can now express the matrix element for a general $M\to 0$ transition, ${\cal A} \equiv \la 0 | \bar{q}_2 \Gamma q_1 |M(p)\ra$, as
\begin{align}
{\cal A} &=\sqrt{N_c} \sum_{h_1,h_2} \int \frac{\d x \d^2{ k}_{\bot}}{(2\pi)^32\sqrt{x\bar{x}}} \psi(x,{ k}_{\bot})\,S_{h_1,h_2}(x,{ k}_{\bot})\,C_{h_1,h_2}(x,{ k}_{\bot}) \,,
\label{eq:A}
\end{align}
where $C_{h_1,h_2}(x,{ k}_{\bot}) \equiv \bar{v}_{h_2}(k_2^{+},-{ k}_{\bot}) \Gamma u_{h_1}(k_1^+,{ k}_{\bot})$. On the other hand, the matrix element for a general $M'\to M''$ transition, ${\cal B} \equiv \la  M''(p'') | \bar{q}''_1 \Gamma q'_1 |M'(p') \ra $, can be written as
\begin{align}
{\cal B} &=\sum_{h'_1,h''_1,h_2} \int  \frac{\d k'^+ \d^2k_{\bot}'}{(2\pi)^3\,2\sqrt{k'^+\,k''^+}}  {\psi''}^{*}({k''}^+, \bar{k}_{\bot}'')\,{\psi'}({k'}^+,k_{\bot}')\nonumber\\
&\qquad\times S''^{\dagger}_{h''_1,h_2}({k''}^+, k_{\bot}'')\,C_{h''_1,h'_1}({k''}^+,k_{\bot}'',{k'}^+,k_{\bot}')\,S'_{h'_1,h_2}({k'}^+,k_{\bot}')\,, 
\label{eq:B}
\end{align}
where $C_{h''_1,h'_1}({k''}^+,k_{\bot}'',{k'}^+,k_{\bot}') \equiv  \bar{u}_{h''_1}({k''}^+,k_{\bot}'') \Gamma  u_{h'_1}({k'}^+,k_{\bot}')$. In practice, we usually work in the $q^+=0$ frame, which leads to $q^2=(p'-p'')^2=-q_\bot^2\leqslant 0$,  implying that the transition form factors are known only for space-like momentum transfer. The transition form factors in the time-like region can be obtained by making an additional $q^2$ extrapolation. In Eq.~\eqref{eq:B}, the incoming-quark momentum, $(k'^+, k_{\bot}')$, has already been given in Eq.~\eqref{eq:momk1} but now with an additional superscript ``$'$'', and the outgoing-quark momentum is given by $(k''^+=k'^+=xp'^+, k_{\bot}''=k_{\bot}'-q_{\bot})$ in the $q^+=0$ frame. In addition, the kinetic invariant mass of the outgoing meson takes the form
\begin{eqnarray}
{M_0''}^2=\frac{{m_1''}^2+ {\bar{k}}_{\bot}''^2}{x}+\frac{m_2^2+{\bar{k}}_{\bot}''^2}{\bar{x}}\,, 
\end{eqnarray}
where ${\bar{k}}_{\bot}''\equiv k_{\bot}'-\bar{x}{ q}_{\bot}$. We shall use the formulas for the matrix elements ${\cal A}$, Eqs.~\eqref{eq:A}, and ${\cal B}$, Eq.~\eqref{eq:B}, to extract the decay constant and weak transition form factors, respectively. 

\subsection{CLF quark model}

\begin{figure}[t]
\begin{center}
\subfigure[]{\includegraphics[scale=0.32]{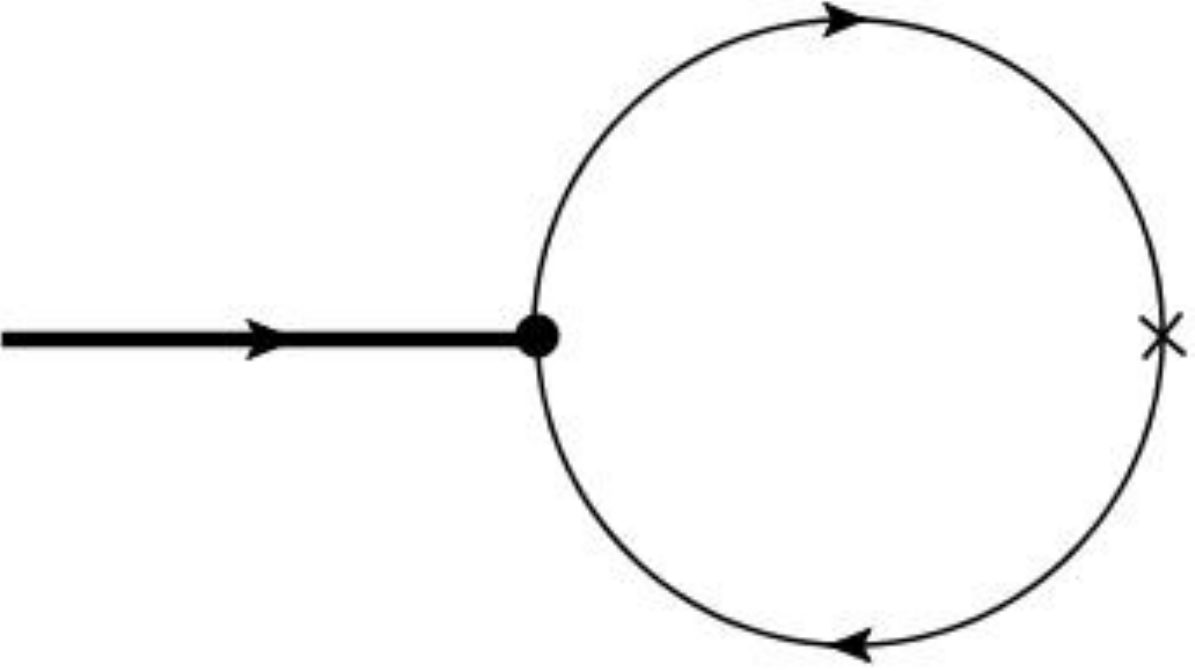}}\qquad\qquad
\subfigure[]{\includegraphics[scale=0.32]{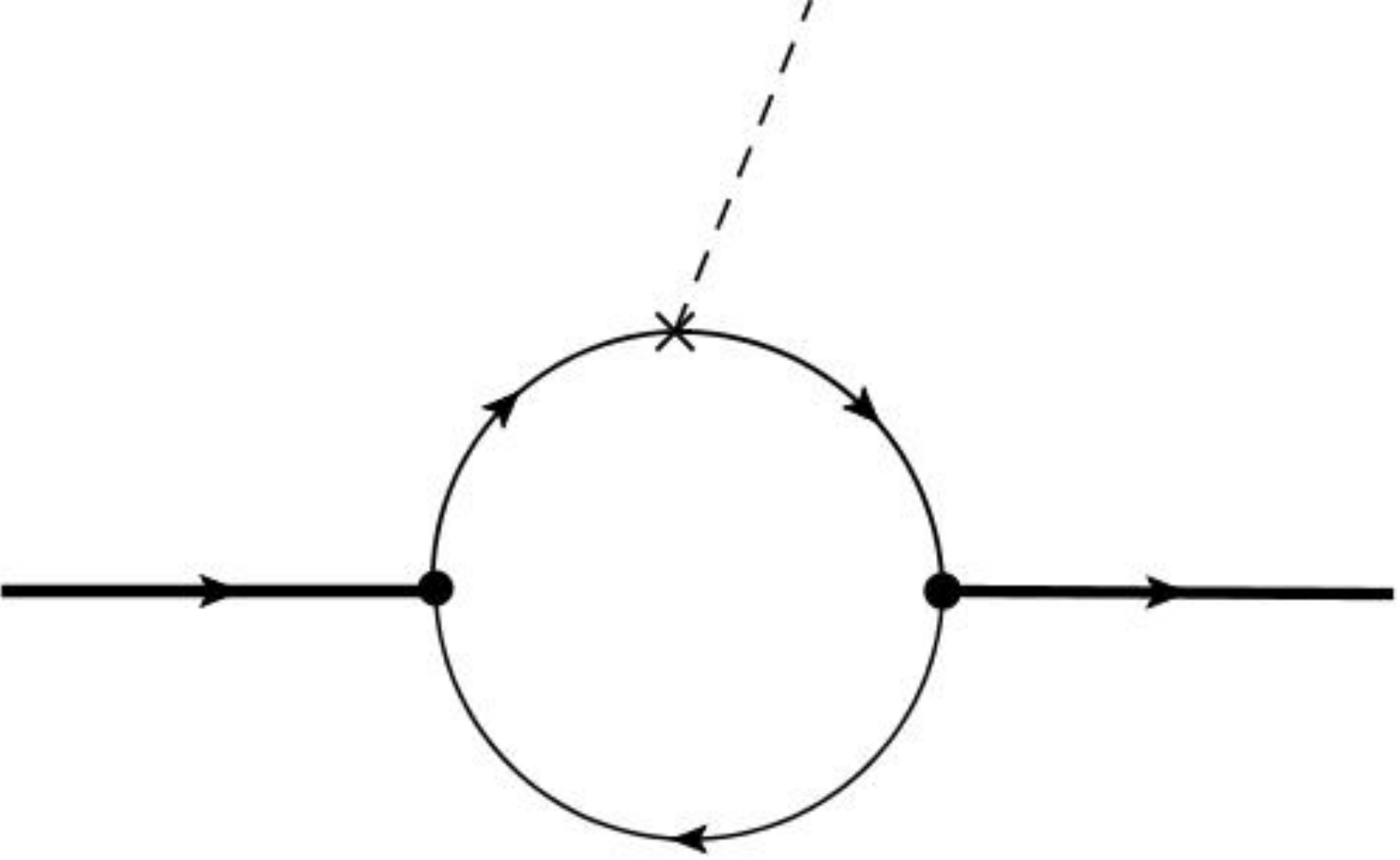}}
\caption{\small \label{fig:fayn} Feynman diagrams for the matrix elements $\cal A$~[Fig.~(a)] and $\cal B$~[Fig.~(b)] in the one-loop approximation.}
\end{center}
\end{figure}

In contrast to the SLF approach, the CLF quark model provides a systematical way of dealing with the zero-mode contribution, and a physical quantity can be calculated in terms of Feynman momentum loop-integrals that are manifestly covariant. In this paper, we shall employ the same formalism proposed by Jaus~\cite{Jaus:1999zv}, Choi and Ji~\cite{Choi:1998nf}, as well as Cheng {\it et al.}~\cite{Cheng:2003sm}. 

The Feynman diagrams for the matrix elements $\cal A$ and $\cal B$ are shown in Fig.~\ref{fig:fayn}(a) and Fig.~\ref{fig:fayn}(b), respectively. From these one-loop diagrams and using the Feynman rules for the meson-quark-antiquark vertices given in Refs.~\cite{Jaus:1999zv,Cheng:2003sm}, we can write $\cal A$ and $\cal B$, respectively, as
\begin{eqnarray}\label{eq:Aclf1}
{\cal A}&=&N_c \int \frac{\d^4 k}{(2\pi)^4} \frac{H_M}{N_1N_2} S_{\cal A}\,,\\[0.2cm]
\label{eq:Bclf1}
{\cal B}&=&N_c \int \frac{\d^4 k'}{(2\pi)^4} \frac{H_{M'}H_{M''}}{N_1'\,N_1''\,N_2}iS_{\cal B}\,,
\end{eqnarray}
where $\d^4 k^{(\prime)}=\frac{1}{2} \d k^{(\prime)-} \d k^{(\prime)+} \d^2 k^{(\prime)}_{\bot}$, and $H_{M, M', M''}$ are the bound-state vertex functions. The trace terms $S_{\cal A}$ and $S_{\cal B}$ associated with the fermion loops are given, respectively, by
\begin{align}
S_{\cal A}&={\rm Tr}\left[\Gamma\, (\not\!k_1+m_1)\,(i\Gamma_M)\,(-\not\!k_2+m_2)\right]\,,\\[0.2cm]
S_{\cal B}&={\rm Tr}\left[\Gamma\, (\not\!k'_1+m'_1)\,(i\Gamma_M')\,(-\not\!k_2+m_2)\,(i\r^0{\Gamma}_M''^{\dag}\r^0) (\not\!k_1''+m_1'')\right]\,,
\end{align}
with explicit forms of the vertex operators $\Gamma_M^{(\prime,\prime\prime)}$ for different types of mesons listed as~\cite{Cheng:2003sm}
\begin{align}
i\Gamma_P&=-i\r_5\,,  & i\Gamma_V&=i\left[\gamma^\mu-\frac{ (k_1-k_2)^\mu}{D_{ V,{\rm con}}}\right]\,,\nonumber\\
i\Gamma_{{}^1\!A}&=i\frac{(k_1-k_2)^\mu}{D_{^1\!A,{\rm con}}} \r_5\,, & i \Gamma_{ {}^3\!A}&=i\left[ \gamma^\mu+\frac{ (k_1-k_2)^\mu}{D_{^3\!A,{\rm con}}} \right]\r_5 \,.\label{eq:voCLF}
\end{align} 

Integrating out the minus component of the loop momentum, one goes from the covariant calculation to the LF one. Assuming that $H_{M, M', M''}$ are analytic within the contour and closing the contour in the upper complex $k^-$~($k'^-$) plane, one picks up a residue at $k_2^2=\hat{k}_2^2=m_2^2$, corresponding to putting the spectator antiquark on its mass shell. This manipulation forces us to make the following replacements in Eqs.~\eqref{eq:Aclf1} and \eqref{eq:Bclf1}~\cite{Jaus:1999zv,Cheng:2003sm}:
\begin{eqnarray}
N_1\to\hat{N}_1=x(M^2-M_0^2)\,,\qquad  N_1^{\prime(\prime\prime)} \to \hat{N}_1^{\prime(\prime\prime)}= \hat{N}_1 \left[ M\to M^{\prime(\prime\prime)}, M_0\to M_0^{\prime(\prime\prime)} \right]\,,
\end{eqnarray}
and 
\begin{eqnarray}\label{eq:type1}
\chi_M = H_M/N_1\to h_M/\hat{N}_1\,,\qquad  D_{M,\rm con}\to  D_{M,\rm LF}\,,\qquad (\text{type-I})
\end{eqnarray}
together with similar replacements for $\chi_{M'}$ and $\chi_{M''}$ as for $\chi_M$. Explicit forms of the factors $D_{M,\rm LF}$ for $P$, $V$, $^1\!A$ and $^3\!A$ mesons have already been given by Eq.~\eqref{eq:vSLF}, and the LF forms of the vertex functions $h_M$ for these types of mesons are given, respectively, as~\cite{Cheng:2003sm} 
\begin{align}
h_P/\hat{N}_1&=h_V/\hat{N}_1=\frac{1}{\sqrt{2N_c}}\sqrt{\frac{\bar{x}}{x}}\frac{\psi_s}{\hat{M}_0}\,, 
\label{eq:vPV}\\[0.2cm]
h_{^1\!A}/\hat{N}_1&=\frac{1}{\sqrt{2N_c}}\sqrt{\frac{\bar{x}}{x}}\frac{\psi_p}{\hat{M}_0}\,,
\label{eq:v1A}\\[0.2cm]
h_{^3\!A}/\hat{N}_1&=\frac{1}{\sqrt{2N_c}}\sqrt{\frac{\bar{x}}{x}}\frac{\hat{M}_0^2}{2\sqrt{2} M_0} \frac{\psi_p}{\hat{M}_0}\,.
\label{eq:v3A}
\end{align}
It is noted that Eq.~\eqref{eq:type1} with $h_V/\hat{N}_1$ given by Eq.~\eqref{eq:vPV} is exactly the same as Eq.~\eqref{eq:type-I}, {i.e.}, the type-I replacement for vector meson. This means that Eq.~\eqref{eq:type1} with  $h_{P,^1\!A,^3\!A}/\hat{N}_1$ given by Eqs.~\eqref{eq:vPV}, \eqref{eq:v1A} and \eqref{eq:v3A} should be the traditional type-I correspondence for $P$, $^1\!A$ and $^3\!A$ mesons. Accordingly, corresponding to Eq.~\eqref{eq:type-II}, we get the generalized type-II correspondence
\begin{eqnarray}\label{eq:type2}
\chi_M = H_M/N_1\to h_M/\hat{N}_1\,,\qquad  M\to  M_0\,.\qquad (\text{type-II})
\end{eqnarray}
For simplicity, our following derivation and theoretical results are given only with the traditional type-I correspondence unless otherwise specified. The ones with the type-II correspondence can be obtained with an additional replacement $M\to M_0$.

After integrating out the $k^-$~($k'^-$) component, we can reduce the matrix elements $\cal A$, Eq.~\eqref{eq:Aclf1}, and $\cal B$, Eq.~\eqref{eq:Bclf1}, to the LF forms,
\begin{eqnarray}\label{eq:Aclf2}
\hat{\cal A}&=&N_c  \int \frac{  \d k^+ \d^2 k_{\bot}}{2(2\pi)^3}   \frac{ -ih_M}{\bar{x}p^+  \hat{N}_1\,}\hat{S}_{\cal A}\,,\\[0.2cm]
\label{eq:Bclf2}
\hat{\cal B}&=&N_c \int \frac{\d k'^+\d^2 k_{\bot}'}{2(2\pi)^3}\frac{h_{M'}h_{M''}}{\bar{x}p'^+ \hat{N}_1'\,\hat{N}_1''\,}\hat{S}_{\cal B}\,.
\end{eqnarray}
As noted already in Refs.~\cite{Jaus:1999zv,Cheng:2003sm}, the LF matrix elements $\hat{\cal A}$ and $\hat{\cal B}$ obtained in this way receive additional spurious contributions proportional to the lightlike four-vector $\omega^\mu=(0,2,0_\bot)$, which can however be eliminated after including the zero-mode contributions. As shown in Ref.~\cite{Jaus:1999zv}, the inclusion of zero-mode contributions to the matrix elements in practice amounts to some proper replacements in $\hat{S}_{\cal A\,,B}$ under the integration. Specifying to the quantities considered in this paper, we need~\cite{Jaus:1999zv,Cheng:2003sm}
\begin{align}
 \hat{k}_1^\u&\to x p^\u\,,\nonumber \\
 \hat{k}_1^\u \hat{k}_1^\v&\to -g^{\u\v} \frac{k_{\bot}^2}{2} +p^\u p^\v x^2+\frac{p^\u\w^\v+p^\v\w^\u}{\w\cdot p} B_1^{(2)}  \,,\nonumber \\
\hat{N}_2&\to Z_2=\hat{N}_1+m_1^2-m_2^2+(\bar{x}-x)M^2\,,
\label{eq:repDC}
\end{align}
for $\hat{\cal A}$, and 
\begin{align}
 \hat{k}_1'^{\mu} &\to P^\u A_1^{(1)}+ q^\u  A_2^{(1)} \,,\nonumber\\
 k_1'^{\mu} \hat{N}_2 &\to q^\u \left[A_2^{(1)}Z_2+\frac{q\cdot P}{q^2}  A_1^{(2)} \right] \,,\nonumber\\
 Z_2&=\hat{N}_1'+m_1'^2-m_2^2+(\bar{x}-x)M'^2+(q^2+q\cdot P)\frac{k_{1\bot}'\cdot q_{\bot}}{q^2}\,,
 \label{eq:repFF}
\end{align}
for $\hat{\cal B}$, where $P=p'+p''$, and the coefficients $A^{(1)}_{1,2}$, $A^{(2)}_{1}$ and $ B_1^{(2)}$ are given, respectively, by~\cite{Jaus:1999zv} 
\begin{align}
 A_1^{(1)}&=  \frac{x}{2}\,, \qquad
 A_2^{(1)}=\frac{x}{2} -\frac{k_{1\bot}' \cdot q_{\bot}}{q^2}\,,\qquad
 A_1^{(2)}=-k_{1\bot}'^2 -\frac{(k_{1\bot}' \cdot q_{\bot})^2}{q^2}\,,\nonumber\\[0.2cm]
 B_1^{(2)} &=A_1^{(1)}C_1^{(1)}-A_1^{(2)}=\frac{x}{2}Z_2+\frac{k_{\bot}^2}{2}\,. 
\end{align}
It should be noted that, although the coefficient $B_1^{(2)}$ does not vanish and is combined with $\w^\u$, there is no zero-mode contribution associated with $B_1^{(2)}$ due to $x\hat{N}_2=0$~\cite{Jaus:1999zv}. However, we shall show later that the contributions related to nonvanishing $B_1^{(2)}$ would result in the self-consistency problem of CLF quark model with the traditional type-I correspondence. 

Using the formulas given above, one can then obtain the full results for $\hat{\cal A}$ and  $\hat{\cal B}$ in the CLF quark model, and further extract the physical quantities like the decay constants and transition form factors. Explicitly, for a given quantity $\cal Q$, its full result (${\cal Q}^{\rm full}$) can be expressed as a sum of the valence (${\cal Q}^{\rm val.}$) and the zero-mode (${\cal Q}^{\rm z.m.}$) contribution~\cite{Choi:2013mda}: ${\cal Q}^{\rm full}={\cal Q}^{\rm val.}+{\cal Q}^{\rm z.m.}$. In order to evaluate the zero-mode effect, one needs to calculate ${\cal Q}^{\rm z.m.}$ or the difference ${\cal Q}^{\rm full}-{\cal Q}^{\rm val.}$. A simple way to calculate ${\cal Q}^{\rm val.}$ is to assume that $k_2^+\neq 0$ and $k_1^+\neq 0$, which ensure that the poles of $N_2$ and $N_1$ are safely located inside and outside, respectively, the contour of $k^-$~($k'^-$) integral~({i.e.} the poles of $N_2$ and $N_1$ are both finite) and imply that the zero-mode contributions are absent. In this case, the replacements for $\hat{k}_1^{\u}$ and $\hat{N}_2$ given above need not be applied anymore. Instead, one just needs to directly use the on-mass-shell condition for the spectator antiquark, $k_2^2=m_2^2$, and the four-momentum conservation at each vertex. For example, after integrating out $k^-$, one should take $\hat{N}_2=0$, and the nonindependent minus component of $\hat{k}_1$ is given by  
\begin{align}
 \hat{k}_1^-=p^--\hat{k}_2^-=p^--\frac{m_2^2+k_{\bot}^2}{\hat{k}_2^+}=p^-\left(1-\frac{m_2^2+k_{\bot}^2}{\bar{x}M^2}\right)\,.
\end{align}
The resulting $[f_V]_{\rm val.}$ obtained in this way is exactly the same as that obtained by Choi and Ji~\cite{Choi:2013mda}, which will be clearly seen in the next section. 

\section{Results and discussions}

\subsection{Decay constants of $P$ and $V$ mesons}

The decay constants of $P$ and $V$ mesons are defined, respectively, by
\begin{align}
\label{eq:dcp}
\la 0 | \bar q_2\r^\u\r_5 q_1|P(p)\ra&=if_P p^\u \,,\\
\label{eq:dcv}
\la 0 | \bar q_2\r^\u q_1|V(p,\lambda)\ra&=f_VM_V \epsilon^\u_{\lambda}\,.
\end{align}
For the $P$ meson, with the theoretical formulas given in the last section, the resulting $f_P$ in the SLF and CLF quark models are given, respectively, as\footnote{In the extraction of $[f_P]_{\rm SLF}$ and $[f_P]_{\rm val.}$, we have to take the $\mu=+$ component, which is the only choice because, on the one hand, $\mu=-$ is not an independent component and, on the other hand, Eq.~\eqref{eq:dcp} would become an identity $0=0$ when taking $\mu=\bot$ in the $P$-meson rest frame. In addition, it has been known that $\mu=+$ is a ``good'' component for calculating $f_P$ as mentioned in the Introduction.}
\begin{align}
[f_P]_{\rm SLF}&=\sqrt{N_c} \int\frac{\d x\,\d^2{ k}_{\bot}}{(2\pi)^3}\frac{\psi_s(x,{k}_{\bot})}{\sqrt{x\bar{x}}} \frac{2}{\sqrt{2}\hat{M}_0}( \bar{x}m_1+xm_2)\,,\\[0.2cm]
[f_P]_{\rm full}&=[f_P]_{\rm val.}=N_c \int \frac{  \d x\, \d^2 k_{\bot}}{(2\pi)^3} \frac{\chi_P  }{\bar{x} }\,2(\bar{x}m_1+xm_2)\,,
\end{align}
which agree with the ones obtained in the previous works, for instance, Refs.~\cite{Jaus:1991cy,Jaus:1999zv,Cheng:2003sm,Choi:2013mda}. The finding $[f_P]_{\rm full}=[f_P]_{\rm val.}$ implies that $f_P$ is free of the zero-mode contribution. This is also the reason why one usually uses $f_P$ to determine the LF vertex function, $h_P$ or $\chi_P$. Finally, it can be easily found that the SLF and CLF results for $f_P$ are also consistent with each other,
\begin{align}
[f_P]_{\rm SLF}=[f_P]_{\rm val.}=[f_P]_{\rm full}\,,
\end{align}
under both the type-I and the type-II correspondence.

\begin{table}[t]
\begin{center}
\caption{\label{tab:fP} \small Available experimental data and LQCD results for the decay constants $f_{P}$ (in unit of ${\rm MeV}$).}
\vspace{0.1cm}
\footnotesize 
\let\oldarraystretch=\arraystretch
\renewcommand*{\arraystretch}{1.1}
\setlength{\tabcolsep}{8.5pt}
\begin{tabular}{lcccccccccccc}
\hline\hline
                 &$f_{\pi}$ &$f_{K}$ &$f_{\eta_s}$
                 &$f_{D}$   &$f_{D_s}$\\\hline

Exp. data      &{$130.50\pm0.13$}\cite{Patrignani:2016xqp}
              &$155.72\pm0.51$\cite{Patrignani:2016xqp}
              &$-$
         &$203.7\pm4.7$\cite{Patrignani:2016xqp}
         &$257.8\pm4.1$\cite{Patrignani:2016xqp}
              \\\hline
LQCD         &$130.2\pm1.4$\cite{Aoki:2016frl}
             &$155.6\pm0.4$\cite{Aoki:2016frl}
             &$181.1\pm0.6$\cite{Dowdall:2013rya}
             &$212.2\pm1.4$\cite{Aoki:2016frl}
             &$248.8\pm1.3$\cite{Aoki:2016frl}
\\\hline\hline
                 &$f_{\eta_c}$ &$f_{B}$ &$f_{B_s}$
                 &$f_{B_c}$   &$f_{\eta_b}$\\\hline
Exp. data         &$335\pm75$\cite{Patrignani:2016xqp,Kher:2018wtv}
             &$188\pm25$\cite{Patrignani:2016xqp}
             &$-$
             &$-$
             &$-$
\\\hline
LQCD         &{$387\pm 7$}\cite{Becirevic:2013bsa}
             &$186\pm4$\cite{Aoki:2016frl}
              &$224\pm5$\cite{Aoki:2016frl}
              &$427\pm6$\cite{McNeile:2012qf}
              &$667\pm6$\cite{McNeile:2012qf}
\\\hline\hline
\end{tabular}
\end{center}
\end{table}

\begin{table}[t]
\begin{center}
\caption{\label{tab:input} \small Fitting results for the parameter $\beta$ (in unit of MeV), where $q=u,\,d$. See text for details.}
\vspace{0.1cm}
\let\oldarraystretch=\arraystretch
\renewcommand*{\arraystretch}{1.1}
\setlength{\tabcolsep}{14.8pt}
\begin{tabular}{lcccccccccc}
\hline\hline
  &$\beta_{q\bar{q}}$    &$\beta_{s\bar{q}}$   &$\beta_{s\bar{s}}$
  &$\beta_{c\bar{q}}$    &$\beta_{c\bar{s}}$ \\
  \hline
this work
  &$314.1_{-0.5}^{+0.5}$ &$342.8_{-1.4}^{+1.3}$ &$365.8_{-1.8}^{+1.2}$ &$464.1_{-10.8}^{+11.2}$ &$537.5_{-8.7}^{+9.0}$\\\hline
  Ref.\cite{Choi:1999nu}
  &$365.9$ &$388.6$ &$412.8$ &$467.9$ &$501.6$ \\\hline\hline
  &$\beta_{c\bar{c}}$    &$\beta_{b\bar{q}}$   &$\beta_{b\bar{s}}$   &$\beta_{b\bar{c}}$   &$\beta_{b\bar{b}}$    \\\hline
this work
  &$654.5_{-132.4}^{+143.3}$ &$547.9_{-10.2}^{+9.9}$
  &$601.4_{-7.3}^{+7.3}$ &$947.0_{-10.9}^{+11.2}$
  &$1391.2_{-48.2}^{+51.6}$
  \\\hline
  Ref.\cite{Choi:1999nu}
  &$650.9$ &$526.6$
  &$571.2$ &$936.9$ &$1145.2$
          \\
\hline\hline
\end{tabular}
\end{center}
\end{table}

Before proceeding to discuss $f_V$, we firstly determine the Gaussian parameter $\beta$ appearing in Eqs.~\eqref{eq:WFs} and \eqref{eq:WFp}, which is the key input for the LF quark models. Thanks to the self-consistencies of the LF quark models for and the available precision data on $f_P$, we shall perform $\chi^2$-fits on $\beta$ by using the data on $f_P$ collected in Table~\ref{tab:fP}\footnote{The lattice QCD (LQCD) results for $f_{\eta_{s},\eta_{b}}$ and $f_{B_{s},B_{c}}$ are used in the fits because of the lack of the corresponding experimental data.}, along with the constituent quark masses $m_{u(d),s,c,b}=(0.25,0.50,1.5,4.8)\,{\rm GeV}$. Our fitting results are given in Table~\ref{tab:input}, from which one can see that these values are generally in agreement with that obtained by the variational principle~\cite{Choi:1999nu}, and will be therefore used in our following numerical calculations. In addition, we assume that the parameter $\beta$ is universal for a given $(q_1\bar{q}_2)$ bound-state system. 

For the $V$ meson, taking the $\lbd=0$ and $\lbd=\pm$ polarization states, respectively\footnote{Because of the same reason as in the case of $f_P$, the component $\u=+$~($\u=\bot$) has to be taken for extracting $[f_V]_{\rm SLF\,, val.}^{\lbd=0(\pm)}$, as well as the decay constants of $^1\!A$ and $^3\!A$ mesons.}, we obtain 
\begin{align}
\label{eq:fVSLF0}
[f_V]_{\rm SLF}^{\lambda=0}
&=\sqrt{ N_c}  \int \frac{\d x \d^2{ k}_{\bot}}{(2\pi)^3} \frac{\psi_s(x,{k}_{\bot})  }{\sqrt{x\,\bar{x}}}\frac{2}{\sqrt{2}\hat{M}_0}
 \left(\bar{x}m_1+xm_2+\frac{2{k}_{\bot}^2}{D_{V,\rm LF}}\right)\,, \\[0.2cm]
 \label{eq:fVSLFpm}
 [f_V]_{\rm SLF}^{\lambda=\pm}
&=\sqrt{ N_c} \int \frac{\d x \d^2{ k}_{\bot}}{(2\pi)^3} \frac{\psi_s(x,{k}_{\bot})  }{\sqrt{x\,\bar{x}}}\frac{2}{\sqrt{2}\hat{M}_0}   
 \left(\frac{\hat{M}^2_0}{2 M_V } -  \frac{{k}_{\bot}^2 }{D_{V,\rm LF}} \frac{M_0}{M_V} \right)\,,
\end{align}
in the SLF quark model, in which $[f_V]_{\rm SLF}^{\lambda=0}$ is usually given, while $[f_V]_{\rm SLF}^{\lambda=\pm}$ is always ignored in previous works due to the traditional bias that $\mu=\bot$ is not a ``good" component as is $\mu=+$. 
By employing the CLF approach, on the other hand, we obtain 
\begin{align}
\label{eq:fVfull0}
[f_V]_{\rm full}^{\lbd=0}=& N_c  \int \frac{  \d x \d^2 k_{\bot}}{(2\pi)^3}   \frac{\chi_V}{\bar{x} } \frac{2}{M_V} \left[ xM_0^2 - m_1(m_1-m_2)  -\left(1-\frac{m_1+m_2}{D_{V,\rm con}}\right)(k_{\bot}^2 -2 B_1^{(2)}) \right]\,,\\
\label{eq:fVfullpm}
[f_V]_{\rm full}^{\lbd=\pm}=& N_c \int \frac{  \d x \d^2 k_{\bot}}{(2\pi)^3}   \frac{\chi_V}{\bar{x}} \frac{2}{M_V} \left[ xM_0^2 - m_1(m_1-m_2)  -\left(1-\frac{m_1+m_2}{D_{V,\rm con}}\right)  k_{\bot}^2 \right]\,,
\end{align}
and 
\begin{align}
\label{eq:fVval0}
[f_V]_{\rm val.}^{\lbd=0}=& N_c  \int \frac{  \d x \d^2 k_{\bot}}{(2\pi)^3}   \frac{\chi_V}{\bar{x} } \frac{2}{M_V} \left[ k_{\bot}^2+x\bar{x}M_V^2+m_1 m_2+\frac{\bar{x}^2M_V^2-m_2^2-k_{\bot}^2}{\bar{x}D_{V,\rm con}}\, \left(\bar{x}m_1 -  xm_2\right) \right]\,,\\
\label{eq:fVvalpm}
[f_V]_{\rm val.}^{\lbd=\pm}=& N_c \int \frac{  \d x \d^2 k_{\bot}}{(2\pi)^3}   \frac{\chi_V}{\bar{x}} \frac{2}{M_V} \left[  \frac{\bar{x}M_V^2+xM_0^2-(m_1-m_2)^2 }{2}-\left(1-\frac{m_1+m_2}{D_{V,\rm con}}\right)  k_{\bot}^2 \right]\,.
\end{align}
Our CLF results given above agree with the ones obtained in the previous works; for instance, Eqs.~\eqref{eq:fVfull0} and \eqref{eq:fVfullpm} have been obtained in Ref.~\cite{Jaus:2002sv} and Refs.~\cite{Jaus:1999zv,Cheng:2003sm}, respectively, while Eqs.~\eqref{eq:fVval0} and \eqref{eq:fVvalpm} have been given in Ref.~\cite{Choi:2013mda}. In order to clearly separate the contributions related to the coefficient $B_1^{(2)}$~( {i.e.} the difference between $[f_V]_{\rm full}^{\lbd=0}$ and $[f_V]_{\rm full}^{\lbd=\pm}$), we define
\begin{align}
\Delta^M_{\rm full}(x) \equiv \frac{\d [f_M]_{\rm full}^{\lbd=0}}{ \d x}-\frac{\d [f_M]_{\rm full}^{\lbd=\pm}}{ \d x}\,,
\end{align}
where $M=V$, $^1\!A$ or $^3\!A$. Specifying to the vector meson, from Eqs.~\eqref{eq:fVfull0} and \eqref{eq:fVfullpm}, we obtain 
\begin{equation}
\Delta^V_{\rm full}(x) = N_c  \int \frac{ \d^2 k_{\bot}}{(2\pi)^3}   \frac{\chi_V}{\bar{x} } \frac{2}{M_V}  \frac{D_{V,\rm con}-m_1-m_2}{D_{V,\rm con}} 2 B_1^{(2)}. 
\end{equation}
Similarly, in order to discuss the difference between $[f_V]_{\rm SLF}^{\lbd=0}$ and $[f_V]_{\rm SLF}^{\lbd=\pm}$, we define
\begin{align}
\Delta^M_{\rm SLF}(x) \equiv \frac{\d [f_M]_{\rm SLF}^{\lbd=0}}{ \d x}-\frac{\d [f_M]_{\rm SLF}^{\lbd=\pm}}{ \d x}\,.
\end{align}

\begin{table}[t]
\begin{center}
\caption{\label{tab:fvdis} \small Numerical results for the decay constants (in unit of ${\rm MeV}$) of $\rho$ and $D^*$ mesons based on Eqs.~\eqref{eq:fVSLF0}--\eqref{eq:fVvalpm} and within the type-I and the type-II scheme.}
\vspace{0.1cm}
\let\oldarraystretch=\arraystretch
\renewcommand*{\arraystretch}{1.1}
\setlength{\tabcolsep}{13.8pt}
\begin{tabular}{lcccccccccccc}

\hline\hline
                 &$[f_{ \rho}]^{\lambda=0}_{\rm SLF}$
                 &$[f_{\rho}]^{\lambda=\pm}_{\rm SLF}$
                 &$[f_{\rho}]^{\lambda=0}_{\rm full}$
                 &$[f_{\rho}]^{\lambda=\pm}_{\rm full}$
                 &$[f_{\rho}]^{\lambda=0}_{\rm val.}$
                 &$[f_{\rho}]^{\lambda=\pm}_{\rm val.}$\\\hline
type-I    &{$211.1$}&$226.9$&$248.7$&$288.9$&$229.1$&$212.1$
              \\\hline
type-II    &{$211.1$}&{$211.1$}&{$211.1$}&{$211.1$}&{$211.1$}&{$211.1$}
\\\hline\hline
                 &$[f_{D^*}]^{\lambda=0}_{\rm  SLF}$
                 &$[f_{D^*}]^{\lambda=\pm}_{\rm SLF}$
                 &$[f_{D^*}]^{\lambda=0}_{\rm full}$
                 &$[f_{D^*}]^{\lambda=\pm}_{\rm full}$
                 &$[f_{D^*}]^{\lambda=0}_{\rm  val.}$
                 &$[f_{D^*}]^{\lambda=\pm}_{\rm  val.}$\\\hline
type-I     &$252.6$&$273.5$&$275.3$&$305.6$&$244.6$&$258.9$
              \\\hline
type-II    &$252.6$&$252.6$&$252.6$&$252.6$&$252.6$&$252.6$
\\\hline\hline
\end{tabular}
\end{center}
\end{table}

\begin{figure}[ht]
	\begin{center}
		\subfigure[]{\includegraphics[scale=0.21]{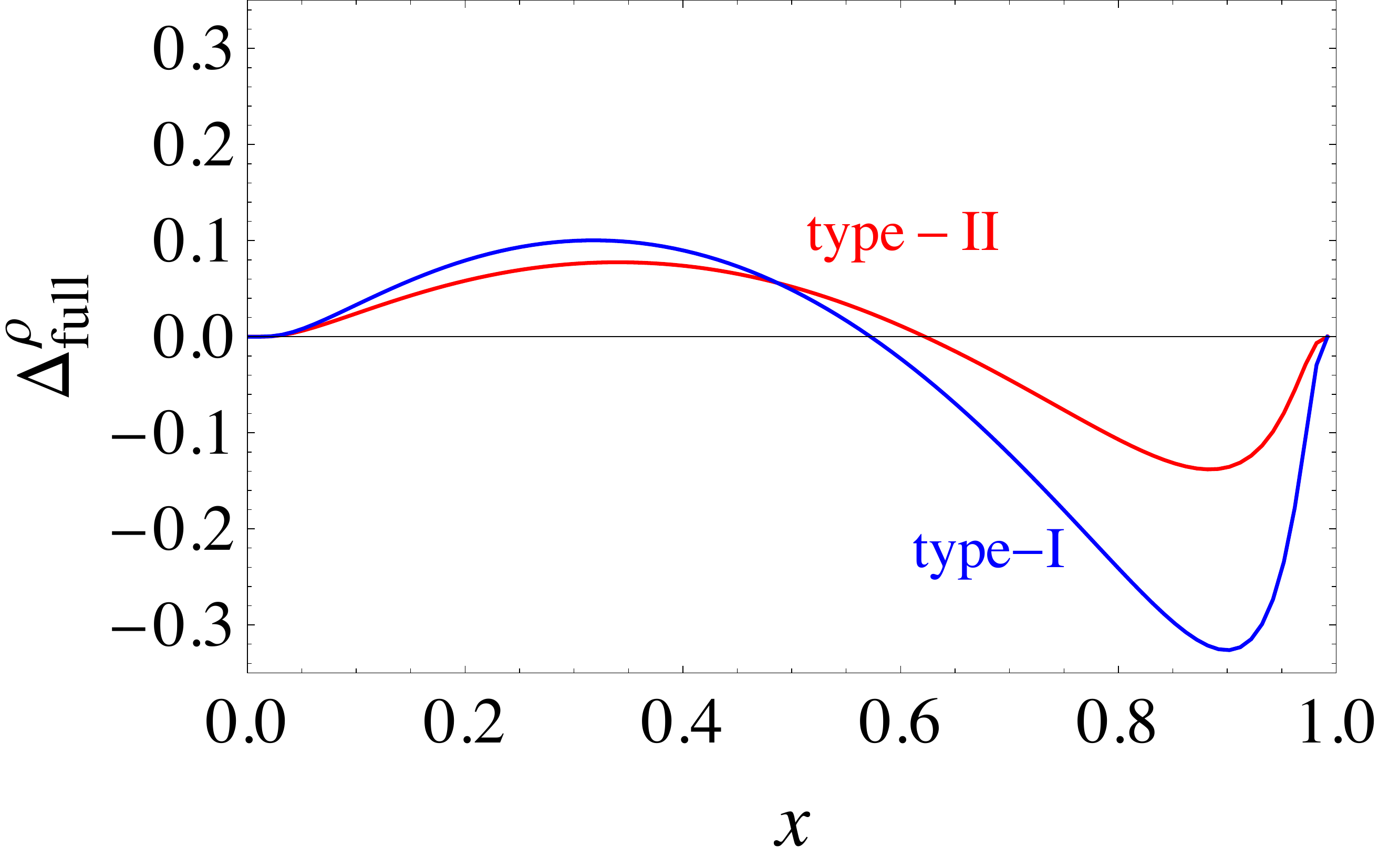}}\qquad\qquad
		\subfigure[]{\includegraphics[scale=0.22]{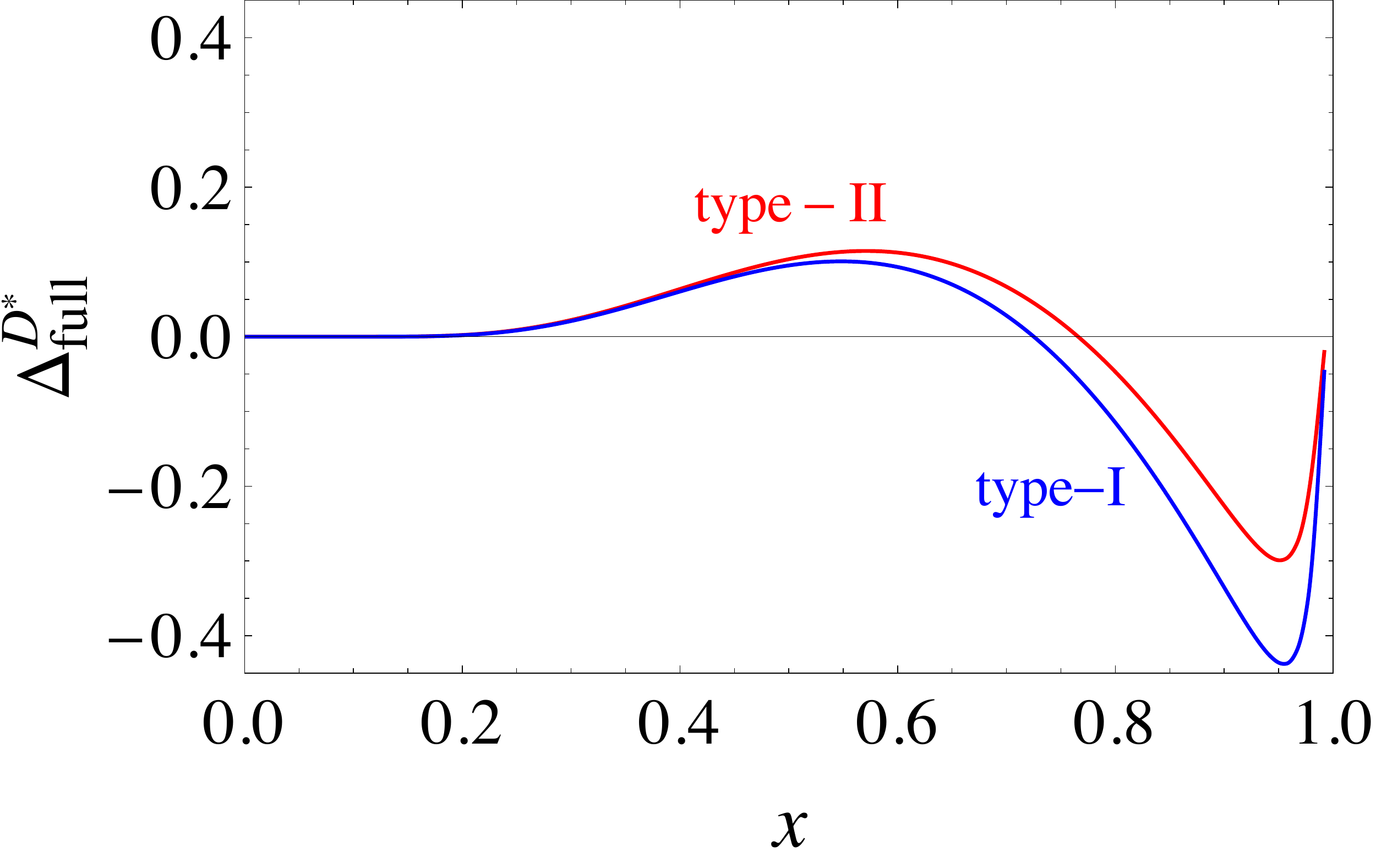}}\\
		\subfigure[]{\includegraphics[scale=0.22]{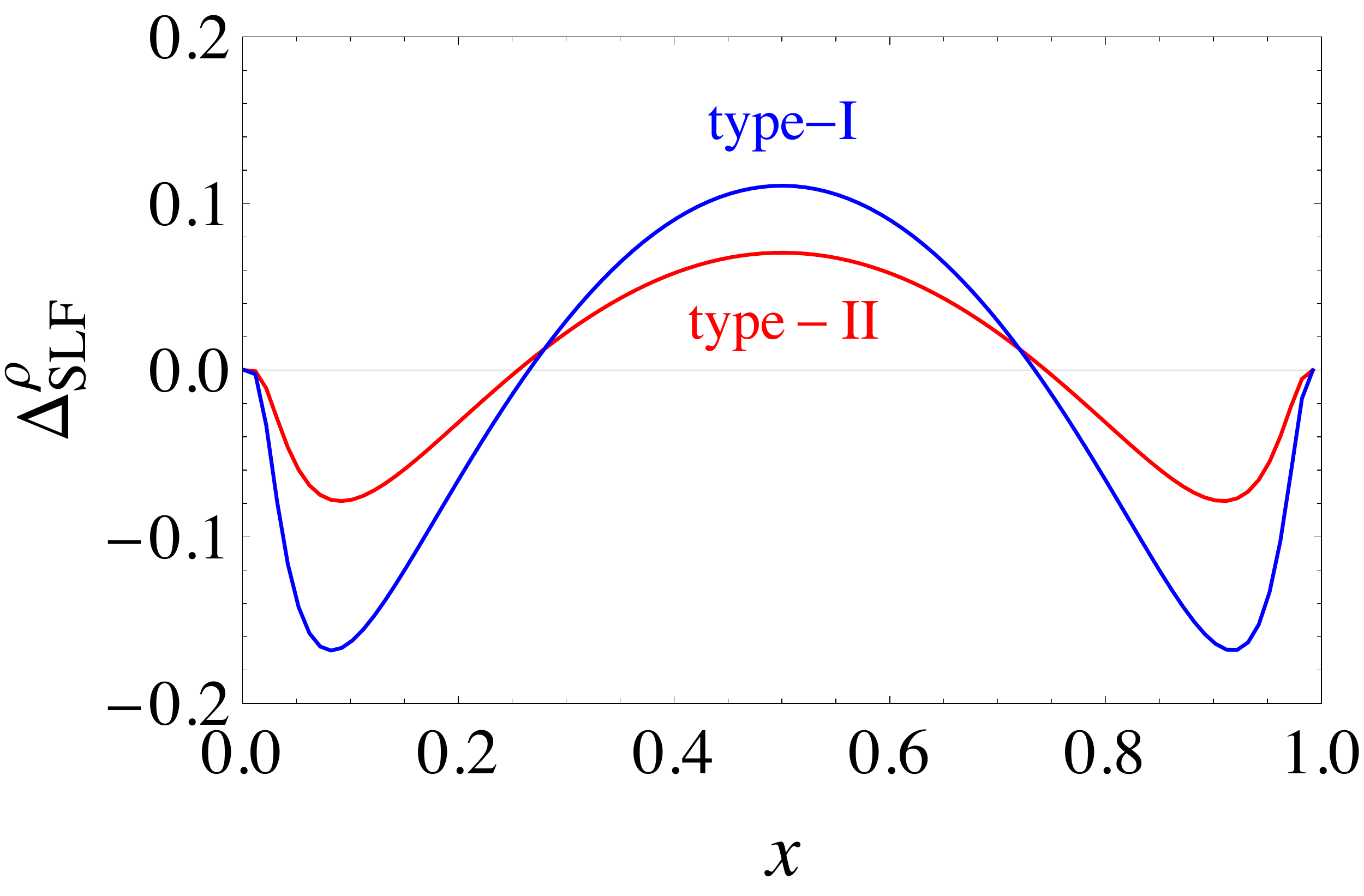}}\qquad\qquad
		\subfigure[]{\includegraphics[scale=0.225]{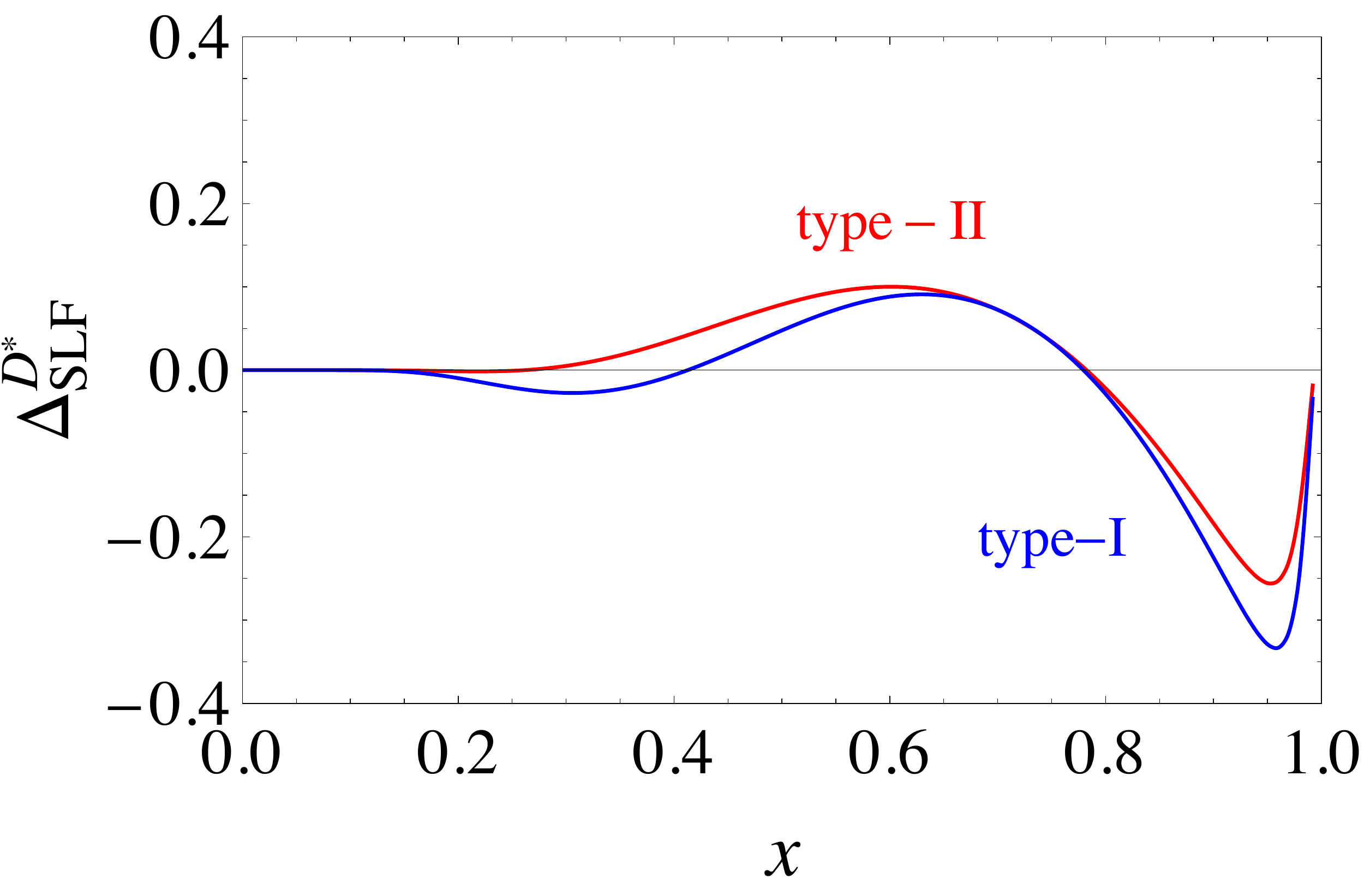}}\\ 
		\subfigure[]{\includegraphics[scale=0.225]{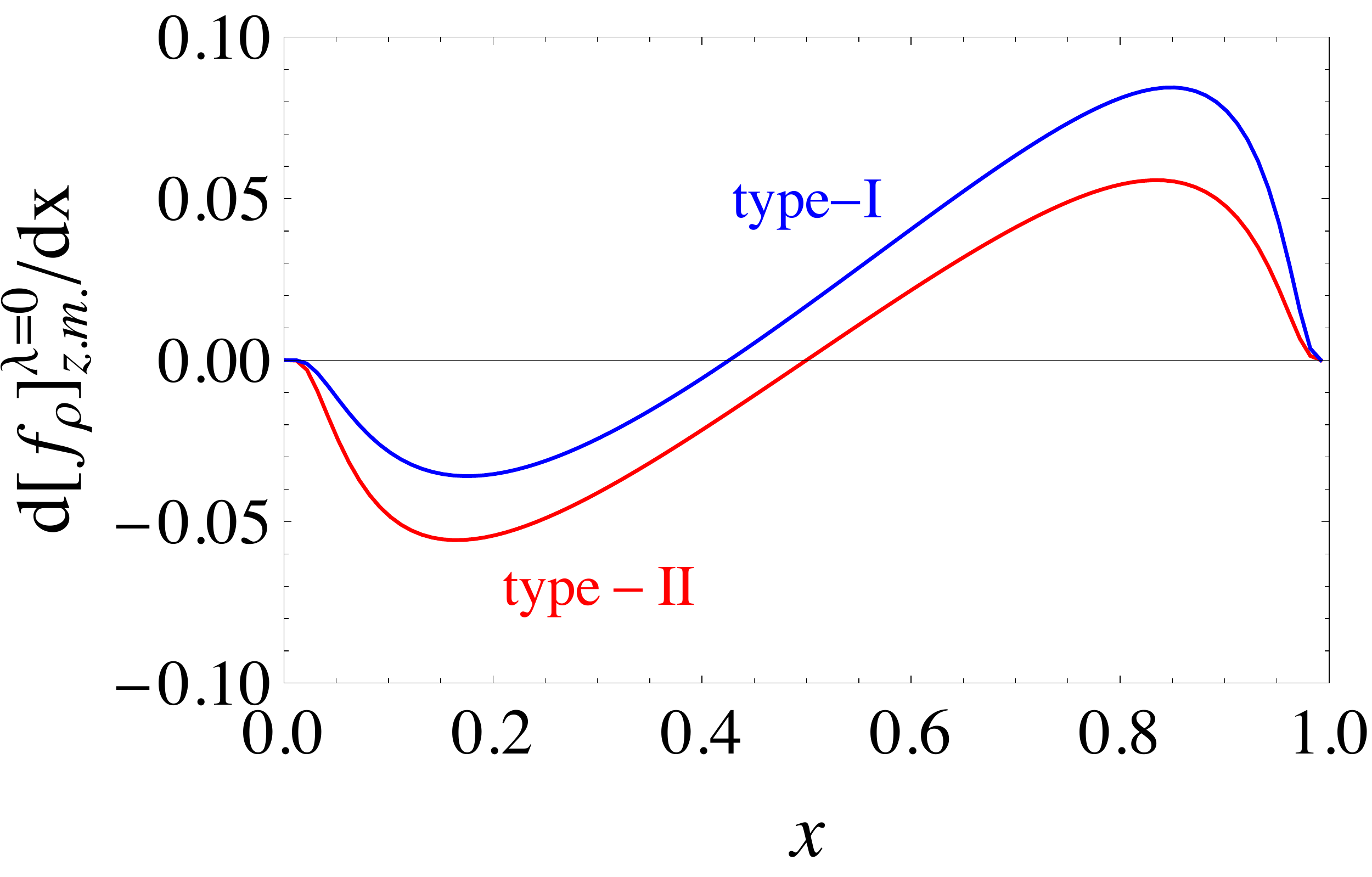}}\qquad\qquad
		\subfigure[]{\includegraphics[scale=0.22]{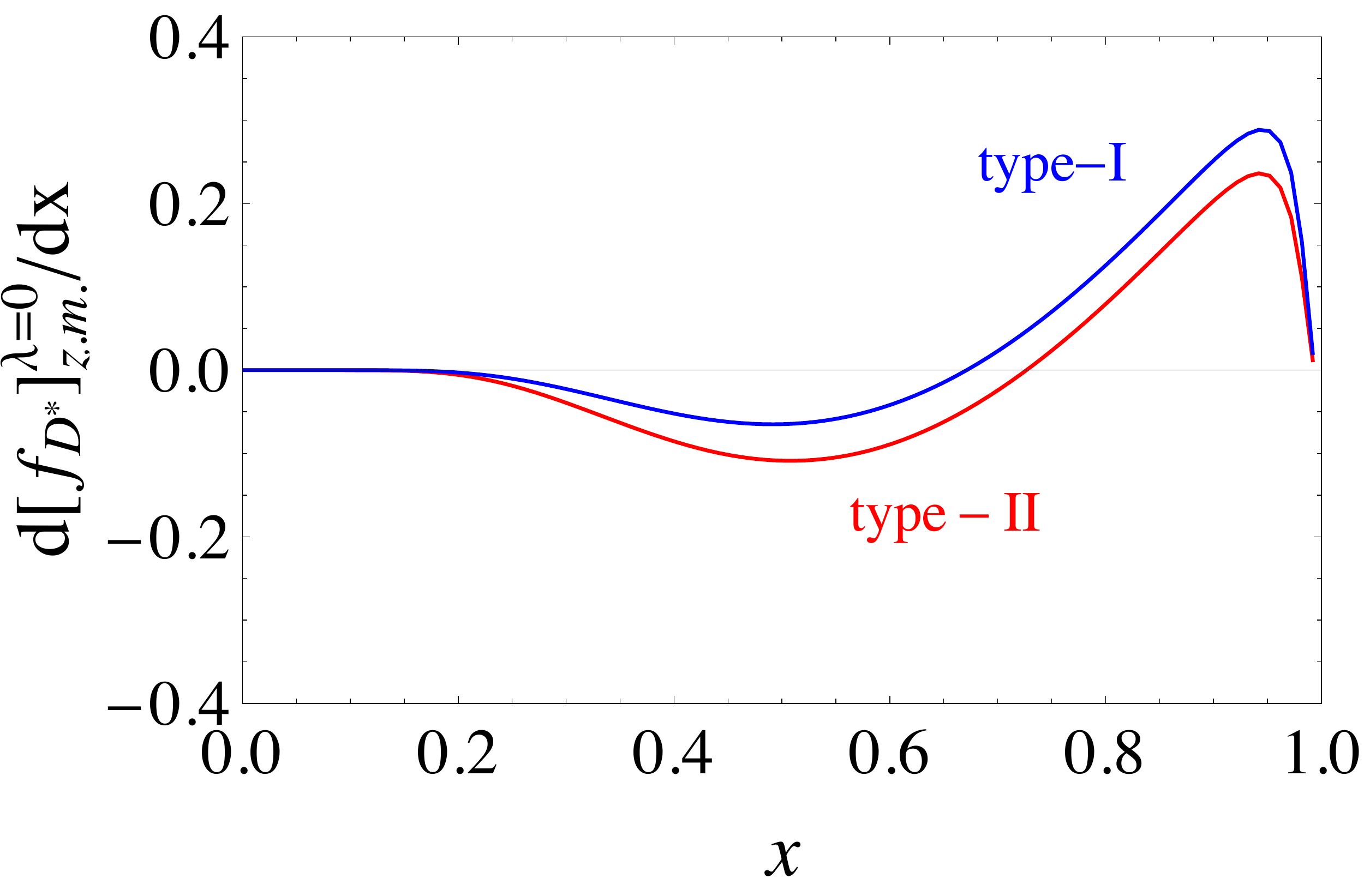}}\\ 
		\subfigure[]{\includegraphics[scale=0.22]{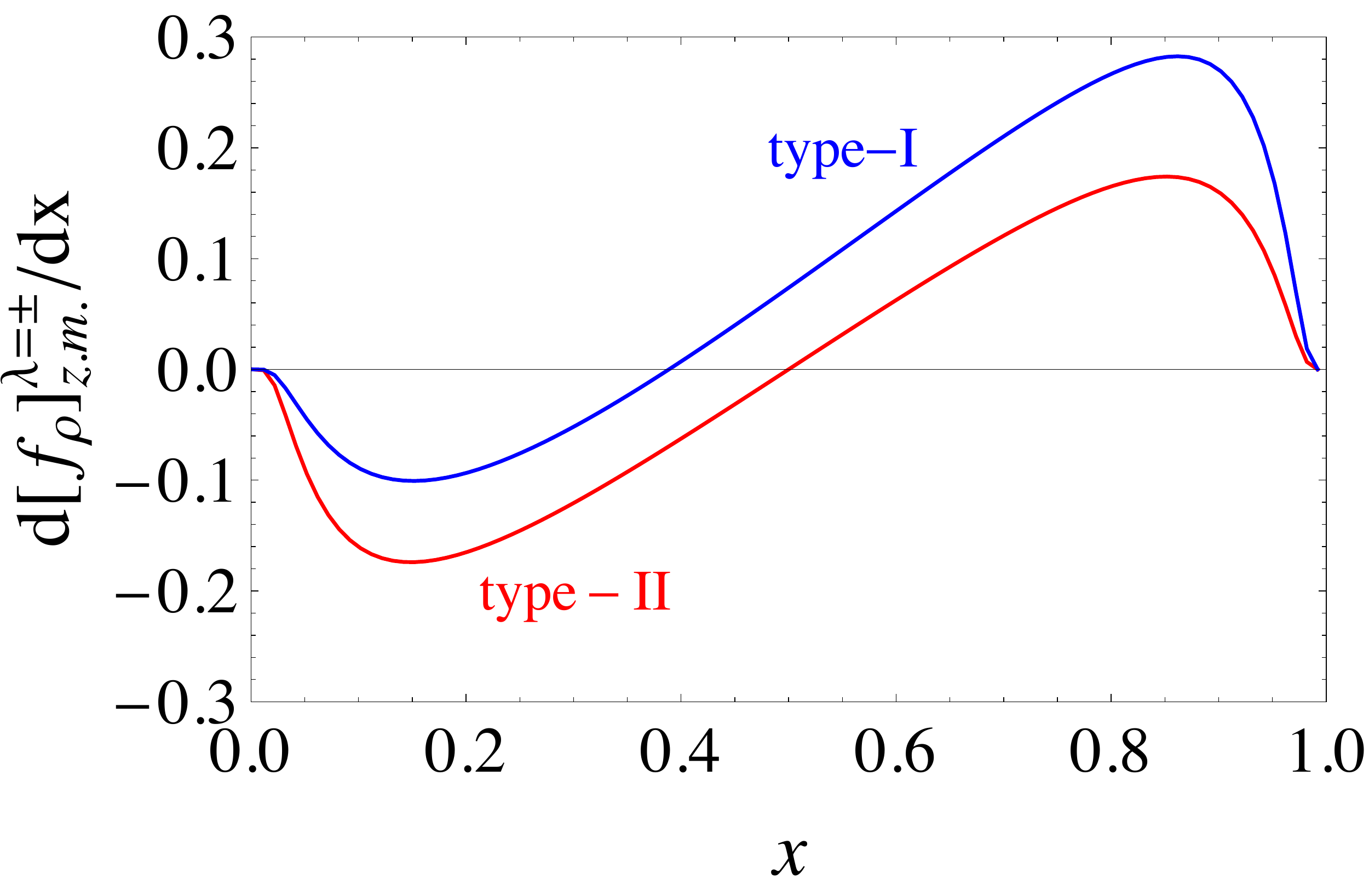}}\qquad\qquad
		\subfigure[]{\includegraphics[scale=0.22]{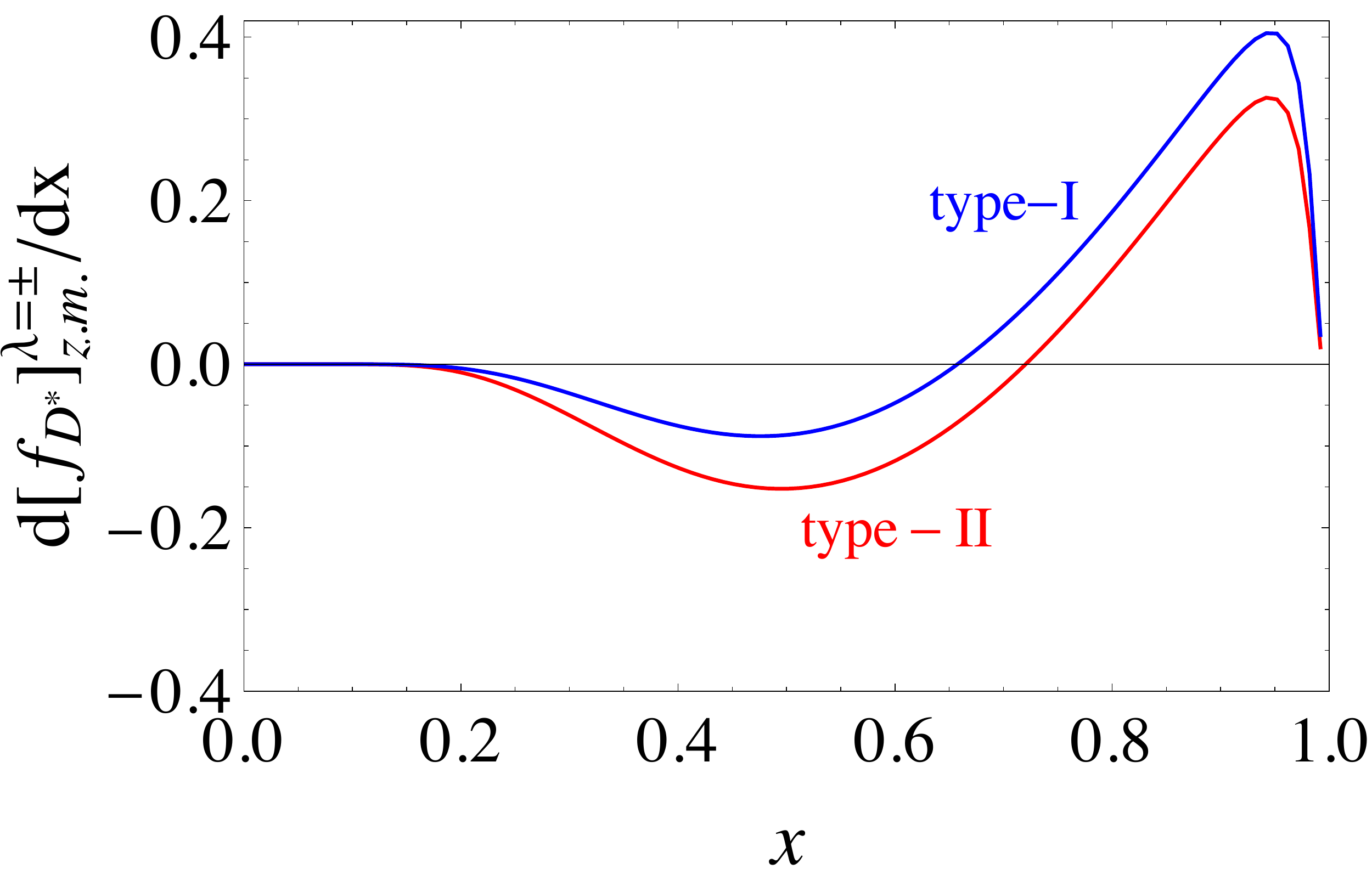}}
		\caption{\label{fig:dep1} \small Dependence of $\Delta^{V}_{\rm full}(x)$, $\Delta^{V}_{\rm SLF}(x)$ and $\d [f_{V}]_{\rm z.m.}^{\lbd=0\,,\pm}/\d x$ on the momentum fraction $x$. See text for details.}
	\end{center}
\end{figure}

For convenience of analyses and discussions about the relations among the decay constants given by Eqs.~\eqref{eq:fVSLF0}--\eqref{eq:fVvalpm}, we take the $\rho$ and $D^*$ mesons as examples, and present our numerical results in Table~\ref{tab:fvdis} by using the best-fit values of inputs given in Table~\ref{tab:input}. In addition, we show in Fig.~\ref{fig:dep1} the dependence of $\Delta^{V}_{\rm full}(x)$, $\Delta^{V}_{\rm SLF}(x)$ and $\d[f_V]_{\rm z.m.}^{\lbd=0}/\d x$~(where $V=\rho$ or $D^*$, and $[f_V]_{\rm z.m.}=[f_V]_{\rm full}-[f_V]_{\rm val.}$) on the momentum fraction $x$. Based on these numerical results and the theoretical formulas given above, the following discussions and findings can be made\footnote{Here it should be emphasized that all these findings based on the $\rho$ and $D^*$ mesons are also applicable for the other vector mesons unless stated otherwise.}:
\begin{itemize}
\item As pointed out by Cheng {\it et al.} in Ref.~\cite{Cheng:2003sm}, the CLF results for $f_V$ extracted via the $\lbd=0$ and $\lbd=\pm$ polarization states are different from each other due to the additional contribution to $[f_V]_{\rm full}^{\lbd=0}$ characterized by the coefficient $B_1^{(2)}$, which can be clearly seen from Eqs.~\eqref{eq:fVfull0} and \eqref{eq:fVfullpm}. Numerically, it is found from Figs.~\ref{fig:dep1}(a,b) and Table~\ref{tab:fvdis} that the $B_1^{(2)}$ term gives a nonzero contribution and results in about ${\cal O}(10\%)$ correction to $[f_V]_{\rm full}$ within the type-I scheme, which means that the CLF approach with the traditional type-I correspondence suffers the self-consistency problem, $[f_V]_{\rm full}^{\lbd=0}\neq [f_V]_{\rm full}^{\lbd=\pm}$ (type-I). However, within the type-II scheme, the positive $\Delta^V_{\rm full}$ at small $x$ and the negative one at large $x$ can exactly cancel each other, resulting in $\int\d x\Delta^V_{\rm full}=0$. This interesting observation can be roughly seen from Figs.~\ref{fig:dep1}(a) and \ref{fig:dep1}(b). As a consequence, we find that 
\begin{equation}
[f_V]_{\rm full}^{\lbd=0}\;\dot{=}\;[f_V]_{\rm full}^{\lbd=\pm}\,,\qquad (\text{type-II})
\end{equation}
where the symbol ``$\dot{=}$'' used throughout this paper denotes that the two quantities are equal to each other only numerically but not formally. This confirms the findings of Ref.~\cite{Choi:2013mda} and implies that the type-II correspondence might be a self-consistency scheme for the CLF approach. 
 
\item As is the case for the CLF quark model with the type-I correspondence, the traditional SLF quark model also encounters the self-consistency problem, $[f_V]_{\rm SLF}^{\lbd=0}\neq [f_V]_{\rm SLF}^{\lbd=\pm}$, which can be easily found by comparing Eqs.~\eqref{eq:fVSLF0} with \eqref{eq:fVSLFpm}. Inspired by the self-consistent results achieved by the type-II replacements in the CLF quark model, we now test whether the self-consistency of the SLF quark model also requires the  replacement $M\to M_0$. In analogous to the discussions in the CLF approach, we name the traditional SLF approach and the one with an additional $M\to M_0$ replacement in the integrand as the type-I and the type-II scheme, respectively. From Figs.~\ref{fig:dep1}(c,d) and Table~\ref{tab:fvdis}, it can be seen that $[f_V]_{\rm SLF}^{\lbd=0}<[f_V]_{\rm SLF}^{\lbd=\pm}$ in the traditional SLF quark model (type-I), while, after making the replacement $M\to M_0$,
\begin{equation}
[f_V]_{\rm SLF}^{\lbd=0}\;\dot{=}\;[f_V]_{\rm SLF}^{\lbd=\pm}\,,\qquad (\text{type-II})
\end{equation}
because $\int\d x\Delta^V_{\rm SLF}=0$. This implies that the replacement $M\to M_0$ is also required by the self-consistency of the SLF quark model.

\item Comparing Eqs.~\eqref{eq:fVSLF0} and \eqref{eq:fVSLFpm} with Eqs.~\eqref{eq:fVval0} and \eqref{eq:fVvalpm}, we do not find any relations between $[f_V]_{\rm SLF}^{\lambda=0\,,\pm}$ and $[f_V]_{\rm val.}^{\lbd=0\,,\pm}$ within the type-I scheme; however, employing the type-II scheme and making some simplifications on these formulas, we find surprisingly that the SLF results are exactly the same as the valence contributions in the CLF approach, 
 \begin{equation}\label{eq:VSLFval}
 [f_V]_{\rm SLF}^{\lbd=0}\,=\,[f_V]_{\rm val.}^{\lbd=0} \quad\text{and}\quad [f_V]_{\rm SLF}^{\lbd=\pm}\, = \,[f_V]_{\rm val.}^{\lbd=\pm}\,, \qquad(\text{type-II})
\end{equation}
which can also been seen from the numerical results given in Table~\ref{tab:fvdis}.

\item The effect of the zero-mode contributions to $f_V$, $[f_V]_{\rm z.m.}$, are shown in Figs.~\ref{fig:dep1}(e--h). It can be found that, within the type-I scheme, the zero-mode effect presents a sizable positive correction to $f_V$, but its contributions to $f_V^{\lbd=0}$ and $f_V^{\lbd=\pm}$ are different from each other, with $[f_V]_{\rm z.m.}^{\lbd=0}<[f_V]_{\rm z.m.}^{\lbd=\pm}$ numerically. Within the type-II scheme, however, although existing formally, the zero-mode contributions vanish numerically, $[f_V]_{\rm z.m.}^{\lbd=0,\pm}\dot{=}0$. A very obvious example is the $\rho$ meson shown in Figs.~\ref{fig:dep1}(e) and \ref{fig:dep1}(g).  This in turn implies that
 \begin{align}
 [f_V]_{\rm full}^{\lbd=0}\;\dot{=}\;[f_V]_{\rm val.}^{\lbd=0} \quad\text{and}\quad [f_V]_{\rm full}^{\lbd=\pm}\;\dot{=}\;[f_V]_{\rm val.}^{\lbd=\pm}\,, \qquad(\text{type-II})
\end{align}
which have also been demonstrated by the numerical results given in Table~\ref{tab:fvdis}. 
\end{itemize}

\begin{table}[t]
\begin{center}
\caption{\label{tab:fvpre} \small Updated predictions for $f_{V}$ (in unit of ${\rm MeV}$) in the LF approach, where the errors are due to the uncertainties of the parameter $\beta$ given in Table~\ref{tab:input} obtained by fitting to the available data on $f_P$. The experimental data as well as the LQCD and QCDSR predictions are also shown for comparision. }
\vspace{0.1cm}
\let\oldarraystretch=\arraystretch
\renewcommand*{\arraystretch}{1.1}
\setlength{\tabcolsep}{18.2pt}
\begin{tabular}{lcccccccccccc}

\hline\hline
                 & data              &LQCD              & QCD SR
                 &this work\\  \hline

$f_{\rho}$   &$210\pm4$\cite{Straub:2015ica} &$199\pm4$\cite{Braun:2016wnx}
                        &$206\pm7$\cite{Ball:2006eu}
                        &{$211\pm1$}
\\ \hline
$f_{K^*}$    &$204\pm7$\cite{Straub:2015ica} &$-$
                        &$222\pm8$\cite{Ball:2006eu}
                        &$223\pm1$
\\
$f_{\phi}$   &$228.5\pm3.6$\cite{Chakraborty:2017hry}
             &$238\pm3$\cite{Chakraborty:2017hry}
             &{$215\pm5$}\cite{Ball:2006eu}
             &$236\pm1$
\\\hline
$f_{D^*}$    &$-$ &$223.5\pm8.4$\cite{Lubicz:2016bbi}
                  &$250\pm8$\cite{Narison:2015nxh}
                  &$253\pm7$
\\
$f_{D_s^*}$  &$301\pm13$\cite{Yu:2015xwa}
                  &$268.8\pm6.6$\cite{Lubicz:2016bbi}
                  &$290\pm11$\cite{Narison:2015nxh}
                  &$314\pm6$
\\
$f_{J/\psi}$ &$411\pm5$\cite{Kher:2018wtv}
             &$418\pm9$\cite{Becirevic:2013bsa}
             &{$401\pm46$}\cite{Becirevic:2013bsa}
             &$382\pm {96}$
\\\hline
$f_{B^*}$   &$-$ &$185.9\pm7.2$\cite{Lubicz:2016bbi}
                 &$210\pm6$\cite{Narison:2015nxh}
                 &$205\pm5$
\\
$f_{B_s^*}$  &$-$ &$223.1\pm5.4$\cite{Lubicz:2016bbi}
                  &$221\pm7$\cite{Narison:2015nxh}
                  &$246\pm4$
\\
$f_{B_c^*}$  &$-$  &$422\pm13$\cite{Colquhoun:2015oha}
                   &{$453\pm20$}\cite{Narison:2015nxh}
                   &$465\pm7$
\\
$f_{\Upsilon(1S)}$ &$708\pm8$\cite{Negash:2015rua}
                   &$-$
                   &$-$
                   &$713\pm34$
\\\hline\hline
\end{tabular}
\end{center}
\end{table}

Combining all the findings given above, we can finally conclude that 
\begin{align}\label{eq:confV}
  [f_V]_{\rm SLF}^{\lbd=0}\;=\;[f_V]_{\rm val.}^{\lbd=0}\;\dot{=}\; [f_V]_{\rm full}^{\lbd=0}\;\dot{=}\;[f_V]_{\rm full}^{\lbd=\pm} \;\dot{=}\; [f_V]_{\rm val.}^{\lbd=\pm}\;=\;[f_V]_{\rm SLF}^{\lbd=\pm}\,,\qquad(\text{type-II})
\end{align}
within the type-II scheme, in which $[f_V]_{\rm SLF}^{\lbd=0}\;\dot{=}\;[f_V]_{\rm SLF}^{\lbd=\pm}$ and $[f_V]_{\rm full}^{\lbd=0}\;\dot{=}\;[f_V]_{\rm full}^{\lbd=\pm}$ reflect the self-consistencies of SLF and CLF quark models, respectively. However, none of these relations holds within the type-I scheme. Finally, using the inputs listed in Table~\ref{tab:input} and employing the self-consistent type-II scheme, we present in Table~\ref{tab:fvpre} our updated predictions for $f_V$ in the LF approach. It can be easily found that our updated results are generally in consistence with the experimental data as well as the theoretical results obtained in the LQCD~\cite{Becirevic:2013bsa,Chakraborty:2017hry,Braun:2016wnx,Lubicz:2016bbi,Colquhoun:2015oha}and QCD sum rules~(QCDSR)~\cite{Becirevic:2013bsa,Ball:2006eu,Narison:2015nxh,Wang:2015mxa,Khodjamirian:2008xt,Gelhausen:2013wia} approaches. 

\subsection{Decay constants of $^1\!A$ and $^3\!A$ mesons}

The decay constants of axial-vector mesons are defined by
\begin{eqnarray}
\label{eq:dcA}
\la 0 | \bar q_2\r^\u\r_5 q_1|^{3(1)}\!A(p,\lbd)\ra = f_{^{3(1)}\!A}\, M_{^{3(1)}\!A}{\e}^{\u}_{\lbd}\,.
\end{eqnarray}
Using the theoretical formulas given in Sec.~\ref{sec:2} and taking the $\lbd=0$ and $\lbd=\pm$ polarization states respectively, we obtain the SLF results:
\begin{align}
\label{eq:f1ASLF0}
[ f_{^1\!A}]_{\rm SLF}^{\lambda=0}=
&-\sqrt{ N_c} \int\frac{\d x \d^2{ k}_{\bot}}{(2\pi)^3}  \frac{\psi_p(x,{k}_{\bot})}{\sqrt{x\,\bar{x}}}     \frac{1}{\sqrt{2}\hat{M}_0}  \frac{2}{M_0} \frac{ (\bar{x}m_1+xm_2) [ (\bar{x}-x)k_{\bot}^2+\bar{x}^2m_1^2-x^2m_2^2]}{x\bar{x}D_{^1\!A, {\rm LF}}} \,,\\
 \label{eq:f1ASLFpm}
 [f_{^1\!A}]_{\rm SLF}^{\lambda=\pm}
=&-\sqrt{ N_c} \int\frac{\d x \d^2{ k}_{\bot}}{(2\pi)^3}  \frac{\psi_p(x,{k}_{\bot})}{\sqrt{x\,\bar{x}}}      \frac{1}{\sqrt{2}\hat{M}_0} \frac{2}{M_{^1\!A}} \frac{m_1-m_2}{D_{^1\!A, {\rm LF}}}k_{\bot}^2\,,
\end{align}
for the $^1\!A$ meson, and
\begin{align}
\label{eq:f3ASLF0}
[ f_{^3\!A}]_{\rm SLF}^{\lambda=0}
=&\sqrt{ N_c} \int \frac{\d x\d^2{ k}_{\bot}}{(2\pi)^3}  \frac{\psi_p(x,{k}_{\bot})}{\sqrt{x\,\bar{x}}}  \,   \frac{1}{\sqrt{2}\hat{M}_0} \,\frac{\hat{M}_0^2}{2\sqrt{2} M_0} \frac{2}{M_0}\bigg\{  2k_{\bot}^2+(m_1-m_2)(\bar{x}m_1-xm_2)\,
\nonumber \\[0.1cm]
&-\frac{ (\bar{x}m_1+xm_2) [ (\bar{x}-x)k_{\bot}^2+\bar{x}^2m_1^2-x^2m_2^2]}{x\bar{x}\,D_{^3\!A, {\rm LF}}} \bigg\}\,,\\[0.2cm]
 \label{eq:f3ASLFpm}
 [f_{^3\!A}]_{\rm SLF}^{\lambda=\pm}
=&\sqrt{ N_c} \int\frac{\d x \d^2{ k}_{\bot}}{(2\pi)^3} \frac{\psi_p(x,{k}_{\bot})}{\sqrt{x\,\bar{x}}}    \,  \frac{1}{\sqrt{2}\hat{M}_0} \,\frac{\hat{M}_0^2}{2\sqrt{2} M_0}\, \frac{2}{M_{^3\!A}} \bigg[ \frac{k_{\bot}^2-2\bar{x}x k_{\bot}^2+(\bar{x}m_1-xm_2)^2}{2\bar{x}x}\nonumber \\[0.1cm]
& -\frac{k_{\bot}^2(m_1-m_2)}{D_{^3\!A, {\rm LF}}}  \bigg ]\,,
\end{align}
for the $^3\!A$ meson. Employing the CLF approach, on the other hand, we obtain 
\begin{align}
\label{eq:f1Afull0}
[f_{^1\!A}]_{\rm full}^{\lbd=0}=&-N_c  \int \frac{  \d x \d^2 k_{\bot}}{(2\pi)^3}   \frac{\chi_{^1\!A}}{\bar{x} } \frac{2}{M_{^1\!A}}  \frac{m_1-m_2}{D_{^1\!A, {\rm con}}}\left( k_{\bot}^2 -2 B_1^{(2)}\right) \,,\\[0.2cm]
\label{eq:f1Afullpm}
[f_{^1\!A}]_{\rm full}^{\lbd=\pm}=&-N_c  \int \frac{  \d x \d^2 k_{\bot}}{(2\pi)^3}   \frac{\chi_{^1\!A}}{\bar{x} } \frac{2}{M_{^1\!A}}  \frac{m_1-m_2}{D_{^1\!A, {\rm con}}} k_{\bot}^2\,,
\end{align}
for the $^1\!A$ meson, and 
\begin{align}
\label{eq:f3Afull0}
[f_{^3\!A}]_{\rm full}^{\lbd=0}=&N_c \int \frac{  \d x \d^2 k_{\bot}}{(2\pi)^3}   \frac{\chi_{^3\!A}}{\bar{x}} \frac{2}{M_{^3\!A}} \bigg\{ xM_0^2 - m_1(m_1{+}m_2)  -\left(1+\frac{m_1{-}m_2}{D_{^3\!A, {\rm con}}}\right) (k_{\bot}^2 -2 B_1^{(2)})  \bigg\} \,,\\[0.2cm]
\label{eq:f3Afullpm}
[f_{^3\!A}]_{\rm full}^{\lbd=\pm}=& N_c \int \frac{  \d x \d^2 k_{\bot}}{(2\pi)^3}   \frac{\chi_{^3\!A}}{\bar{x} } \frac{2}{M_{^3\!A}}  \left[ xM_0^2 - m_1(m_1{+}m_2)  -\left(1+\frac{m_1{-}m_2}{D_{^3\!A, {\rm con}}}\right)   k_{\bot}^2 \right] \,,
\end{align}
for the $^3\!A$ meson. At the same time, the valence contributions in the CLF approach are given, respectively, by
\begin{align}
\label{eq:f1Aval0}
[f_{^1\!A}]_{\rm val.}^{\lbd=0}=&-N_c  \int \frac{  \d x \d^2 k_{\bot}}{(2\pi)^3}   \frac{\chi_{^1\!A}}{\bar{x} } \frac{2}{M_{^1\!A}}  \frac{M_{^1\!A}^2\bar{x}^2-m_2^2-k_{\bot}^2}{\bar{x}D_{^1\!A, {\rm con}}}\, \left(\bar{x}m_1 {+}  xm_2\right) \,,\\[0.2cm]
\label{eq:f1Avalpm}
[f_{^1\!A}]_{\rm val.}^{\lbd=\pm}=&-N_c  \int \frac{  \d x \d^2 k_{\bot}}{(2\pi)^3}   \frac{\chi_{^1\!A}}{\bar{x} } \frac{2}{M_{^1\!A}}  \frac{m_1{-}m_2}{D_{^1\!A, {\rm con}}}k_{\bot}^2\,,
\end{align}
for the $^1\!A$ meson, and 
\begin{align}
\label{eq:f3Aval0}
[f_{^3\!A}]_{\rm val.}^{\lbd=0}=&N_c \int \frac{  \d x \d^2 k_{\bot}}{(2\pi)^3}   \frac{\chi_{^3\!A}}{\bar{x}} \frac{2}{M_{^3\!A}} \left[k_{\bot}^2+x\bar{x}M_{^3\!A}^2{-}m_1 m_2-\frac{M_{^3\!A}^2\bar{x}^2-m_2^2-k_{\bot}^2}{\bar{x}D_{^3\!A, {\rm con}}}\, \left(\bar{x}m_1 {+} xm_2\right)\right] \,,\\
\label{eq:f3Avalpm}
[f_{^3\!A}]_{\rm val.}^{\lbd=\pm}=& N_c \int \frac{  \d x \d^2 k_{\bot}}{(2\pi)^3}   \frac{\chi_{^3\!A}}{\bar{x} } \frac{2}{M_{^3\!A}}  \left[ \frac{\bar{x}M_{^3\!A}^2+xM_0^2-(m_1{+}m_2)^2 }{2}-\left(1{+}\frac{m_1{-}m_2}{D_{^3\!A, {\rm con}}}  \right)  k_{\bot}^2  \right] \,,
\end{align}
for the $^3\!A$ meson.

\begin{table}[t]
\begin{center}
\caption{\label{tab:fA} \small Numerical results for the decay constants (in unit of ${\rm MeV}$) of $^{1(3)}\!A_{(q\bar{q})}$ and $^{1(3)}\!A_{(c\bar{q})}$ mesons given by Eqs.~(\ref{eq:f1ASLF0}--\ref{eq:f3Avalpm}) and within the type-I and the type-II scheme.}
\vspace{0.1cm}
\let\oldarraystretch=\arraystretch
\renewcommand*{\arraystretch}{1.1}
\setlength{\tabcolsep}{8.8pt}
\begin{tabular}{lcccccccccccc}
\hline\hline
                 &$[f_{^1\!A_{(q\bar{q})}}]^{\lambda=0}_{\rm SLF}$
                 &$[f_{^1\!A_{(q\bar{q})}}]^{\lambda=\pm}_{\rm SLF}$
                 &$[f_{^1\!A_{(q\bar{q})}}]^{\lambda=0}_{\rm full}$
                 &$[f_{^1\!A_{(q\bar{q})}}]^{\lambda=\pm}_{\rm full}$
                 &$[f_{^1\!A_{(q\bar{q})}}]^{\lambda=0}_{\rm val.}$
                 &$[f_{^1A_{(q\bar{q})}}]^{\lambda=\pm}_{\rm val.}$\\\hline
type-I     &$0$&$0$&$0$&$0$&$-47.4$&$0$
              \\\hline
type-II     &$0$&$0$&$0$&$0$&$0$&$0$
              \\\hline\hline
                 &$[f_{^1\!A_{(c\bar{q})}}]^{\lambda=0}_{\rm SLF}$
                 &$[f_{^1\!A_{(c\bar{q})}}]^{\lambda=\pm}_{\rm SLF}$
                 &$[f_{^1\!A_{(c\bar{q})}}]^{\lambda=0}_{\rm full}$
                 &$[f_{^1\!A_{(c\bar{q})}}]^{\lambda=\pm}_{\rm full}$
                 &$[f_{^1\!A_{(c\bar{q})}}]^{\lambda=0}_{\rm val.}$
                 &$[f_{^1A_{(c\bar{q})}}]^{\lambda=\pm}_{\rm val.}$\\\hline
type-I     &$-78.5$&$-84.6$&$-78.4$&$-84.6$&$-65.2$&$-84.6$
              \\\hline
type-II    &$-78.5$&$-78.5$&$-78.5$&$-78.5$&$-78.5$&$-78.5$
\\\hline\hline
                 &$[f_{^3\!A_{(q\bar{q})}}]^{\lambda=0}_{\rm SLF}$
                 &$[f_{^3\!A_{(q\bar{q})}}]^{\lambda=\pm}_{\rm SLF}$
                 &$[f_{^3\!A_{(q\bar{q})}}]^{\lambda=0}_{\rm full}$
                 &$[f_{^3\!A_{(q\bar{q})}}]^{\lambda=\pm}_{\rm full}$
                 &$[f_{^3\!A_{(q\bar{q})}}]^{\lambda=0}_{\rm val.}$
                 &$[f_{^3\!A_{(q\bar{q})}}]^{\lambda=\pm}_{\rm val.}$\\\hline
type-I     &$218.7$&$223.6$&$260.6$&$223.6$&$263.1$&$263.1$
              \\\hline
type-II    &$218.7$&$218.7$&$218.7$&$218.7$&$218.7$&$218.7$
\\\hline\hline
                 &$[f_{^3\!A_{(c\bar{q})}}]^{\lambda=0}_{\rm SLF}$
                 &$[f_{^3\!A_{(c\bar{q})}}]^{\lambda=\pm}_{\rm SLF}$
                 &$[f_{^3\!A_{(c\bar{q})}}]^{\lambda=0}_{\rm full}$
                 &$[f_{^3\!A_{(c\bar{q})}}]^{\lambda=\pm}_{\rm full}$
                 &$[f_{^3\!A_{(c\bar{q})}}]^{\lambda=0}_{\rm val.}$
                 &$[f_{^3\!A_{(c\bar{q})}}]^{\lambda=\pm}_{\rm val.}$\\\hline
type-I     &$231.7$&$256.7$&$244.7$&$256.7$&$228.5$&$228.5$
              \\\hline
type-II    &$231.7$&$231.7$&$231.7$&$231.7$&$231.7$&$231.7$
\\\hline\hline

\end{tabular}
\end{center}
\end{table}

\begin{figure}[ht]
\begin{center}
\subfigure[]{\includegraphics[scale=0.21]{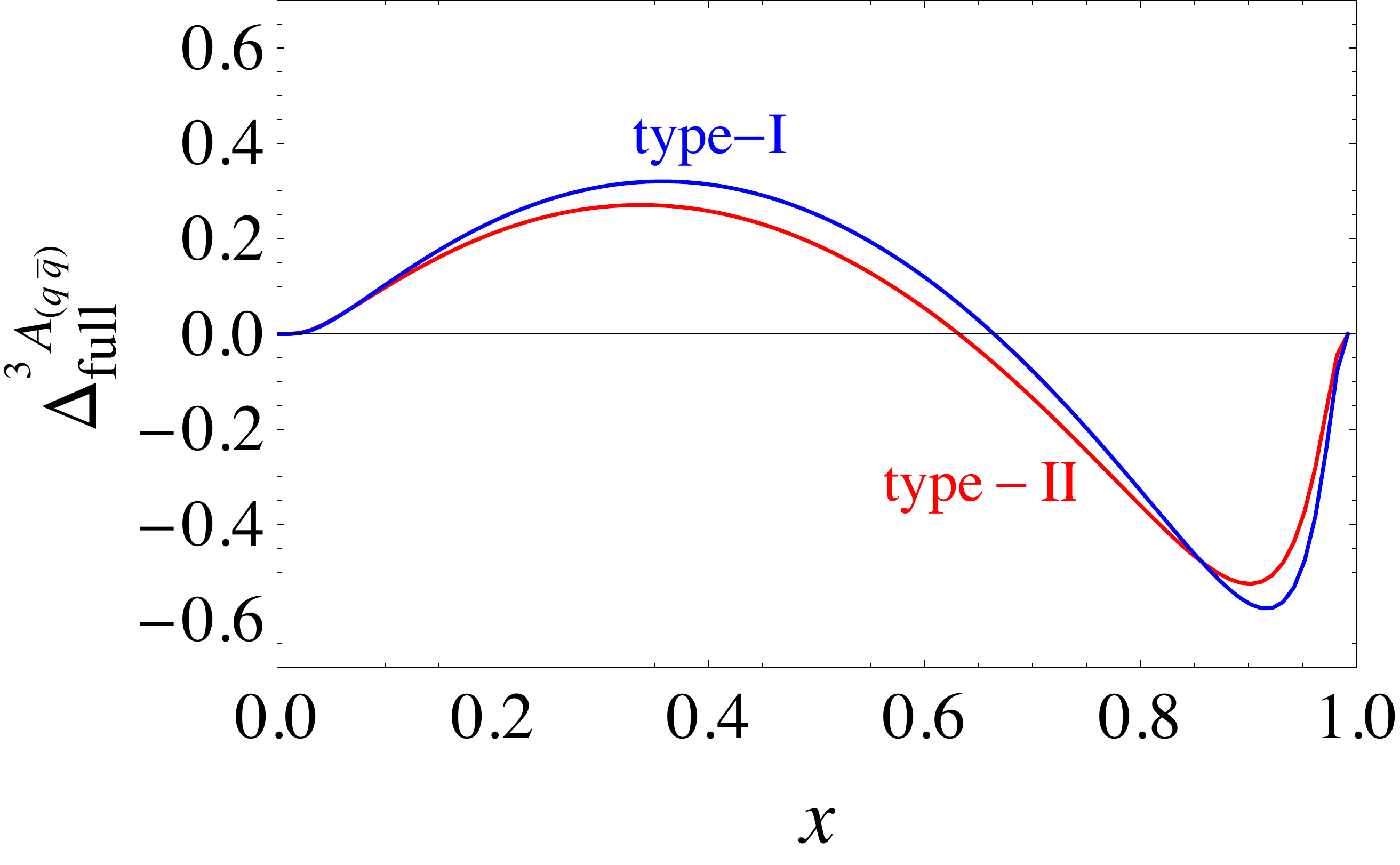}}\qquad\qquad
\subfigure[]{\includegraphics[scale=0.21]{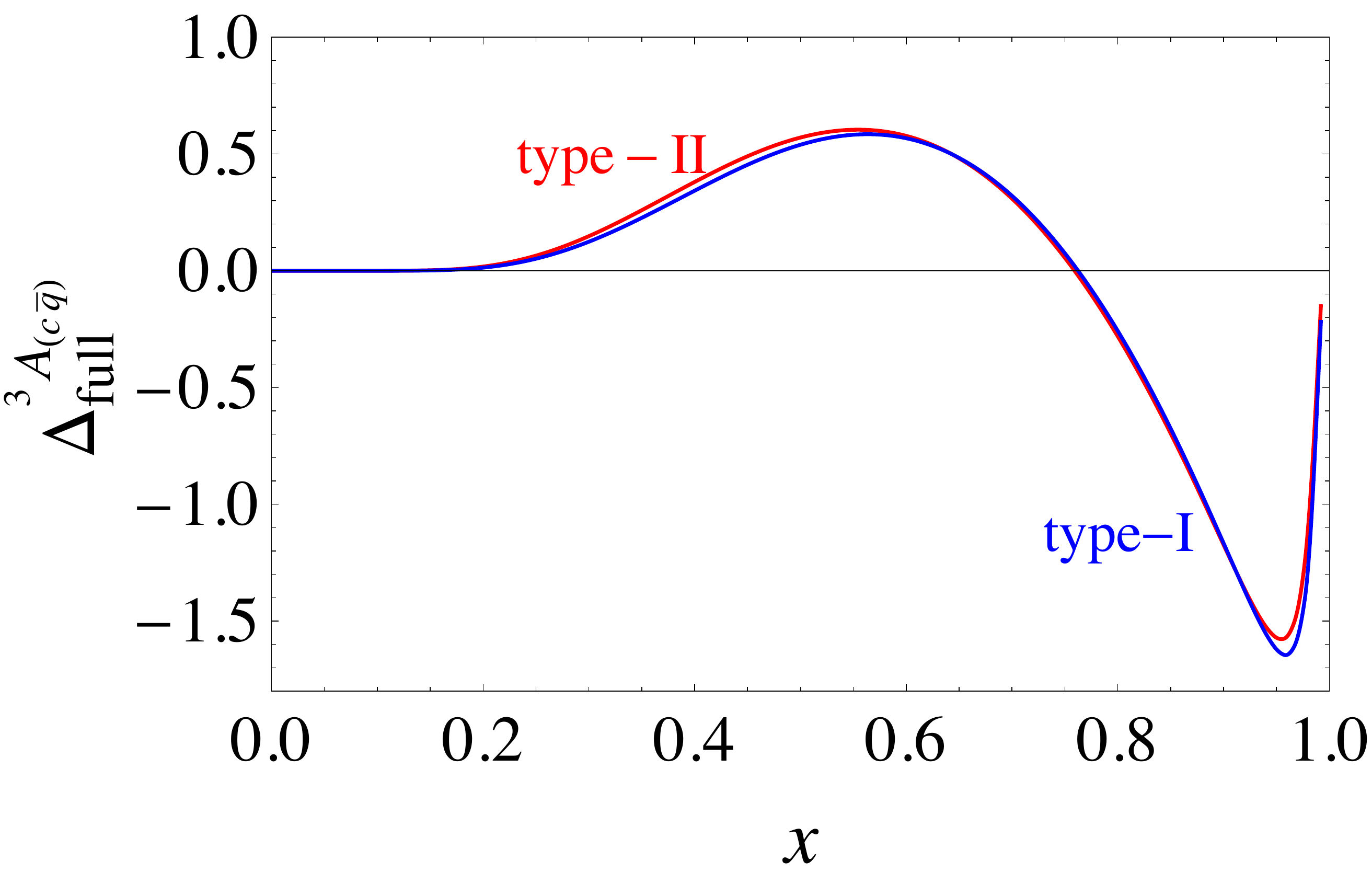}}\\
\subfigure[]{\includegraphics[scale=0.21]{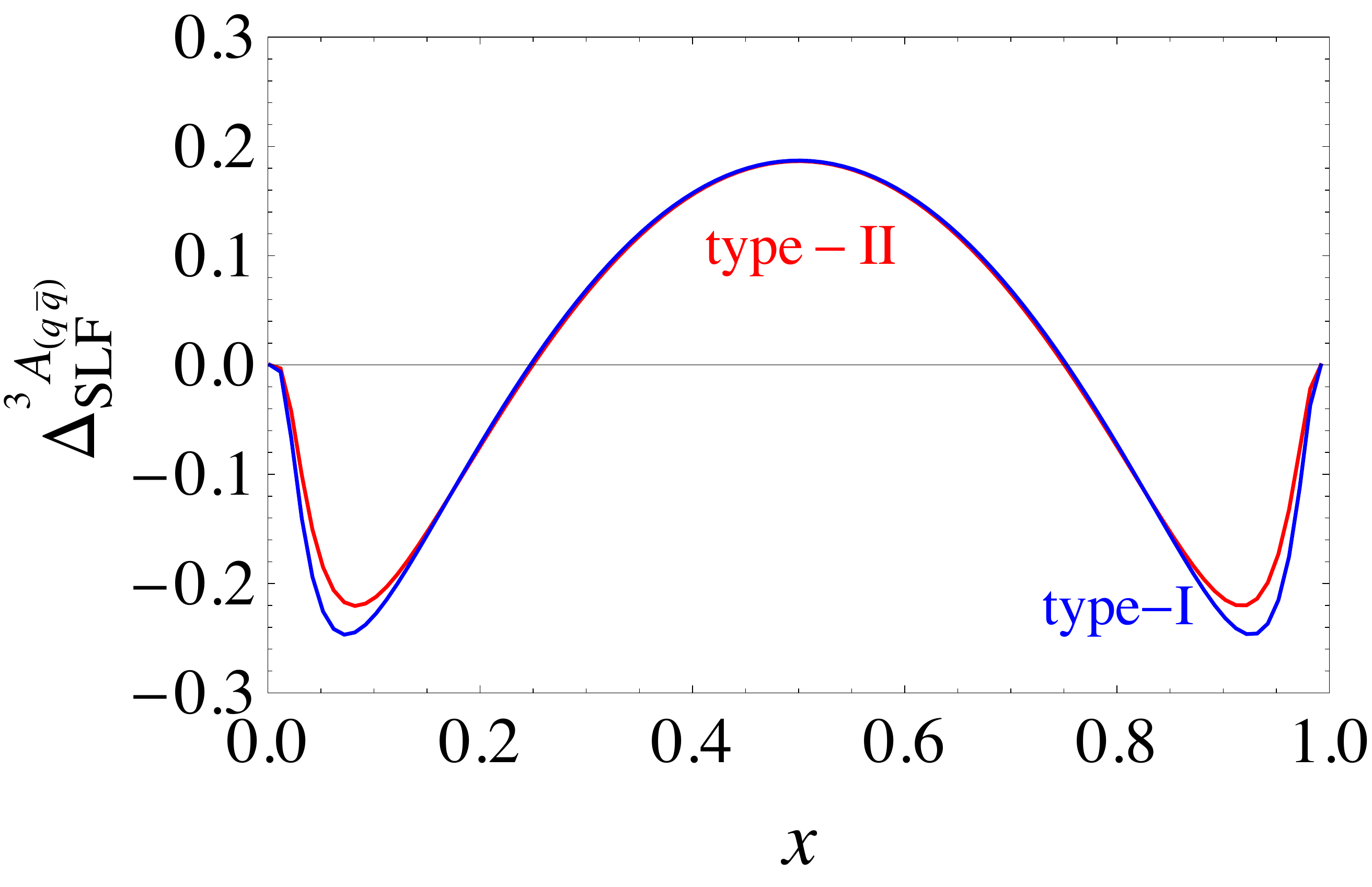}}\qquad\qquad
\subfigure[]{\includegraphics[scale=0.21]{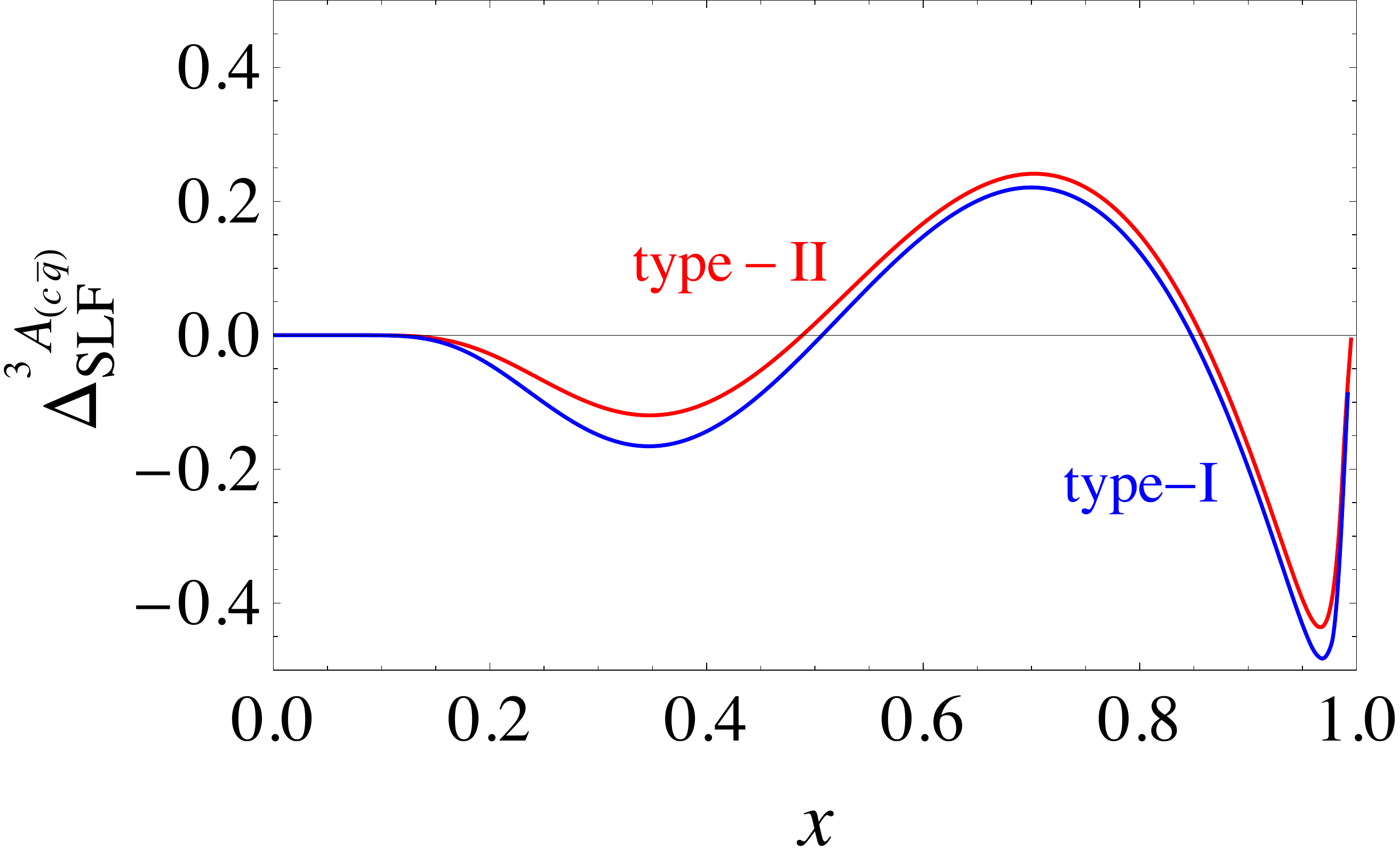}}\\ 
\subfigure[]{\includegraphics[scale=0.21]{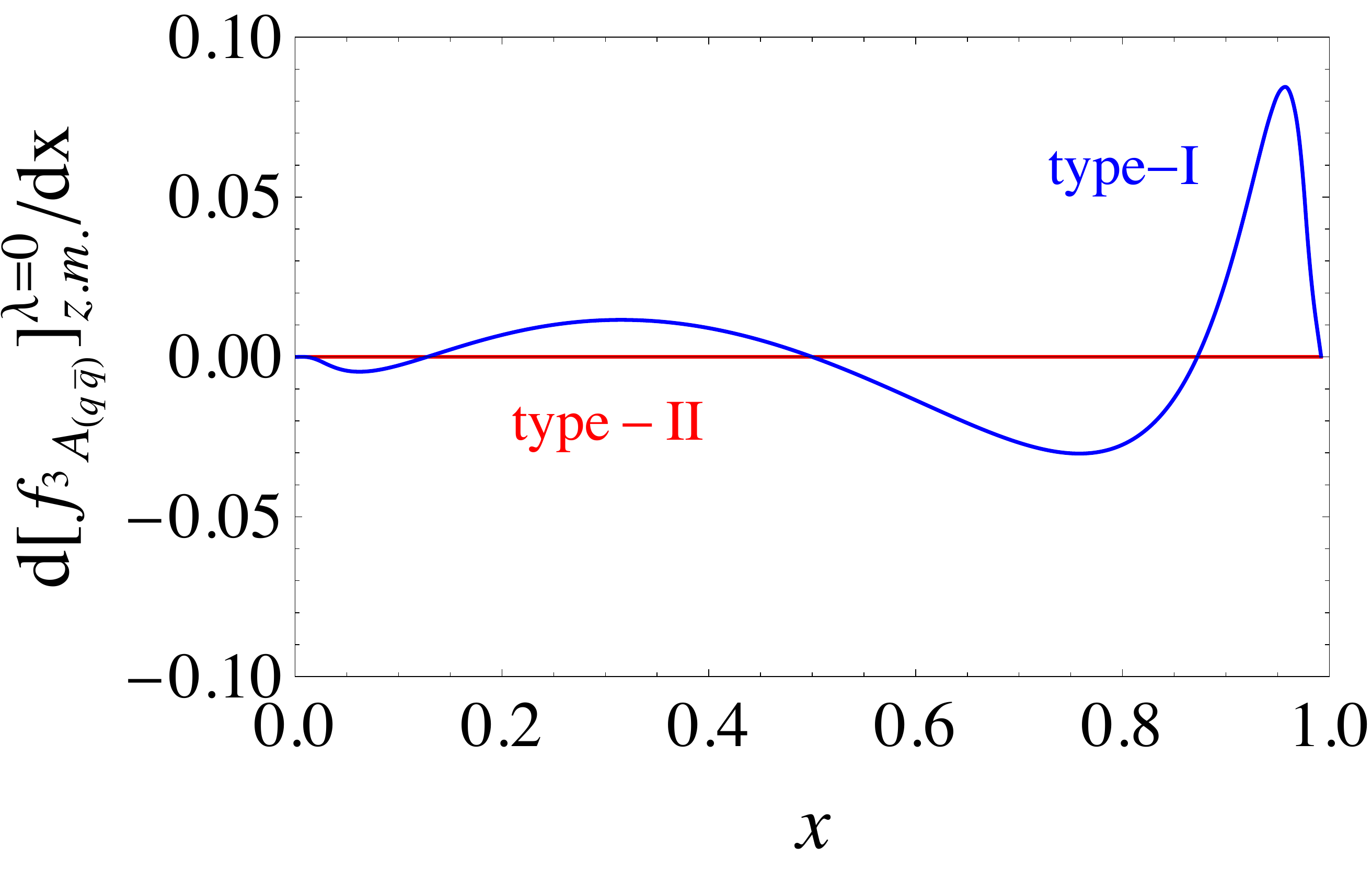}}\qquad\qquad
\subfigure[]{\includegraphics[scale=0.21]{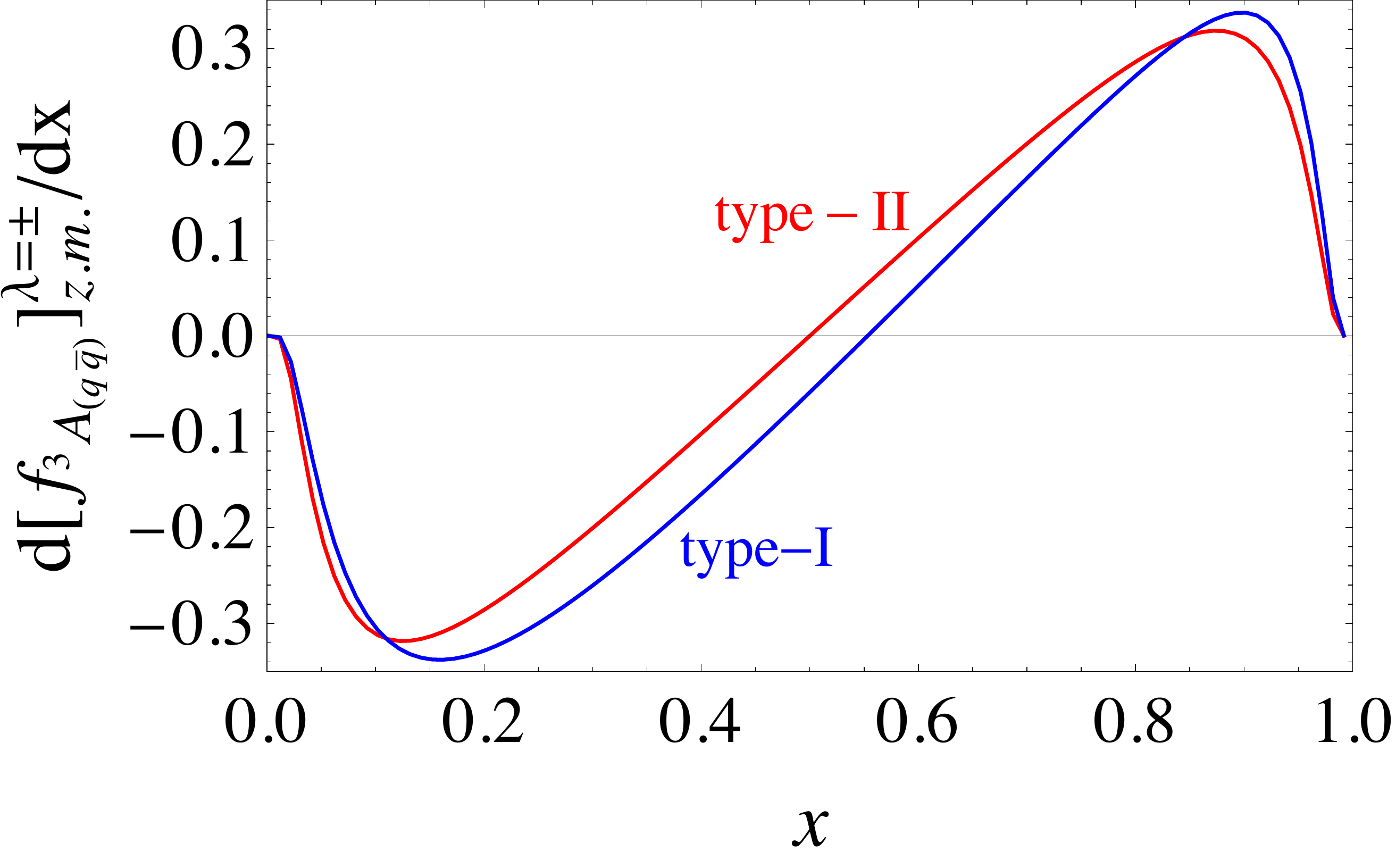}}\\ 
\subfigure[]{\includegraphics[scale=0.21]{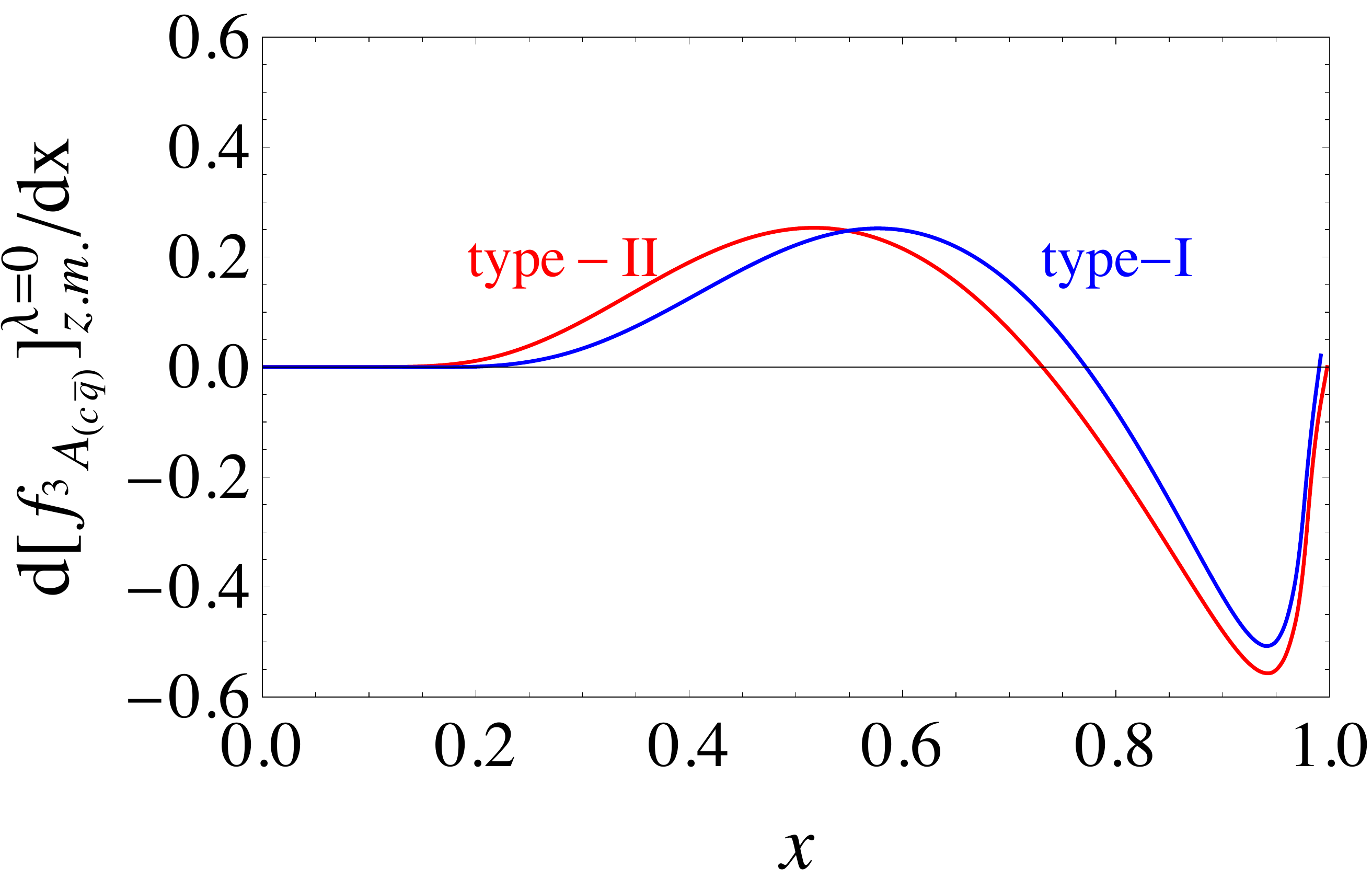}}\qquad\qquad
\subfigure[]{\includegraphics[scale=0.21]{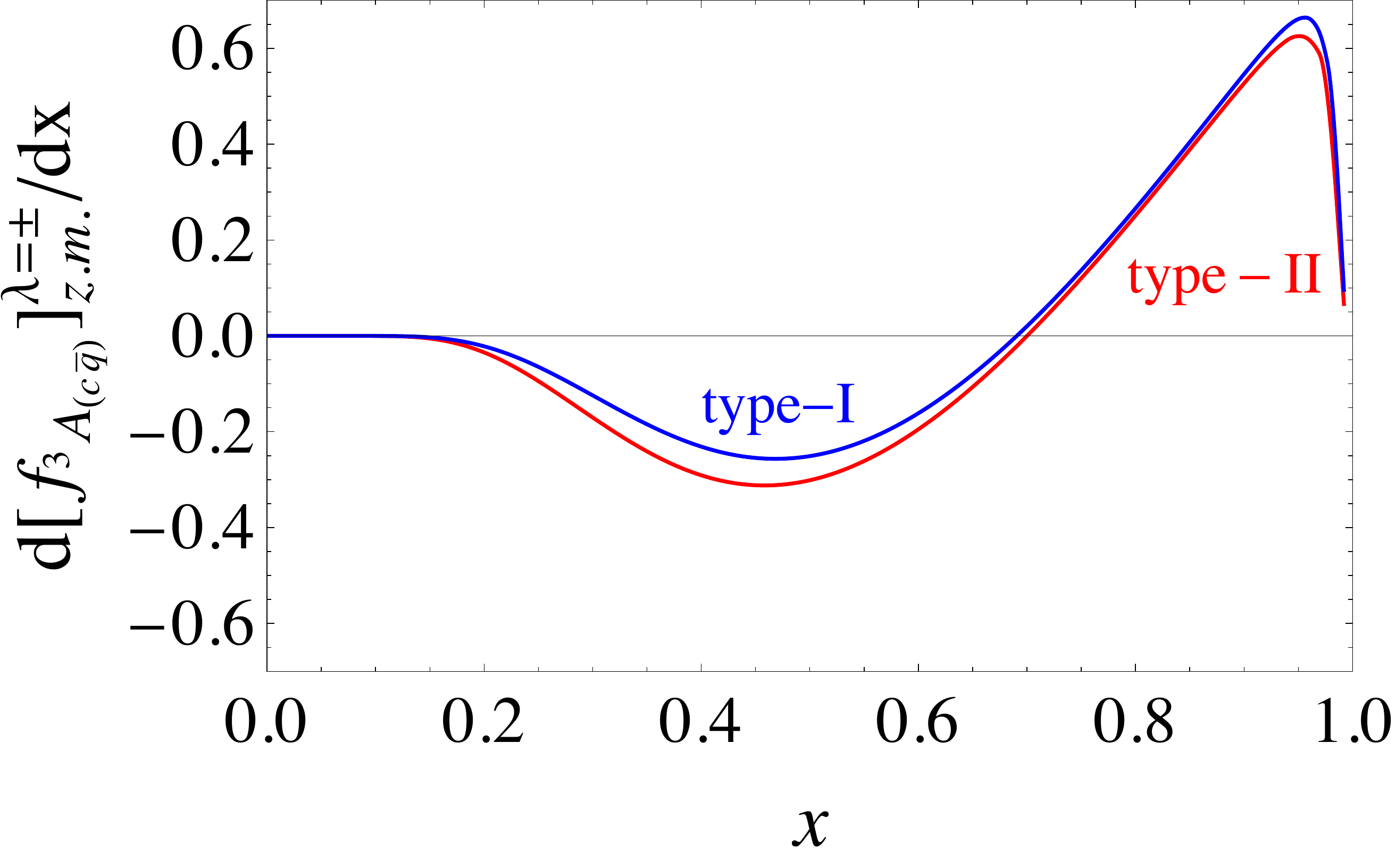}}
\caption{\label{fig:dep2} \small Dependence of $\Delta^{^3\!A}_{\rm full}(x)$, $\Delta^{^3\!A}_{\rm SLF}(x)$ and $\d [f_{^3\!A}]_{\rm z.m.}^{\lbd=0\,,\pm}/\d x$ on the momentum fraction $x$ for $(q\bar{q})$ and $(c\bar{q})$ bound-states. See text for details.}
\end{center}
\end{figure}

\begin{figure}[t]
\begin{center}
\subfigure[]{\includegraphics[scale=0.22]{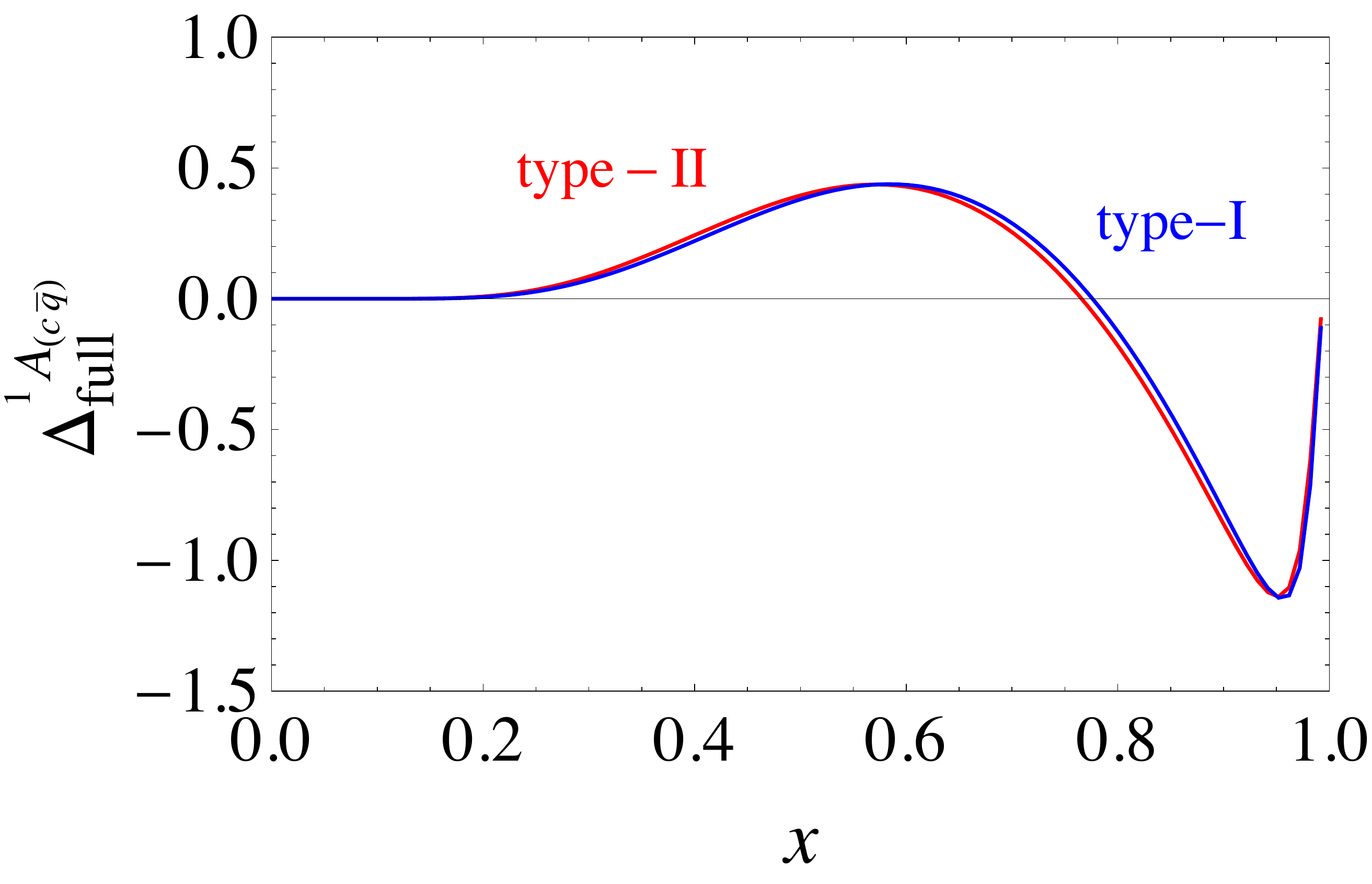}}\qquad\qquad
\subfigure[]{\includegraphics[scale=0.22]{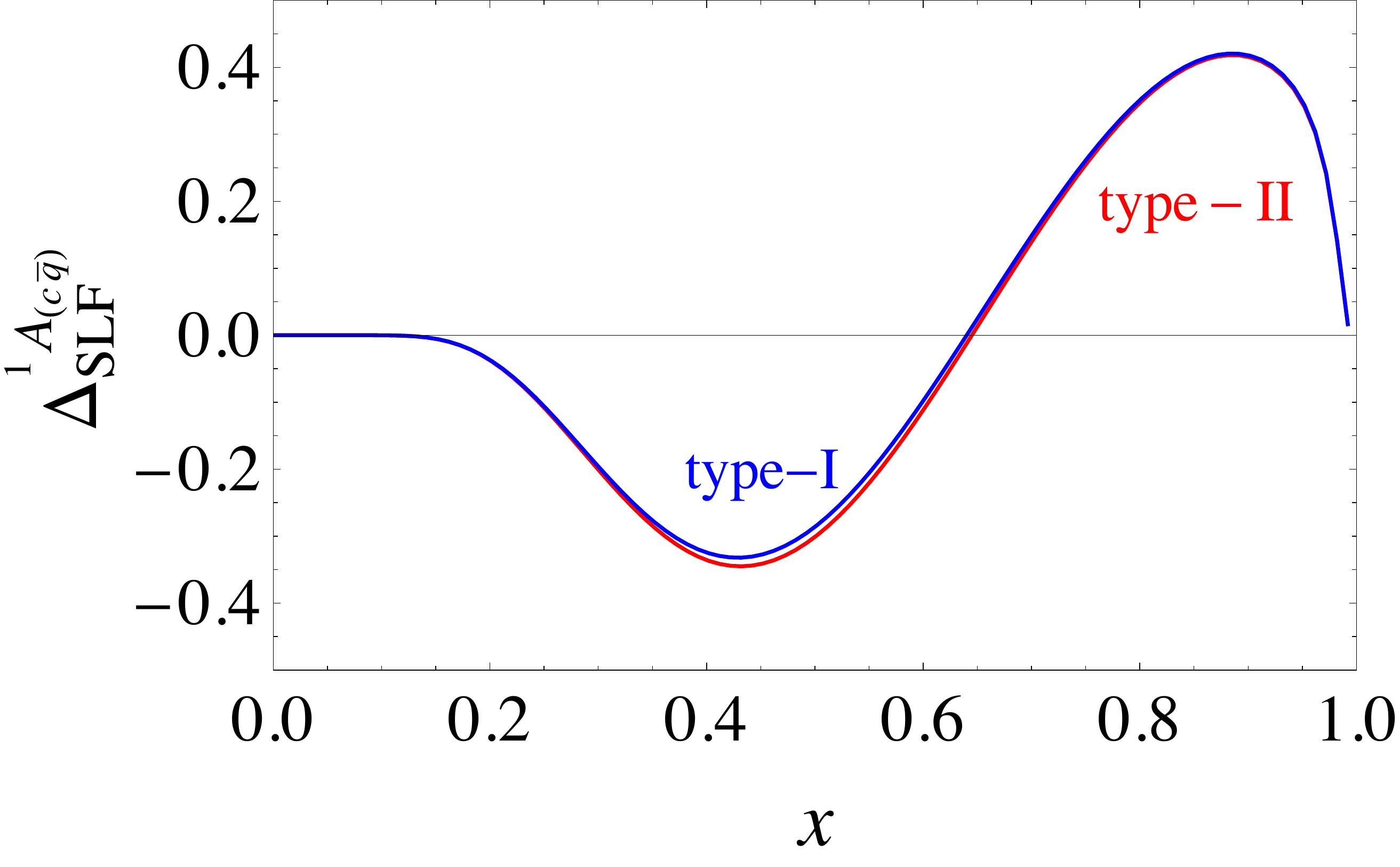}}\qquad\qquad
\subfigure[]{\includegraphics[scale=0.22]{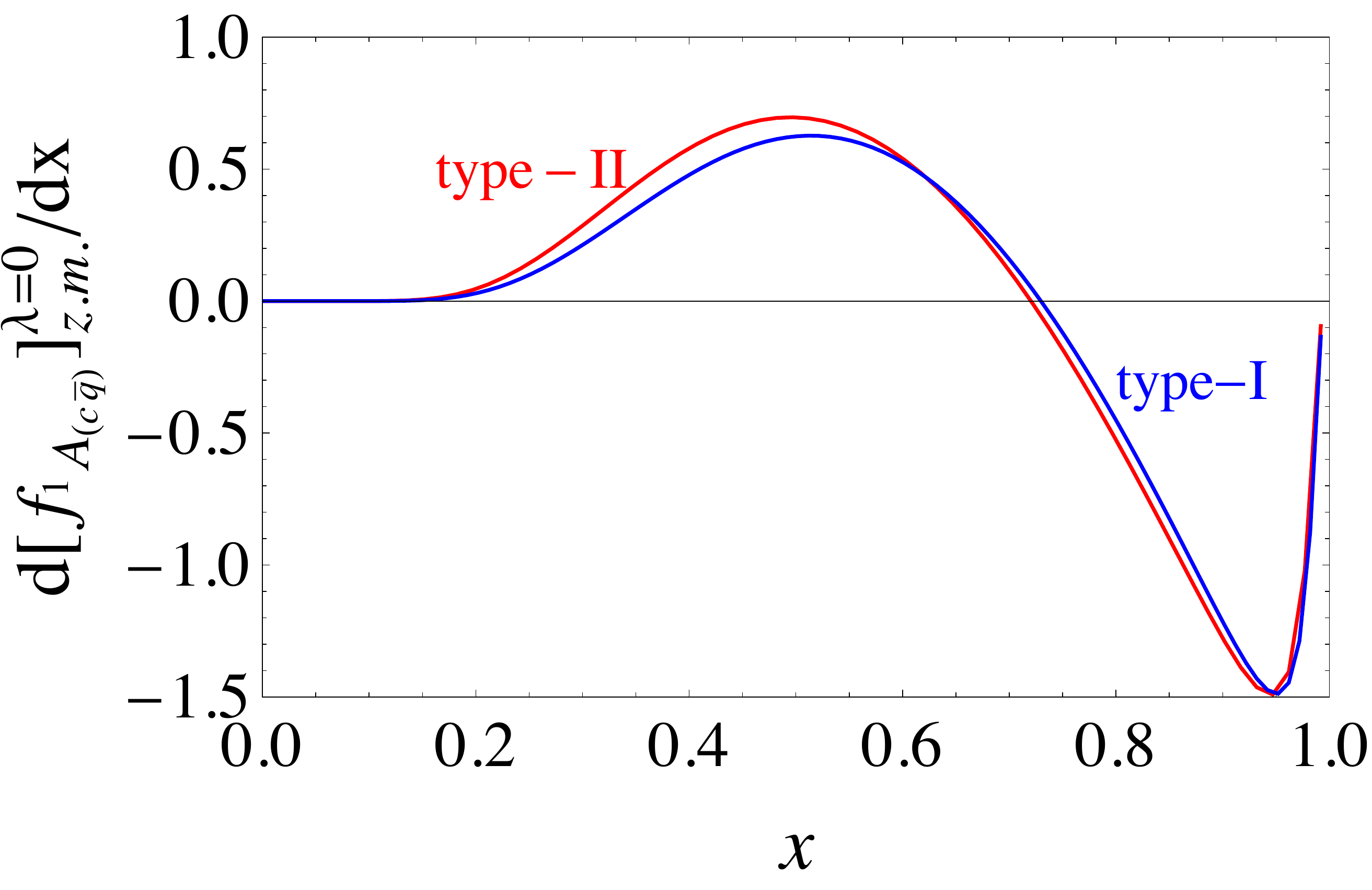}}\qquad\qquad
\subfigure[]{\includegraphics[scale=0.22]{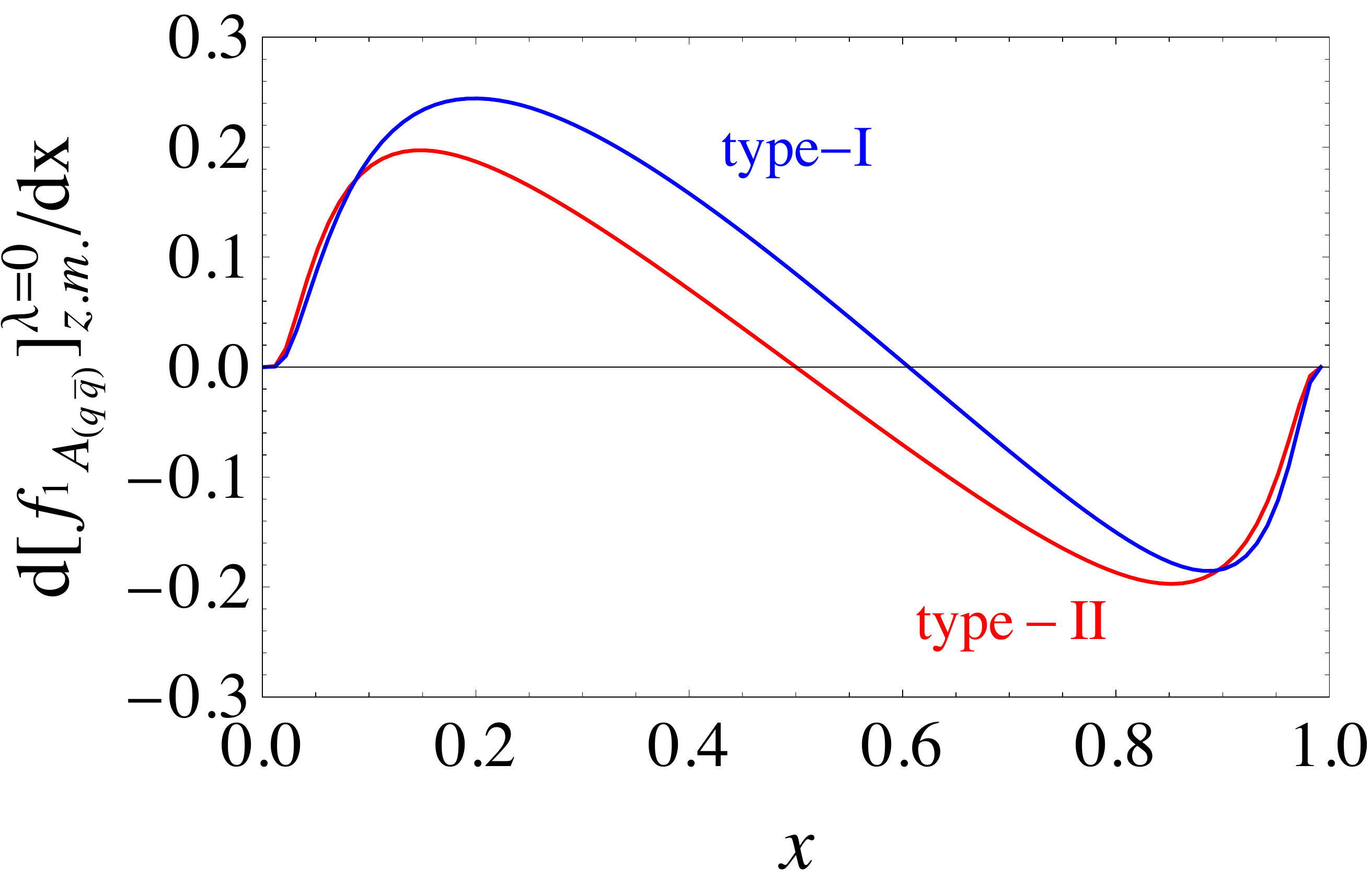}}
\caption{\label{fig:dep3} \small Dependence of $\Delta^{^1\!A_{(c\bar{q})}}_{\rm full}(x)$, $\Delta^{^1\!A_{(c\bar{q})}}_{\rm SLF}(x)$, $\d [f_{^1\!A_{(c\bar{q})}}]_{\rm z.m.}^{\lbd=0}/\d x$ and $\d [f_{^1\!A_{(q\bar{q})}}]_{\rm z.m.}^{\lbd=0}/\d x$ on the momentum fraction $x$. The quantities $\Delta^{^1\!A_{(q\bar{q})}}_{\rm full}(x)$, $\Delta^{^1\!A_{(q\bar{q})}}_{\rm SLF}(x)$, $\d [f_{^1\!A_{(q\bar{q})}}]_{\rm z.m.}^{\lbd=\pm}/\d x$ and $\d [f_{^1\!A_{(c\bar{q})}}]_{\rm z.m.}^{\lbd=\pm}/\d x$ are all equal to zero and hence not shown here. See text for details.}
\end{center}
\end{figure}

Again, for convenience, we take the $(q\bar{q})$ and $(c\bar{q})$~($q=u,\,d$) bound-states as examples to analyze and discuss the relations among the decay constants given above; here, $^1\!A_{(q\bar{q})}$, $^3\!A_{(q\bar{q})}$, $^1\!A_{(c\bar{q})}$ and $^3\!A_{(c\bar{q})}$ are interpreted, respectively, as $b_1(1235)$, $a_1(1260)$, $D_1(2420)$ and $D_1(2430)$ mesons~\cite{Patrignani:2016xqp}. Note that all the following findings and conclusions are also applicable for the other axial-vector mesons. Numerical results for these decay constants obtained with the best-fit values of $\beta$ are collected in Table~\ref{tab:fA}, and the dependences of $\Delta^{A}_{\rm full}(x)$, $\Delta^{A}_{\rm SLF}(x)$ and $\d [f_{A}]_{\rm z.m.}^{\lbd=0\,,\pm}/\d x$ on the momentum fraction $x$ are shown in Figs.~\ref{fig:dep2} and \ref{fig:dep3}. Based on these numerical results and the theoretical formulas given above, we have the following discussions and findings:
\begin{itemize}
\item Because of $m_1=m_2$ in the isospin-symmetry limit, the $^1\!A_{(q\bar{q})}$ meson is not ideal for testing the self-consistencies of LF quark models, as the corresponding decay constants, except $[f_{^1\!A_{(q\bar{q})}}]_{\rm SLF}^{\lbd=0}$ and $[f_{^1\!A_{(q\bar{q})}}]_{\rm val.}^{\lbd=0}$, are all proportional to $m_1-m_2$ and hence identically zero. Moreover, $[f_{^1\!A_{(q\bar{q})}}]_{\rm SLF}^{\lbd=0}$ with $m_1=m_2$ is also equal to zero because its integrand is anti-symmetric under the exchange $x\leftrightarrow \bar{x}$. On the other hand, $[f_{^1\!A_{(q\bar{q})}}]_{\rm val.}^{\lbd=0}$ is nonzero only in the type-I but vanishes numerically in the type-II scheme, which can be found from Table~\ref{tab:fA} and clearly seen from Fig.~\ref{fig:dep3}(d). Thus, in our following analyses for the $^1\!A$ mesons, we shall focus mainly on the general case with $m_1\neq m_2$, such as the $(c\bar{q})$ bound-states.

\item Comparing Eqs.~\eqref{eq:f1Afull0} with \eqref{eq:f1Afullpm} for $^1\!A$ and Eqs.~\eqref{eq:f3Afull0} with \eqref{eq:f3Afullpm} for $^3\!A$, one can easily found that the CLF results for $f_{^{1(3)}\!A}$ extracted via the $\lbd=0$ and $\lbd=\pm$ polarization states are formally different from each other, due to the additional contributions to $[f_{^{1(3)}\!A}]_{\rm full}^{\lbd=0}$ characterized by the coefficient $B_1^{(2)}$. Numerically, it can be found from Table~\ref{tab:fA}, Figs.~\ref{fig:dep2}(a, b) and Fig.~\ref{fig:dep3}(a) that $[f_{^{1(3)}\!A}]_{\rm full}^{\lbd=0}\neq [f_{^{1(3)}\!A}]_{\rm full}^{\lbd=\pm}$ in the type-I scheme, which means that the CLF approach with the type-I correspondence also suffers the self-consistency problem when applied to $f_{^{1(3)}\!A}$; in the type-II scheme, we can however obtain self-consistent results,
 \begin{align}\label{eq:scAfull}
 [f_{^{1(3)}\!A}]_{\rm full}^{\lbd=0}\;\dot{=}\;[f_{^{1(3)}\!A}]_{\rm full}^{\lbd=\pm}\,, \qquad(\text{type-II})
\end{align}
due to $\int\d x\Delta^{^{1(3)}\!A}_{\rm full}=0$. These findings for the $^1\!A$ and $^3\!A$ mesons in the CLF quark model are exactly the same as what we have found for the $V$ meson. The same conclusion is also applied to the SLF quark model; explicitly, we find that $\left|[f_{^{1(3)}\!A}]_{\rm SLF}^{\lbd=0}\right|<\left|[f_{^{1(3)}\!A}]_{\rm SLF}^{\lbd=\pm}\right|$ within the traditional type-I scheme, while  
\begin{align}\label{eq:scASLF}
 [f_{^{1(3)}\!A}]_{\rm SLF}^{\lbd=0}\;\dot{=}\;[f_{^{1(3)}\!A}]_{\rm SLF}^{\lbd=\pm}\,,\qquad(\text{type-II})
\end{align}
within the type-II scheme, due to $\int\d x\Delta^{^{1(3)}\!A}_{\rm SLF}=0$, which can be seen from Table~\ref{tab:fA}, Figs.~\ref{fig:dep2}(c,d) and Fig.~\ref{fig:dep3}(b). Thus, the above findings, Eqs.~\eqref{eq:scAfull} and \eqref{eq:scASLF}, reinforce our conclusion obtained in the $f_V$ case that the replacement $M\to M_0$ is required to give self-consistent results for the decay constants in both the CLF and the SLF quark model. 

\item Employing the type-II scheme and then making some simplifications on Eqs.~\eqref{eq:f1ASLF0}--\eqref{eq:f3ASLFpm} and \eqref{eq:f1Aval0}--\eqref{eq:f3Avalpm}, we find that
 \begin{align}\label{eq:ASLFval}
 [f_{^{1(3)}\!A}]_{\rm SLF}^{\lbd=0}\, = \,[f_{^{1(3)}\!A}]_{\rm val.}^{\lbd=0} \quad\text{and}\quad [f_{^{1(3)}\!A}]_{\rm SLF}^{\lbd=\pm}\, = \,[f_{^{1(3)}\!A}]_{\rm val.}^{\lbd=\pm}\,, \qquad(\text{type-II})
\end{align}
which can also be seen clearly from the numerical results given in Table~\ref{tab:fA}. Among these relations, only $[f_{^1\!A}]_{\rm SLF}^{\lbd=\pm}=[f_{^1\!A}]_{\rm val.}^{\lbd=\pm}$ holds in the type-I scheme (see Eqs.~\eqref{eq:f1ASLFpm} and \eqref{eq:f1Avalpm}). The relations given by Eq.~\eqref{eq:ASLFval} are direct generalizations of Eq.~\eqref{eq:VSLFval} from the vector to the axial-vector mesons. 

\item For the $^1\!A$ meson, it can be found by comparing Eqs.~\eqref{eq:f1Afullpm} with \eqref{eq:f1Avalpm} that  
\begin{align}
[f_{^1\!A}]_{\rm full}^{\lbd=\pm}=[f_{^1\!A}]_{\rm val.}^{\lbd=\pm}\,,
\end{align}
which implies that there is no zero-mode contribution, $[f_{^1\!A}]_{\rm z.m.}^{\lbd=\pm}=0$; while, from Eqs.~\eqref{eq:f1Afull0} and \eqref{eq:f1Aval0}, one can see that the zero-mode contribution, $[f_{^1\!A}]_{\rm z.m.}^{\lbd=0}$, exists formally and, moreover, $[f_{^1\!A}]_{\rm z.m.}^{\lbd=0}\neq 0$ numerically within the type-I scheme, which can be clearly seen from Figs.~\ref{fig:dep3}(c) and \ref{fig:dep3}(d). This means that the existence or absence of $[f_{^1\!A}]_{\rm z.m.}$ depend on the choice of the polarization state $\lambda$. Within the type-II scheme, however, although existing formally, $[f_{^1\!A}]_{\rm z.m.}^{\lbd=0}$ vanishes numerically, and we have therefore again 
\begin{align}
[f_{^1\!A}]_{\rm full}^{\lbd=0}\;\dot{=}\;[f_{^1\!A}]_{\rm val.}^{\lbd=0}\,,~\quad(\text{type-II})
\end{align}
which can be seen from the numerical results given in Table~\ref{tab:fA}. 

\item For the $^3\!A$ meson, from Figs.~\ref{fig:dep2}(e--h) and Table~\ref{tab:fA}, we find that the zero-mode contributions $[f_{^3\!A}]_{\rm z.m.}^{\lbd=0,\pm}$ always exist formally, and do not vanish numerically in the type-I scheme. However, in the type-II scheme, $[f_{^3\!A}]_{\rm z.m.}^{\lbd=0}\propto (m_1-m_2)$ and hence vanishes for the quarkonia, which explains the red line shown in Fig.~\ref{fig:dep2}(e); moreover, the zero-mode contributions vanish numerically for the other cases shown by Figs.~\ref{fig:dep2}(f--h). Therefore, we have
\begin{align}
 [f_{^3\!A}]_{\rm full}^{\lbd=0}\;\dot{=} \;[f_{^3\!A}]_{\rm val.}^{\lbd=0} \quad\text{and}\quad [f_{^3\!A}]_{\rm full}^{\lbd=\pm}\;\dot{=}\;[f_{^3\!A}]_{\rm val.}^{\lbd=\pm}\,, \qquad(\text{type-II})
\end{align}
in which the symbol ``$\dot{=}$'' in the first equation should be replaced by ``$=$'' for the quarkonia. These relations can also be found directly from the numerical results given in Table~\ref{tab:fA}. 
\end{itemize}

\begin{table}[t]
\begin{center}
\caption{\label{tab:fApre} \small Updated predictions for $f_{^1\!A}$ and $f_{^3\!A}$ (in unit of ${\rm MeV}$) in the LF approach. The other captions are the same as in Table~\ref{tab:fvpre}.}
\vspace{0.1cm}
\let\oldarraystretch=\arraystretch
\renewcommand*{\arraystretch}{1.1}
\setlength{\tabcolsep}{21.0pt}
\begin{tabular}{lcccccccccccc}

\hline\hline
                 &$f_{q\bar{q}}$&$f_{s\bar{q}}$&$f_{s\bar{s}}$
                 &$f_{c\bar{q}}$&$f_{c\bar{s}}$\\\hline
$^1\!A$     &$0$&$-27\pm1$&$0$&$-78\pm2$&$-62\pm2$
              \\\hline
$^3\!A$    &$220\pm1$&$219\pm2$&$203\pm2$&$231\pm8$&$257\pm8$
\\\hline\hline
                 &$f_{c\bar{c}}$&$f_{b\bar{q}}$&$f_{b\bar{s}}$
                 &$f_{b\bar{c}}$&$f_{b\bar{b}}$\\\hline
$^1\!A$     &$0$&$-95\pm3$&$-88\pm2$&$-86\pm3$&$0$
              \\\hline
$^3\!A$    &$250\pm90$&$176\pm6$&$180\pm5$&$281\pm7$&$353\pm25$
\\\hline\hline
\end{tabular}
\end{center}
\end{table}

Combining all the above findings for the axial-vector mesons, we can extend the conclusion, Eq.~\eqref{eq:confV}, for the vector meson to the more general form
\begin{align}\label{eq:find1}
  [{\cal Q}]_{\rm SLF}^{\lbd=0}\,=\,[{\cal Q}]_{\rm val.}^{\lbd=0}\,\dot{=}\,[{\cal Q}]_{\rm full}^{\lbd=0}\,\dot{=}\,[{\cal Q}]_{\rm full}^{\lbd=\pm}\,\dot{=}\,[{\cal Q}]_{\rm val.}^{\lbd=\pm}\,=\,[{\cal Q}]_{\rm SLF}^{\lbd=\pm}\,, \qquad(\text{type-II})
\end{align}
where ${\cal Q}=f_V$, $f_{^1\!A}$ and $f_{^3\!A}$, and the first and the last ``$\dot{=}$'' should be replaced by ``$=$'' for the $^3\!A_{(q\bar{q})}$ and $^1\!A $ mesons, respectively. These relations reflect the self-consistencies of the SLF and CLF quark models in the type-II scheme. Finally, using the inputs listed in Table~\ref{tab:input} and employing the self-consistent type-II scheme, we present in Table~\ref{tab:fApre} our updated predictions for $f_{^1\!A}$ and $f_{^3\!A}$ in the LF approach.

\subsection{Form factors in $P\to P$ weak transition}

In the last two subsections, we have tested the self-consistencies of LF quark models via the mesonic decay constants, and found that both the CLF and the SLF quark model within the type-II scheme can give self-consistent results for the decay constants of vector and axial-vector mesons, {i.e.}, $[{\cal Q}]_{\rm full}^{\lbd=0}\;\dot{=}\; [{\cal Q}]_{\rm full}^{\lbd=\pm}$ and $[{\cal Q}]_{\rm SLF}^{\lbd=0}\;\dot{=}\;[{\cal Q}]_{\rm SLF}^{\lbd=\pm}$. More interestingly, we have also found that the results of the SLF quark model are consistent with both the full and the valence result of the CLF quark model, {i.e.},  
\begin{align}\label{eq:relaSCLF}
 [{\cal Q}]_{\rm SLF}\;=\; [{\cal Q}]_{\rm val.}\;\dot{=}\;[{\cal Q}]_{\rm full}\,, \qquad(\text{type-II})
\end{align}
where the second relation is due to the fact that the zero-mode contributions exist only formally but vanish numerically within the type-II scheme. It is known that, besides the decay constant, another ideal quantity for studying the zero-mode effect is the transition form factor. To this end, we shall test in this subsection whether the relations, Eq.~\eqref{eq:relaSCLF}, still hold for the $P\to P$ weak transition form factors. 

The form factors for $P\to P$ weak transition are defined by
\begin{align}\label{eq:defFF}
\la  P''(p'') | \bar{q}''_1 \r^{\u} q'_1 |P'(p') \ra=f_+(q^2)P^{\u}+f_-(q^2)q^{\u}\,,
\end{align}
where $q^{\u}=p'^{\u}-p''^{\u}$ and $P^{\u}=p'^{\u}+p''^{\u}$. In the $q^+=0$ frame,  multiplying both sides of  Eq.~\eqref{eq:defFF} by $\w_\u$ and $q_\u$, respectively, one gets 
\begin{align}
f_+(q^2)=\frac{{\cal B}^{+}}{P^+}\,,\qquad
f_-(q^2)=\frac{q\cdot {\cal B} - (q\cdot P)\,f_+(q^2)}{q^2}\,,
\end{align}
where ${\cal B}^\mu$ denotes the l.h.s of Eq.~\eqref{eq:defFF} and can be calculated in the SLF and CLF approaches. 

Employing the theoretical framework and formulas of the SLF quark model introduced in section~\ref{sec:2}, we finally obtain
\begin{align}
\label{eq:FPSLF}
[f_+(q^2)]_{\rm SLF}= &\int  \frac{\d x\, \d^2{k}_{\bot}'}{(2\pi)^3}  \frac{{\psi_s''}^{*}\,{\psi_s'}}{x\bar{x}} \frac{1}{2\hat{M}'_0\hat{M}''_0} \left[k_{\bot}' \cdot {\bar{k}}_{\bot}''  +(\bar{x}m_1^{\prime}+xm_2)(\bar{x}m_1^{\prime\prime}+xm_2) \right]\,,\\[0.2cm]
\label{eq:FMSLF}
[f_-(q^2)]_{\rm SLF}= &[f_+(q^2)]_{\rm SLF}+ \int  \frac{\d x \,\d^2{ k}_{\bot}'}{(2\pi)^3}   \frac{{\psi_s''}^{*}\,{\psi_s'}}{x\bar{x}} \frac{\bar{x}}{2\hat{M}'_0\hat{M}''_0} \bigg\{  \frac{{ q}_{\bot} \cdot { {k}}_{\bot}'}{{ q}_{\bot}^2}  \left[(m_1'-m_1'')^2+{ q}_{\bot}^2\right]   \nonumber\\[0.1cm]
 &+  \frac{ q_{\bot} \cdot ( k_{\bot}'- q_{\bot} ) }{ q_{\bot}^2}  \frac{ (\bar{x}m_1^{\prime}+xm_2)^2+k_{\bot}'^2}{x\bar{x}}
+\frac{{ q}_{\bot} \cdot   k_{\bot}'}{ q_{\bot}^2}   \frac{ (\bar{x}m_1^{\prime\prime}+xm_2)^2+\bar{k}_{\bot}''^2}{x\bar{x}}    \bigg\}\,.
\end{align}
The form factor $f_+(q^2)$ in the SLF approach, Eq.~\eqref{eq:FPSLF}, has been first obtained in Ref.~\cite{Jaus:1989au}, while $f_-(q^2)$, Eq.~\eqref{eq:FMSLF}, is given for the first time in this paper.   

In the CLF quark model, on the other hand, using the theoretical formulas given in Sec.~\ref{sec:2}, we obtain
\begin{align}
\label{eq:FPfull}
[f_+(q^2)]_{\rm full}=&N_c \int \frac{\d x \, \d^2 k_{\bot}'}{(2\pi)^3}\frac{\chi_{P'}\,\chi_{P''}}{2\bar{x}}  
 \Big[xM_0'^2+xM_0''^2+ \bar{x} q^2-x(m_1'-m_2)^2-x(m_1''-m_2)^2 \nonumber\\
 &-\bar{x}(m_1'-m_1'')^2 \Big]\,, \\[0.2cm]
 \label{eq:FMfull}
[f_-(q^2)]_{\rm full} =&N_c \int \frac{\d x \, \d^2 k_{\bot}'}{(2\pi)^3}\frac{\chi_{P'}\,\chi_{P''}}{2\bar{x}}  \Bigg\{-2x\bar{x}M'^2 - 2 k_{\bot}'^2 -2m_1' m_2 + 2 (m_1'' - m_2)(\bar{x}m_1' + x m_2)\,\nonumber\\
& -2\frac{k_{\bot}' \cdot q_{\bot}}{q^2} \Big[(x-\bar{x})M'^2+M''^2-\bar{x} (q^2+q\cdot P)+2x M_0'^2
 -2 (m_1'+m_1'') (m_1'-m_2)\Big]\,\nonumber\\
 &+4\frac{P\cdot q}{q^2} \left[ k_{\bot}'^2 +2 \frac{(k_{\bot}' \cdot q_{\bot})^2}{q^2} \right]+4 \frac{(k_{\bot}'\cdot q_{\bot})^2}{q^2} \Bigg\}\,,
\end{align}
which agree with the results given in the literatures, for instance,  Refs.~\cite{Jaus:1999zv,Cheng:2003sm}. Furthermore, for the valence contributions in the CLF quark model, we obtain
\begin{align}
\label{eq:FPval}
[f_+(q^2)]_{\rm val.}=&[f_+(q^2)]_{\rm full}\,, \\[0.2cm]
\label{eq:FMval}
[f_-(q^2)]_{\rm val.} =&[f_+(q^2)]_{\rm val.}+ N_c \int \frac{\d x \, \d^2 k_{\bot}'}{(2\pi)^3}\frac{\chi_{P'}\,\chi_{P''}}{2\bar{x}}  \bigg\{\big[ -2\bar{x}M'^2-2xM_0'^2+2 (m_1'-m_2)^2 \big]\nonumber\\
& -2\frac{{k}_{\bot}'\cdot q_{\bot}}{q^2} \big[ M'^2+M''^2-q^2+2(m_1'-m_2)(m_2-m_1'')\big] \bigg\}\,.
\end{align}

\begin{figure}[t]
\begin{center}
\subfigure[]{\includegraphics[scale=0.26]{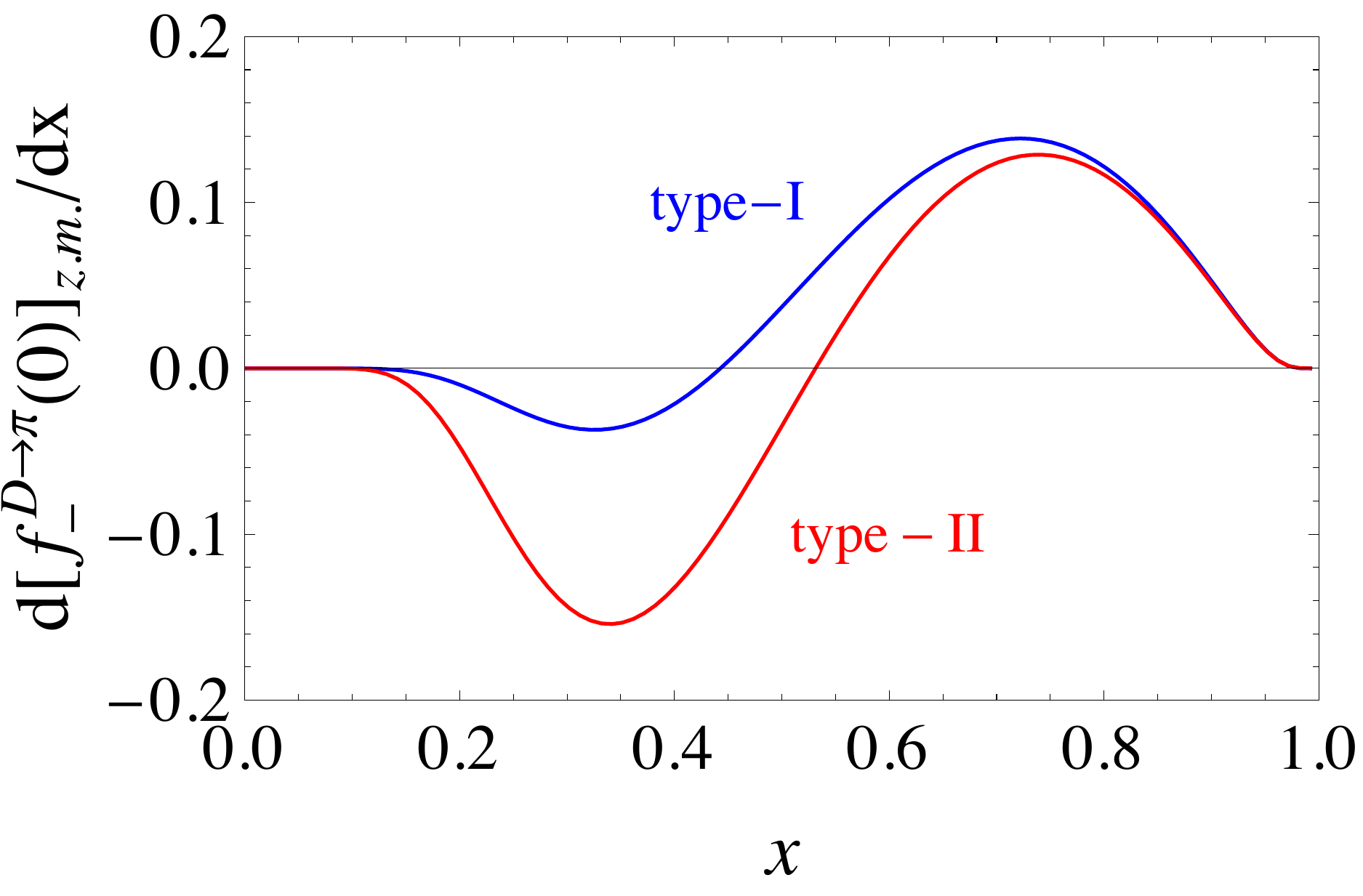}}\qquad
\subfigure[]{\includegraphics[scale=0.26]{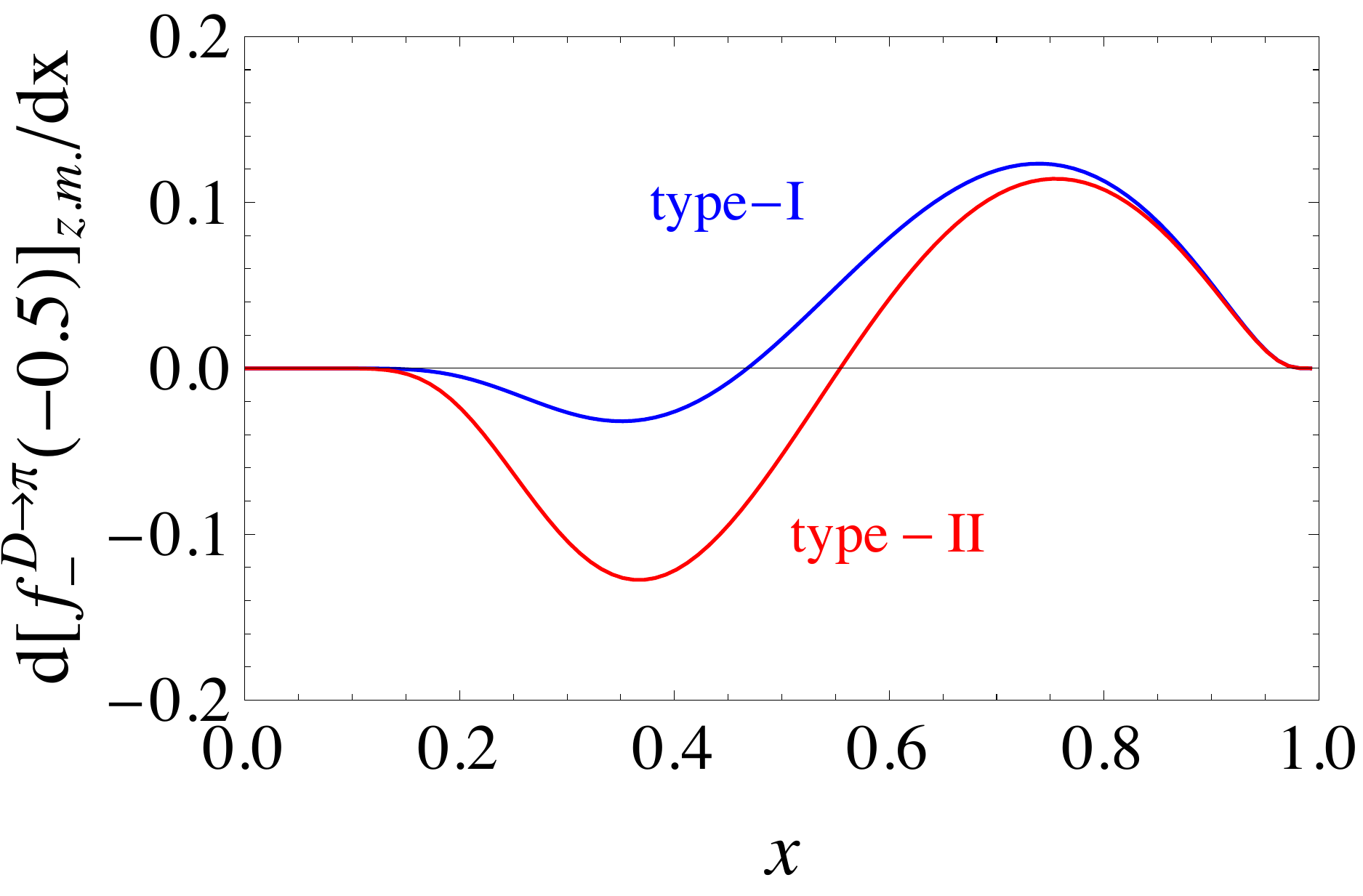}}\qquad
\subfigure[]{\includegraphics[scale=0.26]{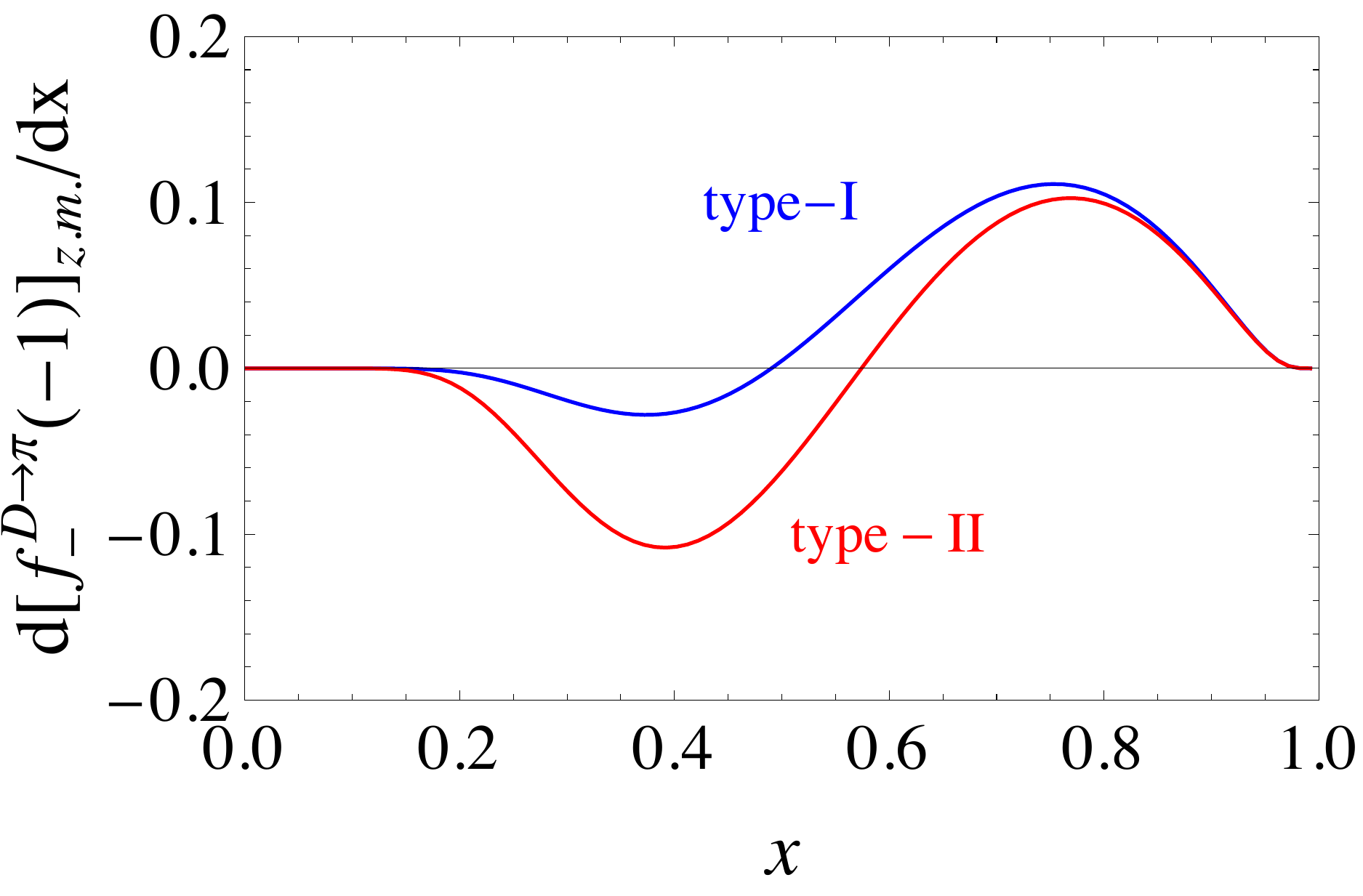}}
\caption{\label{fig:dep4} \small Dependence of $\d [f_-^{D\to\pi}(q^2)]_{\rm z.m.}/\d x$ on the momentum fraction $x$ at $q_\bot^2=0$, $0.5$ and $1~{\rm GeV}^2$, respectively. }
\end{center}
\end{figure}

Based on the results given above, we have the following discussions and findings:
\begin{itemize}
\item For the vector form factor $f_+(q^2)$, as has already been known from the previous studies~\cite{Jaus:1999zv,Cheng:2003sm}, it is free of the zero-mode effect, {i.e.}, $[f_+(q^2)]_{\rm z.m.}=0$, which can be found from Eq.~\eqref{eq:FPval}. Moreover, it is found that, after the replacement $\chi_P\to h_P/\hat{N}_1$ together with some simplifications, the valence contribution, $[f_+(q^2)]_{\rm val.}$, given by Eq.~\eqref{eq:FPval} is exactly the same as the SLF result, $[f_+(q^2)]_{\rm SLF}$, given by Eq.~\eqref{eq:FPSLF}. We can, therefore, conclude that
\begin{align}
[f_+(q^2)]_{\rm SLF}\;=\;[f_+(q^2)]_{\rm val.}\;=\;[f_+(q^2)]_{\rm full}\,,
\end{align}
both in the type-I and in the type-II scheme.

\item For the form factor $f_-(q^2)$, on the other hand, one can easily find that $[f_-(q^2)]_{\rm SLF}\neq [f_-(q^2)]_{\rm val.}$ within the type-I scheme, by comparing Eqs.~\eqref{eq:FMval} with \eqref{eq:FMSLF}; however, taking the type-II scheme and making some further simplifications, we find that
\begin{align}
[f_-(q^2)]_{\rm SLF}\;=\;[f_-(q^2)]_{\rm val.}\,, \qquad(\text{type-II})
\end{align}
which confirms again the first relation in Eq.~\eqref{eq:relaSCLF}. 

\item In contrast to $f_+(q^2)$, the form factor $f_-(q^2)$ obviously suffers the zero-mode effect, which can be found from Eqs.~\eqref{eq:FMval} and \eqref{eq:FMfull}. In order to clearly demonstrate the zero-mode contribution, we take the $D\to \pi$ transition as an example and show in Fig.~\ref{fig:dep4} the dependence of $\d[f_-(q^2)]_{\rm z.m.}/\d x$ on the momentum fraction $x$ at $q_\bot^2=0$, $0.5$ and $1~{\rm GeV}^2$, respectively. It can be seen that the zero-mode contribution to $[f_-(q^2)]$ is sizable in the type-I, but vanishes numerically in the type-II scheme due to $\int\d x\d [f_-(q^2)]_{\rm z.m.}=0$ at the chosen $q_\bot^2$ points. This means that
\begin{align}
[f_-(q^2)]_{\rm val.}\,\dot{=}\,[f_-(q^2)]_{\rm full}\,,\qquad(\text{type-II})
\end{align}
and confirms again the last relation in Eq.~\eqref{eq:relaSCLF}.
\end{itemize}

Therefore, combining the above findings, we can conclude that the relations given by  Eq.~\eqref{eq:relaSCLF} also hold for the $P\to P$ weak transition form factors.

\subsection{Covariance of CLF quark model}

In the last subsections, we have discussed the self-consistencies of LF quark models in detail. In this subsection, we shall test the manifest covariance of the CLF quark model with the type-I and the type-II correspondence.

As has already been shown in Ref.~\cite{Jaus:1999zv}, a peculiar property of the LF matrix elements $\hat{\cal A}$, Eq.~\eqref{eq:Aclf2} and $\hat{\cal B}$, Eq.~\eqref{eq:Bclf2} is that their dependence on the lightlike four vector $\w^\u=(0,2,0_\bot)$, which can be explicitly revealed by their decomposition into four vectors\footnote{In order to treat the complete Lorentz structure of a hadronic matrix element, the authors of Refs.~\cite{Karmanov:1991fv,Karmanov:1994ck,Karmanov:1996un,Carbonell:1998rj} have developed a basically different method to identify and separate the spurious contributions and to determine the physical contributions to the hadronic form factors and coupling constants. Here we shall follow the Jaus' prescription~\cite{Jaus:1999zv} in which a manifestly covariant BS approach is used as a guide to deal with this problem.}. Taking the $V\to 0$ LF matrix element as an example, we have
\begin{align}
\hat{\cal A}_V^\u= M_V(\e^\u f_V+ \w^\u g_V)\,,
\end{align}
where $g_V$ is the unphysical constant related to $\w^\u$. As $\w^\u$ is a fixed vector,  $\hat{\cal A}_V^\u$ is obviously not covariant unless when $g_V=0$. As has been demonstrated in Ref.~\cite{Jaus:1999zv}, after the zero-mode contributions are properly taken into account, the spurious $\w$ dependence of $\hat{\cal A}_V^\u$ can be eliminated, and hence the final result is guaranteed to be covariant. This is an important feature of the CLF quark model. 

Although the main $\w$ dependence is associated with the $C$ coefficients and can be totally eliminated by the zero-mode contributions~\cite{Jaus:1999zv}, there are, however, still some residual $\w$ dependences in $\hat{\cal A}$ and $\hat{\cal B}$ due to the nonvanishing $B$ coefficients. Therefore, the manifest covariance of the CLF results can be claimed only after these residual $\w$ dependences are proven to be spurious too. As the zero-mode effect does not affect terms associated with the $B$ coefficients~\cite{Jaus:1999zv}, we have to invoke a different mechanism to ``neutralize" their effect. It has been found that the remaining $\w$ dependence, although being minimal, does not totally vanish when using the light-front vertex function given by Eq.~\eqref{eq:type1} (type-I)~\cite{Jaus:1999zv}, which implies that the CLF results for some quantities with the traditional type-I correspondence are not strictly covariant. Concerning the quantities discussed in this paper, we find that such an issue is mainly involved in the spin-1 systems, especially for the $\lbd=0$ polarization state. 

In order to illustrate explicitly the $\w$ dependence and search for possible solutions, we still take the $V\to 0$ LF matrix element as an example. After integrating out the $k^-$ component and taking into account the zero-mode contributions, we can decompose the trace term $\hat{S}_{\cal A}$ in the integrand of Eq.~\eqref{eq:Aclf2} as
\begin{align}\label{eq:Shat}
\hat{S}^\u_V=& 4 \left\{ 2 \left(1- \frac{m_1+m_2}{D_{V,{\rm con}}}\right) \frac{\w\cdot\e}{\w\cdot p}\,p^\u\,{B_1^{(2)}}
 +\e^\u\left[\cdots\right] \right\} \,,
\end{align}
where the first term proportional to $B_1^{(2)}$ involves the $\w$-dependent part, while the second term proportional to $\e^\u$ gives the physical contribution to $f_V$. For the $\lbd=\pm$ polarization states, the first term can be dropped directly due to $\w\cdot\e_{\pm}=0$, and hence the $\w$ dependence vanishes in this case. For the $\lbd=0$ polarization state, in order to separate the physical and unphysical contributions, we use the identity~\cite{Jaus:2002sv}  
\begin{align}\label{eq:deco}
p^\u  \frac{\e\cdot \w}{\w\cdot p}=\e^\u-\frac{\w^\u}{\w\cdot p}\left( \e\cdot p-\e\cdot \w \frac{p^2}{\w\cdot p}\right)-\frac{i\lbd}{\w\cdot p}\ve^{\u\v\a\b}\w_\v\e_\a p_\b\,.
 \end{align}
Here, the third term is equal to zero for $\lbd=0$, and the first term gives an additional contribution to $f_V$ that results in the self-consistency problem and has been discussed in Sec. 3.1, while the second term is the residual $\w$-dependent part that contributes to the unphysical constant $g_V$ and may violate the Lorentz covariance. Explicitly, using Eq.~\eqref{eq:deco}, we obtain the unphysical constant, $g_V$, as
\begin{align}\label{eq:gv}
[g_V]^{\lbd=0}&= \frac{N_c }{2} \int \frac{  \d x \d^2 k_{\bot}}{(2\pi)^3}   \frac{\chi_V(x, k_{\bot}^2)}{\bar{x} }  
 4 \left(1-\frac{m_1+m_2}{D_{V,{\rm con}}} \right) \frac{2}{ \w\cdot p}{B_1^{(2)}}\,.
\end{align}
Similarly, for the $^3\!A$ and $^1\!A$ mesons, we obtain
\begin{align}
\label{eq:ga3}
[g_{^3\!A}]^{\lbd=0}&= \frac{N_c }{2} \int \frac{  \d x \d^2 k_{\bot}}{(2\pi)^3}   \frac{\chi_{^3\!A}(x, k_{\bot}^2)}{\bar{x} }  
 4 \left(1+\frac{m_1-m_2}{D_{^3\!A,{\rm con}}} \right) \frac{2}{ \w\cdot p}{B_1^{(2)}}\,,\\[0.2cm]
 \label{eq:ga1}
 [g_{^1\!A}]^{\lbd=0}&= \frac{N_c }{2} \int \frac{  \d x \d^2 k_{\bot}}{(2\pi)^3}   \frac{\chi_{^1\!A}(x, k_{\bot}^2)}{\bar{x} }  
 4 \frac{m_1-m_2}{D_{^1\!A,{\rm con}}} \frac{2}{ \w\cdot p}{B_1^{(2)}}\,.
\end{align}
 
Based on the theoretical results given above, we have the following discussions and findings:
\begin{itemize}
\item The manifest covariance of the CLF quark model requires that $[g_{V}]^{\lbd=0}=[g_{^{3(1)}\!A}]^{\lbd=0}=0$, which are equivalent to the conditions that
\begin{align}\label{eq:cc}
 \int   \d x \d^2 k_{\bot}  \frac{\chi_M(x, k_{\bot}^2)}{\bar{x} }  {B_1^{(2)}}=0\,,\qquad
 \int   \d x \d^2 k_{\bot}  \frac{\chi_M(x, k_{\bot}^2)}{\bar{x} }  \frac{B_1^{(2)}}{D_{M,\rm con}}=0 \,,
\end{align}
where $M=V$ or $^{3(1)}\!A$. The first equation is exactly the covariance condition presented in Ref.~\cite{Jaus:1999zv}, but has been found to be violated when the LF vertex function is used~({i.e.}, when the type-I replacement is applied)~\cite{Jaus:1999zv}. At the same time, we find that the second condition is also violated in this scheme. But interestingly, both of them can be satisfied by taking the additional $M\to M_0$ replacement, which implies that the covariance can be recovered within the type-II scheme. This can be clearly seen from the numerical examples shown in the next item. 

\item From Eqs.~\eqref{eq:gv}--\eqref{eq:ga1}, it can be found that $[g_{V}]^{\lbd=0}$ and $[g_{A}]^{\lbd=0}$ are both proportional to $1/(\w\cdot p)=1/p^+$, and hence their values, if being nonzero, would be reference-frame dependent. This implies that the size of covariance violation within the type-I scheme is in fact out of control. For convenience of numerical analyses, we take here the rest frame of mesons, which gives $p^+=M$. Then, comparing $[g_{V,A}]^{\lbd=0}$ with $[f_{V,A}]_{\rm full}^{\lbd=0,\pm}$~(or $\Delta^M_{\rm full}(x)$),  one can easily find that 
\begin{align}
[g_{V,A}]^{\lbd=0}= [f_{V,A}]_{\rm full}^{\lbd=0}-[f_{V,A}]_{\rm full}^{\lbd=\pm}=\int \d x \,\Delta^{V,A}_{\rm full}(x)\,,
\end{align}
From the figures for $\Delta^{V,A}_{\rm full}(x)$ shown in the last subsections, one can easily judge whether $[g_{V,A}]^{\lbd=0}=0$ within the type-I and the type-II scheme. Explicitly, using the results given in Tables~\ref{tab:fvdis} and \ref{tab:fA}, one can also obtain the following numerical results: 
\begin{align}\label{eq:ntI}
[g_{\rho,\,D^*,\,^1\!A_{(c\bar{q})},\,^3\!A_{(q\bar{q})},^3\!A_{(c\bar{q})}}]^{\lbd=0}&= (-40.2,\,-30.3,\,6.2,\,37.0,\,-12.0)\,{\rm MeV}\,\neq\, 0\,,\quad\text{(type-I)}\\
\label{eq:ntII}
[g_{\rho,\,D^*,\,^1\!A_{(c\bar{q})},\,^3\!A_{(q\bar{q})},^3\!A_{(c\bar{q})}}]^{\lbd=0}&=0\,,\quad \text{ (type-II)}
\end{align}
which confirms the findings presented in the last item. Here, it should be emphasized that the value of $[g_{V,A}]^{\lbd=0}$ within the type-I scheme is reference-frame dependent but is nonzero; while, the result $[g_{V,A}]^{\lbd=0}=0$ within the type-II scheme is independent of the reference frame. From the above analyses and numerical results, we can conclude that the problems of self-consistency and covariance of the CLF quark model within the type-I scheme might have the same origin, and both of them can be resolved simultaneously within the type-II scheme.  

\item In the heavy-quark limit, $M\sim m_Q \gg m_{\bar{q}}$, the momentum fractions carried by the constituents can be taken as $x\sim m_Q/M$ and $\bar{x} \sim m_q/M$. Keeping further in mind that the contributions to $f_{V,A}$ and $g_{V,A}$ are dominated by the momentum region $|k_{\bot}|\lesssim 1~{\rm GeV}$, one can find that $M_0 \simeq M $ would be a good approximation. Therefore, in the heavy-quark limit, the covariance of CLF results within the type-I scheme can be recovered, which has been found in Ref.~\cite{Jaus:1999zv}, and moreover, the CLF results in the type-I scheme would approach to that in the type-II scheme, which could be roughly inferred by comparing the numerical results $[g_{\rho}]^{\lbd=0}$ with $[g_{D^*}]^{\lbd=0}$, or $[g_{^3\!A_{(q\bar{q})}}]^{\lbd=0}$ with $[g_{^3\!A_{(c\bar{q})}}]^{\lbd=0}$, given in Eq.~\eqref{eq:ntI}.
\end{itemize}

The self-consistency problem of the CLF quark model is also correlated with the ambiguous decomposition of the hadronic matrix element. For the $\lbd=\pm$ polarization states, instead of using $\w\cdot\e_{\pm}=0$, one can also decompose the first term of Eq.~\eqref{eq:Shat} by using Eq.~\eqref{eq:deco} in the same manner as for the $\lbd=0$ mode. In this case, the self-consistency problem does not appear, but at the expense of introducing more unphysical decay constants related to the second and the last term in Eq.~\eqref{eq:deco}. Such an ambiguous decomposition becomes trivial only when the integrals related to the $B$ functions are zero, which can be achieved only in the type-II scheme.

From the above findings, one may find that the replacement $M\to M_0$ in the type-II scheme plays an important role in dealing with the problems of self-consistency, strict covariance and ambiguous decomposition of the hadronic matrix element. Possible reasons underlying such a replacement have been discussed in Refs.~\cite{Choi:2013mda,Choi:2014ifm,Choi:2017uos,Choi:2017zxn}. In the CLF quark model~\cite{Jaus:1999zv}, a manifestly covariant BS approach is used to guide the corresponding light-front calculation, but still using the same vertex functions as employed in the SLF quark model, because it is quite difficult to determine these functions by solving the QCD bound-state equation. A significant difference between the covariant BS approach and the SLF quark model is that the constituent quarks of a bound-state are allowed to be off mass-shell in the former, but must be on their respective mass-shell in the latter. In addition, the vertex operator used in the CLF quark model is in fact the Dirac structure of the spin-orbit WF in the SLF quark model, which can be clearly seen by comparing Eqs.~\eqref{eq:voCLF} with \eqref{eq:vSLF}. Note that the spin-orbit WF in the SLF quark model is obtained via the Melosh transformation and by assuming that the ``free'' and ``dressed'' constituents are on their respective mass-shell, implying therefore only the invariant mass squared of the constituents $M_0^2$ appears in the spin-orbit WF. Thus, once the vertex functions and operators mentioned above are used in the CLF quark model, it is quite reasonable to employ the replacement $M\to M_0$ for consistence, which reflects the fact that all the constituents are required to be on their respective mass-shell.

In other words, after the replacement $M\to M_0$ is applied, the CLF quark model with the LF vertex operators given by Eq.~\eqref{eq:voCLF} can be regarded as a covariant expression for the SLF quark model but with the zero-mode contributions taken properly into account. If this statement is correct, the valence contribution of the CLF quark model should be the same as the SLF one not only numerically but also formally, ${\cal O}_{\rm val.}={\cal O}_{\rm SLF}$, which is exactly confirmed in this paper. Besides of the reasons discussed above, the nontrivial role of $M\to M_0$ in improving the covariance and self-consistency of CLF quark model is also a strong but indirect evidence for the validity of this replacement. Finally, we should point out that the LF vertex operators given in Eq.~\eqref{eq:voCLF} as well as the corresponding vertex functions employed in this work are not the only option for the CLF quark model. Explicit forms of the vertex operators are generally model-dependent, and hence the replacement $M\to M_0$ would possibly be less or even not essential in other cases.

\section{SUMMARY}

In this paper, we have studied the self-consistency and covariance of SLF and CLF quark models via the decay constants of pseudoscalar, vector and axial-vector mesons, as well as the $P\to P$ weak transition form factors. For the CLF quark model, the type-I and the type-II correspondence [denoted by Eqs.~\eqref{eq:type1} and \eqref{eq:type2}, respectively] between the manifestly covariant approach and the LF quark model have been tested in detail. The main difference between these two schemes resides in whether the replacement $M\to M_0$ is applied only in the $D$ factor or in each and every term in the integrand. Our main findings and conclusions can be summarized as follows: 
\begin{itemize}
\item In the traditional type-I scheme, the CLF predictions for the decay constants of vector and axial-vector mesons suffer the self-consistency problem, which means that the results obtained via the $\lbd=0$ and $\lbd=\pm$ polarization states are different from each other, due to the additional contributions characterized by the coefficient $B_1^{(2)}$. A similar problem also exists in the traditional SLF quark model, with the findings that $ [f_{V,^1\!A,^3\!A}]_{\rm SLF}^{\lbd=0}\,\neq\,[f_{V,^1\!A,^3\!A}]_{\rm SLF}^{\lbd=\pm}$ (type-I). 

\item The CLF quark model with the type-II correspondence can, however, give self-consistent results for the decay constants, $ [f_{V,^1\!A,^3\!A}]_{\rm full}^{\lbd=0}\;\dot{=}\;[f_{V,^1\!A,^3\!A}]_{\rm full}^{\lbd=\pm}$ and  $ [f_{V,^1\!A,^3\!A}]_{\rm val.}^{\lbd=0}\;\dot{=}\; [f_{V,^1\!A,^3\!A}]_{\rm val.}^{\lbd=\pm}$ (type-II), because the integrations over terms associated with the coefficient $B_1^{(2)}$ vanish numerically after taking the replacement $M\to M_0$. At the same time, the same replacement is also required to obtain self-consistent results in the SLF quark model, $[f_{V,^1\!A,^3\!A}]_{\rm SLF}^{\lbd=0}\;\dot{=}\; [f_{V,^1\!A,^3\!A}]_{\rm SLF}^{\lbd=\pm}$ (type-II). 

\item For the decay constants and form factors studied in this paper, it is found that $f_{P}$ and $f_+(q^2)$ are free of the zero-mode contaminations; $f_{V,^3\!A}$ and $f_-(q^2)$ always receive the zero-mode contributions; but for $f_{^1\!A}$, the existence or absence of the zero-mode contributions depend on the choice of the polarization state $\lambda$. 

\item In the type-II scheme, the SLF quark model always gives identical results as the valence contributions of the CLF approach, $[{\cal Q}]_{\rm SLF} = [{\cal Q}]_{\rm val.}$. The zero-mode contributions to $f_{V,^1\!A,^3\!A}$ and $f_-(q^2)$ exist only formally but vanish numerically, leading therefore to $[{\cal Q}]_{\rm val.}\dot{=} [{\cal Q}]_{\rm full}$.

\item The manifest covariance of the CLF quark model is violated within the type-I scheme due to the same reason as for the self-consistency problem, but remarkably, can be recovered by taking the type-II correspondence. These two schemes are also consistent with each other in the heavy-quark limit.
\end{itemize}

The main findings in the type-II scheme mentioned above have been clearly summarized by Eqs.~\eqref{eq:find1} and \eqref{eq:relaSCLF}, as well as $[g_{V,A}]^{\lbd=0,\pm}=0$. It is expected that such a self-consistent and covariant scheme can be applied to discuss the other properties of hadrons, which will be exploited in the future.

\section*{Acknowledgment}

This work is supported by the National Natural Science Foundation of China (Grant Nos. 11875122, 11475055, 11675061, 11775092, 11521064 and 11435003), and the Program for Innovative Research Team in University of Henan Province (Grant No.19IRTSTHN018). X.L. is also supported in part by the self-determined research funds of CCNU from the colleges' basic research and operation of MOE~(CCNU18TS029).

\end{document}